\documentclass[11pt,a4paper]{article}
\pdfoutput=1

\usepackage[colorlinks=true, linkcolor=black!50!blue, urlcolor=blue, citecolor=blue, anchorcolor=blue]{hyperref}
\usepackage[font=small,labelfont=bf,margin=0mm,labelsep=period,tableposition=top]{caption}
\usepackage[a4paper,top=1.5cm,bottom=2cm,left=1.5cm,right=1.5cm,bindingoffset=0mm]{geometry}

\usepackage{placeins,setup,cite}
\usepackage{graphicx}
\usepackage{float}
\usepackage{afterpage}
\usepackage{epsfig}
\usepackage{amssymb}
\usepackage{amsmath, bm}
\usepackage{bm}
\usepackage{multirow}
\usepackage{url}
\usepackage{xcolor}
\usepackage{ulem}
\usepackage{url}
\usepackage{booktabs,multirow,colortbl}
\usepackage[symbol]{footmisc}
\usepackage[colorinlistoftodos]{todonotes}
\usepackage{listings}
\definecolor{CodeBlack}{HTML}{101018}
\definecolor{CodeGrey}{HTML}{404040}
\definecolor{CodeRed}{HTML}{bf212a}
\definecolor{CodeOrange}{HTML}{d08770}
\definecolor{CodeGreen}{HTML}{0b6500}
\definecolor{CodeBlue}{HTML}{4360a8}

\newcommand\YAMLcolonstyle{\color{red}\mdseries}
\newcommand\YAMLkeystyle{\color{black}\bfseries}
\newcommand\YAMLvaluestyle{\color{blue}\mdseries}

\makeatletter

\newcommand\language@yaml{yaml}

\expandafter\expandafter\expandafter\lstdefinelanguage
\expandafter{\language@yaml}
{
  keywords={true,false,null,y,n},
  keywordstyle=\color{darkgray}\bfseries,
  sensitive=false,
  comment=[l]{\#},
  morecomment=[s]{/*}{*/},
  commentstyle=\color{purple}\ttfamily,
  stringstyle=\YAMLvaluestyle\ttfamily,
  moredelim=[l][\color{orange}]{\&},
  moredelim=[l][\color{magenta}]{*},
  moredelim=**[il][\YAMLcolonstyle{:}\YAMLvaluestyle]{:},   
  morestring=[b]',
  morestring=[b]",
  literate =    {---}{{\ProcessThreeDashes}}3
                {>}{{\textcolor{red}\textgreater}}1     
                {|}{{\textcolor{red}\textbar}}1 
                {\ -\ }{{\mdseries\ -\ }}3,
}

\lst@AddToHook{EveryLine}{\ifx\lst@language\language@yaml\YAMLkeystyle\fi}
\makeatother

\newcommand\ProcessThreeDashes{\llap{\color{cyan}\mdseries-{-}-}}

\makeatletter
\def\thickhline{%
             \noalign{\ifnum0 =`}\fi\hrule \@height \thickarrayrulewidth \futurelet
             \reserved@a\@xthickhline}
\def\@xthickhline{\ifx\reserved@a\thickhline
                \vskip\doublerulesep
                \vskip -\thickarrayrulewidth
                \fi
                \ifnum0 =`{\fi}}
\makeatother
\newlength{\thickarrayrulewidth}
\usepackage{tikz}
\usetikzlibrary{positioning}
\usetikzlibrary{arrows.meta}
\usepackage{varwidth}
\usepackage{xcolor}
\definecolor{mtplotlib1}{HTML}{1f77b4}
\definecolor{mtplotlib2}{HTML}{ff7f0e}
\definecolor{mtplotlib3}{HTML}{2ca02c}
\definecolor{mtplotlib4}{HTML}{d62728}

\tikzset{%
  >={Latex[width=2mm,length=2mm]},
            base/.style = {rectangle, rounded corners, draw=black,
                           minimum width=4cm, minimum height=1cm,
                           text centered}, 
            mystyle/.style={rectangle, rounded corners, draw=black,
            minimum width=12cm, minimum height=1cm,
            text centered}, 
    col0/.style = {base, fill=white!30},
    col1/.style = {base, fill=mtplotlib1!30},
    col11/.style = {mystyle, fill=mtplotlib1!30},
    col2/.style = {base, fill=mtplotlib2!30},
    col3/.style = {base, fill=mtplotlib3!30},
    col4/.style = {base, minimum width=2.5cm, fill=mtplotlib4!15,}
}

\newcommand{\be}{\begin{equation}}
\newcommand{\ee}{\end{equation}}
\newcommand{\bea}{\begin{eqnarray}}
\newcommand{\eea}{\end{eqnarray}}
\newcommand{\bi}{\begin{itemize}}
\newcommand{\ei}{\end{itemize}}
\newcommand{\ben}{\begin{enumerate}}
\newcommand{\een}{\end{enumerate}}
\newcommand{\la}{\left\langle}
\newcommand{\ra}{\right\rangle}
\newcommand{\lc}{\left[}
\newcommand{\rc}{\right]}
\newcommand{\lp}{\left(}
\newcommand{\rp}{\right)}

\def\frac#1#2{{{#1}\over {#2}}}
\def\gsim{\mathrel{\rlap{\lower4pt\hbox{\hskip1pt$\sim$}}
    \raise1pt\hbox{$>$}}}       
\def\lsim{\mathrel{\rlap{\lower4pt\hbox{\hskip1pt$\sim$}}
    \raise1pt\hbox{$<$}}}

\newcommand{\art}{\mathrm{art}}
\newcommand{\rep}{\mathrm{rep}}

\newcommand{\draft}[1]{}

\def\beq{\begin{equation}}
\def\eeq{\end{equation}}

\numberwithin{equation}{section}
\numberwithin{figure}{section}
\numberwithin{table}{section}

\usepackage{tabularx}
\newcolumntype{C}[1]{>{\centering\arraybackslash}p{#1}}

\definecolor{darkblue}{rgb}{0.0,0,0.5}
\definecolor{darkgreen}{rgb}{0.0,0.3,0.0}
\definecolor{redish}{rgb}{0.675,0,0.2}
\definecolor{red}{rgb}{0.8,0,0}
\definecolor{green}{rgb}{0,0.6,0}
\definecolor{bluish}{rgb}{0.2,0.2,0.675}
\definecolor{mygrey}{rgb}{0.6,0.6,0.6}

\newcolumntype{Y}{>{\centering\arraybackslash}X}

\begin{document}
\vspace{-2.0cm}
\begin{flushright}
Edinburgh 2022/27 \\
Nikhef 2022-014 \\
TIF-UNIMI-2023-5 \\
\end{flushright}
\vspace{0.3cm}

\begin{center}
{\Large \bf Neutrino Structure Functions from GeV to EeV Energies  }\\
  \vspace{1.1cm}
  {\small
    Alessandro Candido$^1$, Alfonso Garcia$^{2,3}$, Giacomo Magni$^{4,5}$,
    Tanjona Rabemananjara$^{4,5}$,\\[0.1cm] Juan Rojo$^{4,5}$, and Roy Stegeman$^6$ 
  }\\
  
\vspace{0.7cm}

{\it \small
	~$^1$Tif Lab, Dipartimento di Fisica, Universit\`a di Milano and INFN, Sezione di Milano, \\
	Via Celoria 16, I-20133 Milano, Italy\\[0.1cm]
   	~$^2$Department of Physics and Laboratory for Particle Physics and Cosmology,\\
   	Harvard University, Cambridge, MA 02138, USA \\[0.1cm]
	~$^3$Instituto de Física Corpuscular (IFIC), Universitat de València (UV), 46980 Paterna, València, Spain.\\[0.1cm]
 	~$^4$Department of Physics and Astronomy, Vrije Universiteit, NL-1081 HV Amsterdam\\[0.1cm]
  	~$^5$Nikhef Theory Group, Science Park 105, 1098 XG Amsterdam, The Netherlands\\[0.1cm]
  	~$^6$The Higgs Centre for Theoretical Physics, University of Edinburgh, \\ JCMB, KB, Mayfield Rd, Edinburgh EH9 3JZ, Scotland\\[0.1cm]
 }

\vspace{1.0cm}

{\bf \large Abstract}

\end{center}

The interpretation of present and future neutrino experiments
requires accurate theoretical predictions for neutrino-nucleus scattering rates.
Neutrino structure functions
can be reliably evaluated in  the deep-inelastic scattering regime  within the
perturbative QCD (pQCD) framework.
At  low momentum transfers ($Q^2 \lsim {\rm few}$ GeV$^2$),
inelastic structure functions
are however affected by large uncertainties
 which distort event rate predictions for neutrino energies $E_\nu$ up to the TeV scale.
Here we present a determination of
neutrino inelastic structure functions valid for the complete range
of energies relevant for phenomenology, from the GeV region entering
oscillation analyses to the multi-EeV region accessible
at neutrino telescopes.
Our  {\sc\small NNSF$\nu$} approach combines a machine-learning parametrisation
of experimental data with pQCD calculations based on state-of-the-art analyses of proton and nuclear 
parton distributions (PDFs).
We compare our determination to other calculations,
in particular to the popular Bodek-Yang model.
We provide
updated predictions for inclusive cross sections for a range
of energies and target nuclei, including those relevant for LHC far-forward neutrino
experiments such as
FASER$\nu$, SND@LHC, and the 
Forward Physics Facility.
The {\sc\small NNSF$\nu$} determination is made available as
fast interpolation {\sc\small LHAPDF} grids, and
it can be accessed both
through an independent driver code and directly
interfaced to neutrino event generators such as {\sc\small GENIE}.

\clearpage

\tableofcontents

\section{Introduction}
\label{sec:introduction}

Precise and reliable theoretical predictions for the scattering rates of (anti-)neutrinos on proton and
nuclear targets~\cite{NuSTEC:2017hzk,Balantekin:2022jrq}
constitute a central ingredient for the interpretation of
a wide variety of ongoing and future neutrino experiments.
These include, first of all, oscillation measurements
carried out with reactor, accelerator,
and atmospheric neutrinos at facilities such as
KamLAND~\cite{KamLAND:2008dgz}, DUNE~\cite{DUNE:2015lol}, and
DeepCore/IceCube-Upgrade~\cite{IceCube:2014flw,IceCube-Gen2:2019fet} and KM3NET-ORCA~\cite{KM3NeT:2021ozk}
respectively.
Second, neutrino scattering experiments taking place at the CERN complex, from 
FASER$\nu$~\cite{Feng:2017uoz,FASER:2023zcr}, SND@LHC~\cite{SHiP:2020sos}, and the Forward
Physics Facility (FPF)~\cite{Anchordoqui:2021ghd,Feng:2022inv} using LHC neutrinos
to SPS beam dump experiments such as SHiP~\cite{Alekhin:2015byh}.
And third, astroparticle physics analyses involving high- and ultra-high-energy
 energies~\cite{Ackermann:2022rqc} at neutrino telescopes such as IceCube~\cite{IceCube:2013low}, KM3NET-ARCA~\cite{KM3Net:2016zxf}, GRAND~\cite{GRAND:2018iaj}, and POEMMA~\cite{POEMMA:2020ykm}.

Depending on the neutrino energies $E_\nu$ involved, different regimes are relevant
in the corresponding theoretical calculations.
In the sub-GeV region, the dominant interaction process is 
charged current quasielastic scattering (e.g. $\bar{\nu}_\mu p\to \mu^+ n$),
and then as $E_\nu$ is increased resonance scattering processes (e.g.
$\bar{\nu}_\mu n\to \mu^+ \Lambda^- \to  \mu^+ n \pi^-$) become the leading
contribution.
At higher energies and above the resonance region, starting at $E_\nu=\mathcal{O}(10)$ GeV
and final state invariant masses of $W\gsim 2$ GeV, inelastic scattering dominates
(e.g. $\bar{\nu}_\mu p\to \mu^+ X$,
with $X$ being the hadronic final state).
Inelastic neutrino scattering is further divided into shallow inelastic scattering (SIS)
and deep-inelastic scattering (DIS) involving momentum transfers $Q^2$
below and above the threshold $Q^2 \simeq {\rm few}$ GeV$^2$ which separates
the non-perturbative and perturbative regions, respectively.
Other interaction processes, typically subdominant but relevant in specific phase
space regions, include (in)elastic scattering off the photon field of nucleons, coherent scattering off the photon field of nuclei, and scattering on atomic electrons via the Glashow resonance.
Theoretical models of neutrino scattering are implemented in various 
neutrinos event generators~\cite{Mosel:2019vhx}, such as GENIE~\cite{Andreopoulos:2009rq,Andreopoulos:2015wxa} and its high-energy module HEDIS~\cite{Garcia:2020jwr}, GiBUU~\cite{Buss:2011mx}, and NuWro~\cite{Juszczak:2005zs}.
Most of these generators are
tailored to specific energies  and setups and cannot be straightforwardly applied
to the whole range of experiments listed above.

Differential cross sections in inelastic neutrino-nucleon scattering
are decomposed~\cite{Mangano:2001mj,Conrad:1997ne} in terms
of structure functions, $F^{\nu A}_i(x,Q^2)$,
with $x$ being the Bjorken variable and $Q^2$ the momentum transfer squared  between
the neutrino and the target nucleon.
In the DIS regime, these structure functions can be expressed
in the framework of perturbative QCD
as the factorised convolution of parton distribution functions
(PDFs)~\cite{Gao:2017yyd,Ethier:2020way,Kovarik:2019xvh} and hard-scattering
partonic cross sections.
Their state-of-the-art calculation is based
on PDFs and hard-scattering coefficient functions evaluated at 
next-to-next-to-leading order (NNLO) in the strong coupling $\alpha_s$ expansion,
with partial
and exact~\cite{Moch:2007rq,Moch:2008fj,Moch:2021qrk} results also available
one perturbative order higher (N$^3$LO) and used in~\cite{McGowan:2022nag} to extract the proton PDFs.
Furthermore, heavy quark (charm, bottom, and top) mass effects can be accounted for
by means of general-mass variable-flavour-number (GM-VFN)
schemes~\cite{Forte:2010ta,Guzzi:2011ew,Thorne:1997ga,Ball:2015tna}.
In addition, the applicability of fixed-order
perturbative QCD calculations can be extended to the large (small) $x$
kinematic region by means of all-order
threshold~\cite{Bonvini:2015ira}~(BFKL\cite{Bonvini:2016wki,Ball:2017otu}) 
resummation.

While DIS dominates inclusive
neutrino-nucleon event rates for energies  $E_{\nu }\gsim {\rm few}$ TeV,
at lower energies these rates receive a significant contribution from the SIS region.
For instance, at $E_{\nu} \sim 100$ GeV up to 20\% of the inclusive
cross section can arise from the  $Q \lsim 2$ GeV region~\cite{Bertone:2018dse}.
Theoretical predictions of neutrino structure functions in the SIS region
are therefore affected by much larger uncertainties than their DIS counterparts, given
that they are sensitive to low momentum transfers,
$Q^2 \lsim {\rm few~GeV}^2$~\cite{Reno:2006hj}, where the QCD perturbative and twist expansions break down
and the factorisation theorems stop being applicable.

In order to bypass the limitations of  perturbative QCD
in the SIS region,  phenomenological models
of  low-$Q^2$ neutrino structure functions have been developed
and implemented in various neutrino event generators.
One of the most popular is the Bodek-Yang (BY)
model~\cite{Yang:1998zb,Bodek:2002vp,Bodek:2003wd,Bodek:2004pc,Bodek:2010km,Bodek:2021bde}, 
based on structure functions
with effective leading-order (LO)
PDFs from the GRV98 analysis~\cite{Gluck:1998xa} with
modified scaling variables and $K$-factors to approximate mass and higher-order QCD corrections.
Drawbacks of the Bodek-Yang approach include the reliance on an obsolete set of
PDFs that neglects constraints on the proton and nuclear
structure obtained in the last 25 years, ignoring available
higher-order QCD calculations, and the lack of a systematic estimate of the 
uncertainties associated to their predictions.
Another restriction of the BY structure functions is that they
cannot be consistently matched to calculations
of high-energy  neutrino
scattering based on  modern PDFs and higher-order QCD 
calculations~\cite{Connolly:2011vc,Bertone:2018dse,Cooper-Sarkar:2007zsa,Cooper-Sarkar:2011jtt,Xie:2023suk},
introducing an unnecessary separation between the modelling
of neutrino interactions for experiments sensitive to different energy regions.

Here we present the first determination
of inelastic neutrino-nucleon
structure functions which is valid for the full range of  momentum
transfers $Q^2$ relevant for phenomenology, from oscillation measurements
involving  multi-GeV neutrinos 
to ultra-high energy scattering experiments at EeV energies.
This determination is based on the NNSF$\nu$ approach, which combines a
data-driven machine-learning parametrisation of inelastic structure functions at low 
and intermediate $Q^2$  matched to perturbative QCD calculations at larger $Q^2$ values.
Following the NNPDF methodology~\cite{DelDebbio:2007ee,NNPDF:2017mvq,Ball:2008by,Ball:2010de,Ball:2012wy,NNPDF:2014otw,NNPDF:2021njg,NNPDF:2021uiq}, already
applied to neutral-current structure functions in~\cite{DelDebbio:2004xtd,Forte:2002fg},
neural networks are adopted as universal 
unbiased interpolants and trained on available accelerator data on neutrino structure functions
and differential cross sections, with
the Monte Carlo replica method to estimate and propagate uncertainties.
The perturbative QCD calculations are provided by {\sc\small YADISM}, a new framework
for the evaluation of DIS structure functions from the {\sc\small EKO}~\cite{Candido:2022tld}
family.
Their input is the nNNPDF3.0 determination of nuclear PDFs~\cite{AbdulKhalek:2022fyi},
which reduces in the $A=1$ limit to the proton NNPDF3.1 fit~\cite{NNPDF:2017mvq}, specifically to
the variant with LHCb $D$-meson data included~\cite{Bertone:2018dse}.
This {\sc\small YADISM} perturbative calculation is applicable up to EeV neutrino energies
and accounts for exact top mass effects in charged-current scattering.

The  NNSF$\nu$ determination hence provides a parametrisation
of the  structure functions $F_i^{\nu A}(x,Q^2,A)$ ($i=2,3,L$) reliable for arbitrary values
of the three inputs $x$, $Q^2$, and $A$ together with a comprehensive uncertainty
estimate.
Our strategy provides a robust estimate of all relevant sources of
uncertainty, including those related to experimental 
errors, functional forms, (nuclear) PDFs, and
missing higher order (MHO) uncertainties in the perturbative QCD calculation.
We demonstrate the stability of  NNSF$\nu$ with respect to variations of both
the input data and methodological settings, and show how it correctly inter- and extrapolates
for $A$ values not directly constrained in the fit.
Upon integration of the structure functions over the kinematically allowed ranges
in $x$ and $Q^2$, we obtain predictions for inclusive inelastic cross sections,
together
with the associated uncertainties, without
restrictions on the values of the neutrino energy $E_\nu$ and  mass number $A$
of the target nuclei.

We compare the NNSF$\nu$ structure functions and inclusive
cross sections to related calculations available in the literature, in particular to
the Bodek-Yang (as implemented in GENIE), BGR18~\cite{Bertone:2018dse},
and 
 CSMS11~\cite{Cooper-Sarkar:2011jtt} predictions.
 We assess the dependence of our results with respect to the values of $E_\nu$ and $A$
 and provide dedicated predictions for the energy ranges and target materials
 relevant for the LHC far-forward neutrino scattering experiments, namely FASER$\nu$, SND@LHC, and the FPF,
  quantifying in each case the role played by
 the various sources of uncertainty.
 The  NNSF$\nu$  determination is made available
 as stand-alone fast interpolation grids in the {\sc\small LHAPDF}~\cite{Buckley:2014ana}
 format which can
  be accessed either through an independent driver code and or directly interfaced to neutrino
 event generators such as GENIE.
 We also provide look-up tables for the inclusive cross sections and their
 uncertainties as a function of $E_\nu$ and $A$.

The outline of this paper is the following.
In Sect.~\ref{sec:theory} we review the theoretical
formalism underlying neutrino-nucleus inelastic scattering on proton and nuclear targets,
assess the perturbative stability of the QCD calculation,
 and compare different existing predictions among them.
Sect.~\ref{sec:fitting} describes the NNSF$\nu$  approach to construct a data-driven parametrisation
of the neutrino structure functions matched to pQCD calculations.
Sect.~\ref{sec:results} presents the NNSF$\nu$ determination,
verifies that it reproduces the input data and theory calculations,
studies its robustness and stability, compares it with existing analyses, and
evaluates the Gross-Llewellyn Smith sum rule.
Predictions for inclusive neutrino cross sections are provided in Sect.~\ref{sec:inclusive_xsec},
in particular
we study their $E_\nu$ and $A$ dependence, the agreement with
experimental data, and the sensitivity to kinematic cuts,
and we conclude by outlining some possible future developments
in Sect.~\ref{sec:summary}.

Technical details are collected in three appendices.
App.~\ref{app:delivery} describes the software framework
underlying the  NNSF$\nu$ determination
and how to access and use its results for the  inelastic structure functions
and cross sections.
App.~\ref{app:yadism} outlines
the main features of the {\sc\small YADISM} code
used  to calculate neutrino structure functions in perturbative QCD
and reports the outcome of representative benchmark comparison with {\sc\small APFEL}.
App.~\ref{app:nuclear} summarises the impact of nuclear effects in neutrino inelastic scattering
at the level of the parton distributions provided by nNNPDF3.0.

\section{Neutrino inelastic structure functions}
\label{sec:theory}

In this section we summarise the theoretical formalism
underpinning the evaluation of neutrino-nucleon inelastic scattering cross sections
in terms of structure functions and the calculation of the latter in the framework of
 perturbative QCD.
We then present results  neutrino structure functions evaluated with
{\sc\small YADISM}, quantify their perturbative stability
and their dependence of the input PDFs, 
and compare them with the Bodek-Yang and BGR18 predictions.

\subsection{DIS structure functions in perturbative QCD}
\label{sec:general_formalism}

The double-differential cross section for neutrino-nucleus scattering
can be decomposed
in terms of three independent structure functions $F_i^{\nu A}(x,Q^2)$ with $i=1,2,3$.
Focusing on the charged-current (CC) scattering case mediated
by the exchange of a $W^+$ weak boson, this differential
 cross section reads
\be
\label{eq:neutrino_DIS_xsec}
\frac{d^2\sigma^{\nu A}(x,Q^2,y)}{dxdy} =  \frac{G_F^2s/2\pi}{\lp 1+Q^2/m_W^2\rp^2}\lc (1-y)F^{\nu A}_2(x,Q^2) + y^2xF^{\nu A}_1(x,Q^2) + y\lp 1-\frac{y}{2}\rp xF^{\nu A}_3(x,Q^2)\rc  \, ,
\ee
where $s=2m_n E_\nu$ is the neutrino-nucleon center of mass energy squared, $m_n$ the nucleon mass,
$A$ the atomic mass number of the target nucleus,
$E_\nu$ the incoming neutrino energy,
and the inelasticity $y$ is defined as
\be
y=\frac{Q^2}{2x m_n E_{\nu}}=\frac{Q^2}{xs} \, .
\ee
An analogous expression holds for  antineutrino 
scattering, mediated now
by the exchange of a $W^-$ weak boson, with the only difference being a sign change
in front of the parity-violating structure function $xF_3$,
\be
\label{eq:antineutrino_DIS_xsec}
\frac{d^2\sigma^{\bar{\nu} A}(x,Q^2,y)}{dxdy} =  \frac{G_F^2s/2\pi}{\lp 1+Q^2/m_W^2\rp^2}\lc (1-y)F^{\bar{\nu} A}_2(x,Q^2) + y^2xF^{\bar{\nu} A}_1(x,Q^2) - y\lp 1-\frac{y}{2}\rp xF^{\bar{\nu} A}_3(x,Q^2)\rc  \, .
\ee
While the differential cross sections are a function of three kinematic variables, $(x,Q^2,y)$,
the structure functions themselves depend only on $x$ and $Q^2$.
Furthermore, both the cross sections and the structure functions depend
on the atomic mass number $A$ of the target nucleus only through the nuclear modifications of the free-nucleon
structure functions.
Kinematic considerations indicate that inelastic structure functions vanish
in the elastic limit $x\to 1$, that is,
\be
F_i^{\nu A}(x=1,Q^2)=F_i^{\bar{\nu} A}(x=1,Q^2)=0 \, , \qquad \forall\,\, i \, .
\ee

Alternatively, Eq.~(\ref{eq:neutrino_DIS_xsec}) can be expressed in terms
of the longitudinal structure function  $F_L^{\nu A}(x,Q^2)$ defined
by $F_L = F_2-2xF_1$, leading to
\be
\label{eq:neutrino_DIS_xsec_FL}
\frac{d^2\sigma^{\nu A}(x,Q^2,y)}{dxdy} =  \frac{G_F^2s/4\pi}{\lp 1+Q^2/m_W^2\rp^2}\lc Y_+F^{\nu A}_2(x,Q^2) - y^2F^{\nu A}_L(x,Q^2) +Y_- xF^{\nu A}_3(x,Q^2)\rc  \, ,
\ee
where $Y_\pm = 1 \pm (1-y)^2$ and with the counterpart expression for anti-neutrino scattering,
\be
\label{eq:antineutrino_DIS_xsec_FL}
\frac{d^2\sigma^{\bar{\nu} A}(x,Q^2,y)}{dxdy} =  \frac{G_F^2s/4\pi}{\lp 1+Q^2/m_W^2\rp^2}\lc Y_+F^{\bar{\nu} A}_2(x,Q^2) - y^2F^{\bar{\nu} A}_L(x,Q^2) -Y_- xF^{\bar{\nu} A}_3(x,Q^2)\rc  \, ,
\ee
Expressing the differential cross section as in
Eqns.~(\ref{eq:neutrino_DIS_xsec_FL})-(\ref{eq:antineutrino_DIS_xsec_FL}) is advantageous
because in the parton model (and in perturbative QCD at leading order) the longitudinal
structure function vanishes, and hence $F_L^{\nu A}(x,Q^2)\ne 0$ starting only at NLO.
The combination of neutrino and antineutrino measurements makes it possible to disentangle
the different structure functions, for example the cross-section difference
\be
\label{eq:neutrino_DIS_xsec_FL_diff}
\frac{d^2\sigma^{\nu A}(x,Q^2,y)}{dxdy} - \frac{d^2\sigma^{\bar{\nu} A}(x,Q^2,y)}{dxdy}=  \frac{G_F^2s Y_-}
{4\pi \lp 1+Q^2/m_W^2\rp^2}\lc    xF^{\nu A}_3(x,Q^2)+  xF^{\bar{\nu} A}_3(x,Q^2)\rc \, ,
\ee
is proportional to the parity-violating structure function $xF_3$ averaged over neutrinos
and antineutrinos.

As discussed in the introduction,
depending on the values of the momentum transfer squared $Q^2$ and of the hadronic final-state
invariant mass $W$,
\be
W^2 = m_N^2 + Q^2 \frac{(1-x)}{x} \, ,
\ee
different processes contribute to these neutrino structure functions.
In this work we consider only inelastic scattering, defined by the condition
that the hadronic state invariant mass satisfies $W^2 \gsim 3.5~\rm{GeV}^2$ to avoid the resonance
region.
In the DIS regime, where $Q^2 \gsim {\rm few~GeV}^2$,
neutrino structure functions can be evaluated in perturbative QCD in terms of a factorised
convolution of process-dependent partonic scattering cross sections and
of process-independent parton distribution functions,
\be
\label{eq:sfs_pqcd}
 F^{\nu A}_i(x,Q^2) = \sum_{j=q,\bar{q},g}\int_x^1 \frac{dz}{z}\, C_{i,j}^{\nu N}(z,\alpha_s(Q^2))f^{(A)}_j\lp \frac{x}{z},Q^2\rp \, , \quad i = 2,3,L \, ,
 \ee
 where $j$ is an index that runs over all possible partonic initial states,
 $C_{i,j}^{\nu N}$ is the process-dependent (but target-independent) coefficient function,
 and $f_j^{(A)}$ indicates the PDFs of the average nucleon bounded into a nucleus with mass number $A$.

 DIS coefficient functions can be expressed  as a series expansion in powers of the strong
 coupling $\alpha_s(Q^2)$,
 \be
 \label{eq:coeff_fun_expansion}
 C_{i,j}^{\nu N}(z,\alpha_s(Q^2)) = \sum_{k=0}^m \lp \alpha_s(Q^2) \rp^{k} C_{i,j}^{\nu N(k)}(z) \, .
 \ee
 The leading-order ($k=0$)  term in the coefficient function expansion  Eq.~(\ref{eq:coeff_fun_expansion})
 is independent of $\alpha_s$ for $i=2,3$ since the Born scattering
 is mediated by the weak interaction.
 For massless quarks, charged-current neutrino DIS coefficient functions have been evaluated up to N3LO
 (third-order, $m=3$) in~\cite{Moch:2007rq,Moch:2008fj}.
 For massive quarks, the calculation of strange-to-charm transitions
 with charm mass effects has been performed at NNLO
 (second-order, $m=2$) in~\cite{Gao:2017kkx}.
 Mass effects can be incorporated in the massless calculation by means
 of a general-mass variable-flavour-number scheme~\cite{Forte:2010ta,Guzzi:2011ew,Thorne:1997ga,Ball:2015tna}.
 For neutrino structure functions the  expansion Eq.~(\ref{eq:coeff_fun_expansion})
 displays good perturbative converge unless either  $Q^2$ approaches
 the boundary of the non-perturbative region, $Q^2 \simeq 1$ GeV$^2$, or the Bjorken-$x$ variable
 becomes small enough to be sensitive to BFKL corrections.
 As well know, both PDFs and coefficient function are scheme-dependent objects and only
 their combination Eq.~(\ref{eq:sfs_pqcd}) is scheme-independent (up to higher orders).
  
 Each of the neutrino and anti-neutrino structure functions in Eq.~(\ref{eq:sfs_pqcd})
 depends on a different combination of quark and antiquark PDFs,
 bringing in  unique
 sensitivity to quark flavour separation in nucleons and nuclei.
 To illustrate this, if we consider a LO calculation on a proton target with
$n_f=4$ active quark flavours, 
 neglect heavy quark mass effects, and assume a diagonal CKM matrix,
 one can express the $F_2^{\nu p}$ and $xF_3^{\nu p}$ structure functions as
 \bea
 F_2^{\nu p}(x,Q^2) &=& 2x\lp f_{\bar{u}} + f_{d} + f_{s} + f_{\bar{c}} \rp(x,Q^2) \, , \nonumber  \\
 F_2^{\bar{\nu} p}(x,Q^2) &=& 2x\lp f_u + f_{\bar{d}} + f_{\bar{s}} + f_c \rp(x,Q^2) \, , \label{eq:neutrinoSFs} \\
 xF_3^{\nu p}(x,Q^2) &=& 2x\lp -f_{\bar{u}} + f_d +f_s - f_{\bar{c}}\rp(x,Q^2)  \, , \nonumber\\
 xF_3^{\bar{\nu} p}(x,Q^2) &=& 2x\lp f_u - f_{\bar{d}} -f_{\bar{s}} + f_{c}\rp(x,Q^2) \, , \nonumber
 \eea
 where $f_{q}$ indicates the proton PDFs.
 The corresponding expressions for a neutron target and or isoscalar target are obtained
 from isospin symmetry, for instance the
 neutrino-neutron structure functions are expressed in terms of the proton PDFs as
 \bea
 F_2^{\nu n}(x,Q^2) &=& 2x\lp f_{\bar{d}} + f_{u} + f_{s} + f_{\bar{c}} \rp(x,Q^2) \, , \nonumber  \\
 F_2^{\bar{\nu} n}(x,Q^2) &=& 2x\lp f_d + f_{\bar{u}} + f_{\bar{s}} + f_c \rp(x,Q^2) \, , \label{eq:antineutrinoSFs} \\
 xF_3^{\nu n}(x,Q^2) &=& 2x\lp -f_{\bar{d}} + f_u +f_s - f_{\bar{c}}\rp(x,Q^2)  \, , \nonumber\\
 xF_3^{\bar{\nu} n}(x,Q^2) &=& 2x\lp f_d - f_{\bar{u}} -f_{\bar{s}} + f_{c}\rp(x,Q^2) \, . \nonumber
 \eea
Different combinations of the neutrino structure functions 
in Eqns.~(\ref{eq:neutrinoSFs}) and~(\ref{eq:antineutrinoSFs}) are sensitive
to different PDF combinations.
For instance, assume 
an isoscalar target and neglect nuclear corrections.
In such scenario one has that
\bea
\label{eq:xf3_isoscalar}
xF_3^{\nu A}&=&\lp xF_3^{\nu p}+xF_3^{\nu n}\rp /2 = x\lp f_{u_V} + f_{d_V} + f_{s_V} + f_{c_V} \rp \, ,  \\
xF_3^{\bar{\nu} A}&=&\lp xF_3^{\bar{\nu} p}+xF_3^{\bar{\nu} n}\rp/ 2 =  x\lp f_{u_V} + f_{d_V} + f_{s_V} + f_{c_V} \rp
\nonumber\, ,
\eea
expressed in terms of the valence PDF combinations, e.g. $f_{u_V}=f_u - f_{\bar{u}}$,
 and hence the cross-section difference Eq.~(\ref{eq:neutrino_DIS_xsec_FL_diff}) yields
 \be
\label{eq:neutrino_DIS_xsec_FL_diff_LO}
\frac{d^2\sigma^{\nu A}(x,Q^2,y)}{dxdy} - \frac{d^2\sigma^{\bar{\nu} A}(x,Q^2,y)}{dxdy}=
 \frac{G_F^2s Y_-}
{2\pi \lp 1+Q^2/m_W^2\rp^2}\lc xf_{u_V} + xf_{d_V}+xf_{s_V} + xf_{c_V}   \rc\, .
\ee
The same result is obtained if the target is not isoscalar but rather a purely hydrogen target,
\be
\frac{d^2\sigma^{\nu p}(x,Q^2,y)}{dxdy} - \frac{d^2\sigma^{\bar{\nu} p}(x,Q^2,y)}{dxdy}=  \frac{G_F^2s Y_-}
{2\pi \lp 1+Q^2/m_W^2\rp^2}\lc xf_{u_V} + xf_{d_V}+xf_{s_V} + xf_{c_V}   \rc \, ,
\ee
indicating how the difference between neutrino and antineutrino parity-violating
structure functions $xF_3$ in Eq.~(\ref{eq:neutrino_DIS_xsec_FL_diff}) is a sensitive probe
of the valence quark content in protons and nuclei.

The neutrino structure function $xF_3^{\nu A}$ must also satisfy the
Gross-Llewellyn Smith (GLS) sum rule~\cite{Gross:1969jf}
calculable in  perturbative QCD.
For an isoscalar  target the GLS sum rule is given by
\be
\label{eq:GLS_sumrule}
\int_0^1 \frac{dx}{x} xF_3^{\nu A}(x,Q^2) = 3\lp 1+ \sum_{k=1}^3 \lp \frac{\alpha_s(Q^2)}{\pi}\rp^k c_k(n_f)\rp \, ,
\ee
where $n_f$ is the number of active flavours at the scale $Q^2$ and the coefficients $c_k$
have been computed.
The same expression holds for the anti-neutrino counterpart.
The leading-order contribution to Eq.~(\ref{eq:GLS_sumrule}) follows from the
partonic decomposition of the isoscalar $xF_3^{\nu A}$ in terms of the valence quark
PDFs, Eq.~(\ref{eq:xf3_isoscalar}).

In this work we do not impose the GLS sum rule
in the  data-driven fit and instead verify {\it a posteriori}
that it is satisfied within uncertainties
 in the region of applicability of perturbative QCD.
We note that experimentally one cannot access the $x\to 0$ region,
and hence the evaluation of Eq.~(\ref{eq:GLS_sumrule}) depends
on the modelling of the small-$x$ extrapolation region for the neutrino structure functions.

\subsection{PDF dependence and perturbative stability}
\label{sec:dis-sf}

Here we study the PDF dependence and perturbative stability
of neutrino DIS structure functions.
We focus on the $x$ region relevant for scatterings
involving neutrino energies of  $E_{\nu} \lsim 1$ TeV
and momentum transfers of $Q \gsim 2$ GeV.
Recall that from  DIS kinematics, for given values of $Q^2$ and $E_{\nu}$
the Bjorken-$x$ variable
satisfies
\be
x \ge \frac{Q^2}{2m_n E_{\nu}} \, ,
\ee
and hence it suffices to consider $x\gsim  10^{-3}$.
We compare the following structure function calculations:

\begin{itemize}

\item {\sc\small YADISM}.
  The {\sc\small YADISM} 
  package described in  App.~\ref{app:yadism} evaluates DIS charged-lepton and neutrino
  inclusive and heavy quark
structure functions up to NNLO, and up to N$^3$LO whenever available.
{\sc\small YADISM} has been benchmarked with  {\sc\small APFEL}~\cite{Bertone:2013vaa}
finding good agreement.
Heavy quark mass effects are implemented in various schemes,
including the zero-mass variable-flavour-number (ZM-VFN) scheme,
the fixed-flavour-number (FFN) scheme, and
the FONLL general-mass
variable-flavour-number scheme~\cite{Forte:2010ta,Ball:2011mu}.

For the purpose of the benchmarking comparisons shown in this section,
{\sc\small YADISM} inclusive neutrino structure functions 
are evaluated using either LO, NLO, and NNLO coefficient functions and 
in all cases
NNPDF4.0 NNLO as input PDF set, with FONLL to account for heavy quark mass effects at NLO accuracy.
We denote these calculations as {\sc\small YADISM}-LO, {\sc\small YADISM}-NLO,
and {\sc\small YADISM}-NNLO respectively in the following.

As will be discussed in Sect.~\ref{sec:fitting}, for the
NNSF$\nu$ determination of neutrino structure functions the perturbative QCD
baseline calculation from {\sc\small YADISM} will instead use NLO coefficient functions,
nNNPDF3.0 NLO as input PDF sets for all targets including hydrogen,
and a 5FNS where top quark mass effects
are accounted for exactly while charm and bottom mass effects are neglected.

\item {\sc\small BGR18}.
  This is the calculation of neutrino DIS structure functions first presented in~\cite{Bertone:2018dse}
  in the context of predictions for UHE neutrino-nucleus cross sections
  and then updated in~\cite{Garcia:2020jwr} when evaluating attenuation rates for
  UHE neutrinos propagating within Earth matter.
  {\sc\small BGR18} is based on {\sc\small APFEL} with either NNPDF3.1~\cite{NNPDF:2017mvq}
  or NNPDF3.1+LHCb~\cite{Gauld:2016kpd} as input PDF sets.
  Nuclear corrections from the nNNPDF2.0 determination~\cite{AbdulKhalek:2020yuc} were used in~\cite{Garcia:2020jwr}
  to account for deviations with respect to the free-nucleon calculation.
  Several variants of the BGR18 calculation are available, 
  both at fixed-order QCD (NLO and NNLO) and with
  BFKL resummation (NLO+NLL$x$ and NNLO+NLL$x$).
  
  In this work we consider the variant of {\sc\small BGR18} based on
  NLO coefficient functions and NNPDF3.1 NLO as input PDF,
  as implemented in the {\sc\small HEDIS} module of  {\sc\small GENIE}.
  Since NNPDF3.1 and NNPDF4.0 are consistent within PDF uncertainties,
  one expects agreement between {\sc\small BGR18} and {\sc\small YADISM}-NLO.

\item {\sc\small Bodek-Yang}.
  The BY
  calculation~\cite{Yang:1998zb,Bodek:2002vp,Bodek:2003wd,Bodek:2004pc,Bodek:2010km,Bodek:2021bde}
  is a phenomenological model for inelastic neutrino- and electron-nucleon scattering cross sections based on effective leading order PDFs which can be applied from intermediate-$Q^2$
  in the DIS region down to
  the photo-production
  region $Q^2\simeq 0$.
  The starting point is the GRV98 LO PDF set~\cite{Gluck:1998xa}
  evaluated at a modified scaling variable $\xi_w$ replacing the standard Bjorken-$x$,
  which allows BY to be extended to the low-$Q^2$ non-perturbative region.
  Several phenomenological corrections are applied
  to approximate NLO QCD and nuclear effects.
  In the following, we display the BY structure functions
  as implemented in the {\sc\small GENIE} event generator.

\item {\sc\small LO-SF}.
  This calculation is defined by the LO expression of
  neutrino DIS structure functions on a proton target, Eq.~(\ref{eq:neutrinoSFs}),
  with  PDFs accessed
  directly from the {\sc\small LHAPDF} interface~\cite{Buckley:2014ana}.
  We consider two variants.
  First, {\sc\small LO-SF-NNPDF4.0}, which uses NNPDF4.0 NNLO as input
  and that should coincide with the {\sc\small YADISM}-LO calculation.
  Second, {\sc\small LO-SF-GRV98}, which adopts GRV98LO as input
  PDF set and that should reproduce  the Bodek-Yang calculation
  in the large $Q^2$ limit where its scaling variable $\xi_w$ reduces
  to $x$.
  The {\sc\small LO-SF-NNPDF4.0} and  {\sc\small LO-SF-GRV98} calculations
  are only meant for benchmarking purposes
  and will not be used beyond this section.
  
\end{itemize}

The settings of the
inelastic structure functions calculations that we just  described
are summarised in Table~\ref{tab:dis_sf}.
In each case we indicate the input PDF set used, the perturbative accuracy
of the DIS coefficient functions, the software tool used for its evaluation,
the treatment of heavy quark mass effects and its  region of applicability.
Mass effects and higher-order
QCD corrections are approximated in the {Bodek-Yang}
calculation by means of phenomenological model parameters.

\begin{table}[t]
  \centering
  \small
  \renewcommand{\arraystretch}{1.60}
\begin{tabularx}{\textwidth}{Xccccc}
\toprule
Calculation &  PDF set  &  QCD Accuracy &  Code  & Mass effects  & Validity \\
\midrule
    {\sc\small YADISM}-LO  &  NNPDF4.0~NNLO    &  LO & {\sc\small YADISM}    &  FONLL     &  $Q \ge 1.65$ GeV  \\
    {\sc\small YADISM}-NLO  &  NNPDF4.0~NNLO    &  NLO & {\sc\small YADISM}    &  FONLL     &  $Q \ge 1.65$ GeV  \\
    {\sc\small YADISM}-NNLO  &  NNPDF4.0~NNLO    &  NNLO & {\sc\small YADISM}    &  FONLL     &  $Q \ge 1.65$ GeV  \\
    \midrule
        {\sc\small BGR18} &  NNPDF3.1~NLO  & NLO & {\sc\small APFEL}~({\sc\small GENIE})  &  FONLL     &  $Q \ge 1.65$ GeV  \\
         \midrule
       {\sc\small Bodek-Yang} &  GRV98~LO  & LO & {\sc\small GENIE}  &  pheno model     &  $Q \gsim 0$  \\
         \midrule
 {\sc\small LO-SF-NNPDF4.0} &  NNPDF4.0~NNLO  & LO & {\sc\small LHAPDF}  &  no     &  $Q \ge 1.65$ GeV   \\
    {\sc\small LO-SF-GRV98} &   GRV98~LO  & LO & {\sc\small LHAPDF}  &  no     &  $Q \ge 0.9$ GeV   \\
  \bottomrule
\end{tabularx}
\vspace{0.2cm}
\caption{\small Settings of the
  calculations of inelastic neutrino structure functions on a proton target considered
  in this section.
  We indicate the input PDF set used, the perturbative accuracy
  of the DIS coefficient functions, the software tools used for its evaluation,
  the treatment of heavy quark mass effects, and its region of applicability.
  The {\sc\small Bodek-Yang} and {\sc\small BGR18} calculations are obtained from  their
  implementation
  in {\sc\small GENIE}.
  \label{tab:dis_sf}
}
\end{table}

\begin{figure}[t]
 \centering
 \includegraphics[width=0.95\linewidth]{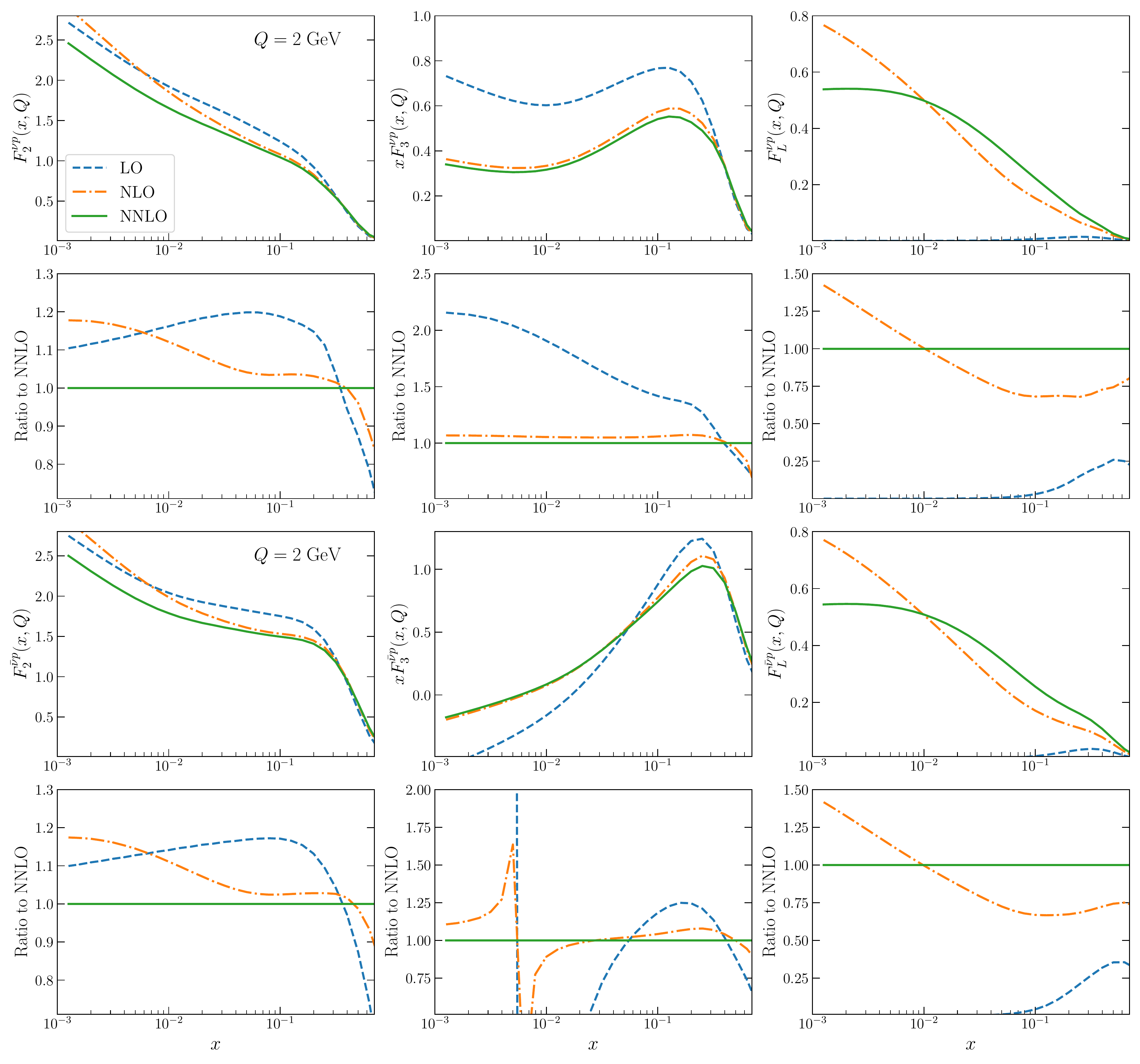}
 \caption{\small Comparison of {\sc\small YADISM}-LO, {\sc\small YADISM}-NLO,
   and {\sc\small YADISM}-NNLO calculations of the $F_2^{\nu p}$, $xF_3^{\nu p}$,
   and $F_L^{\nu p}$ structure functions and their antineutrino counterparts as a function of $x$
   for $Q=2$ GeV.
   The three calculations use the central set of NNPDF4.0 NNLO as input
   and do not display the PDF uncertainties.
   We show both the absolute structure functions and their ratios
   to the NNLO calculation.
 }    
 \label{fig:StructureFunction-PerturbativeStab-xdep-allcomp_q2gev}
\end{figure}

\begin{figure}[t]
 \centering
 \includegraphics[width=0.95\linewidth]{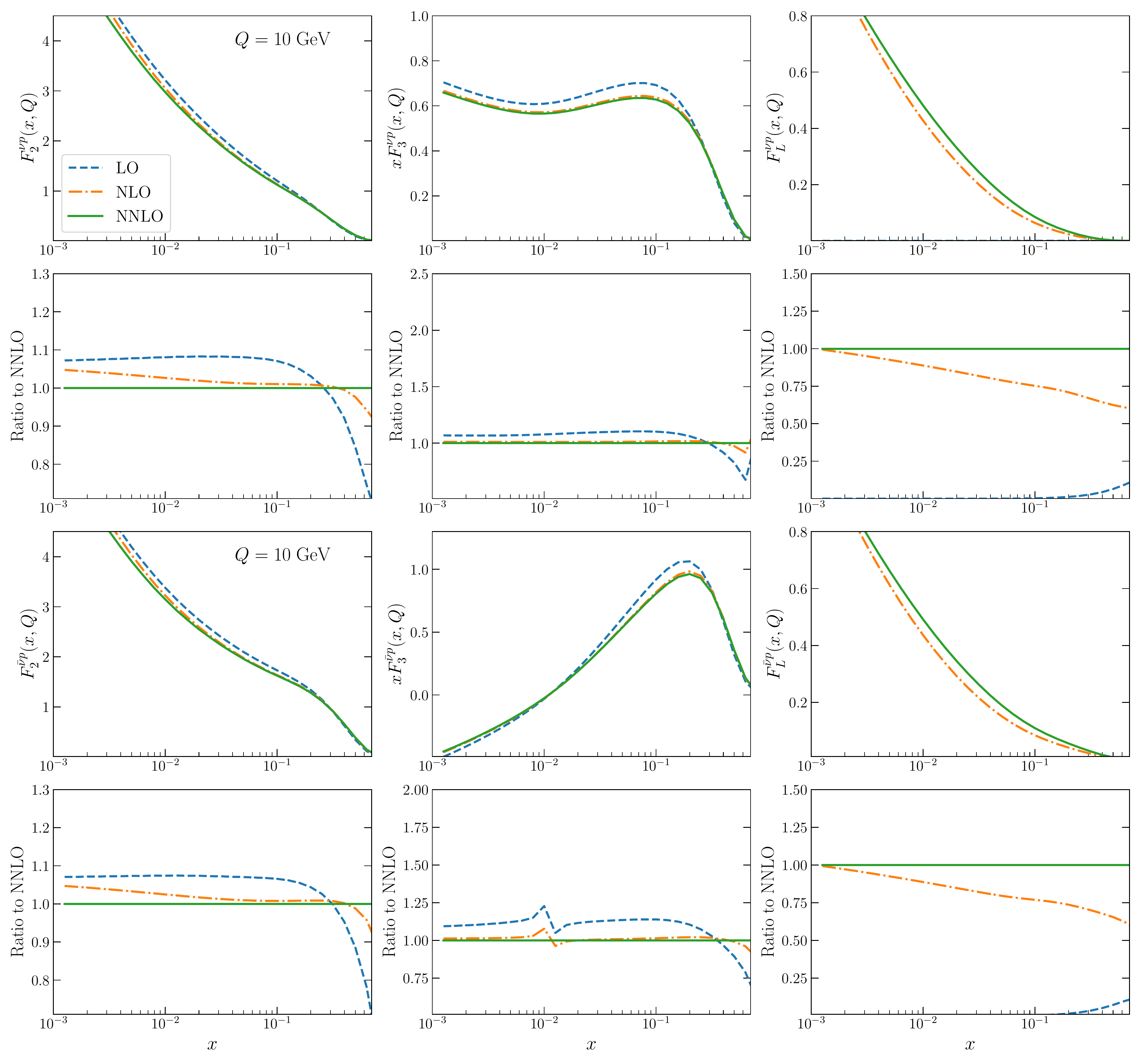}
 \caption{\small Same as Fig.~\ref{fig:StructureFunction-PerturbativeStab-xdep-allcomp_q2gev}
   for $Q=10$ GeV.
 }    
 \label{fig:StructureFunction-PerturbativeStab-xdep-allcomp_q10gev}
\end{figure}

\paragraph{Perturbative stability.}
Figs.~\ref{fig:StructureFunction-PerturbativeStab-xdep-allcomp_q2gev}
and~\ref{fig:StructureFunction-PerturbativeStab-xdep-allcomp_q10gev}
display
the {\sc\small YADISM}-LO, {\sc\small YADISM}-NLO,
and {\sc\small YADISM}-NNLO calculations of $F_2^{\nu p}$, $xF_3^{\nu p}$,
and $F_L^{\nu p}$ and their antineutrino counterparts as a function of $x$
for $Q=2$ GeV and 10 GeV respectively.
As indicated in Table~\ref{tab:dis_sf}, in all cases the common PDF
set NNPDF4.0 NNLO is used.
We display both the absolute structure functions and their ratios
to the NLO calculation, and focus on the $x$ region relevant
for DIS structure functions with $E_\nu \lsim 1$ TeV.

It is found that for $Q=2$ GeV higher-order QCD corrections
are in general significant and exhibit a  similar pattern both for neutrinos and for antineutrinos.
For the dominant $F_2$ structure function, the LO calculation underestimates
at large-$x$ the NNLO result for up to 25\%, while for $x\lsim 0.2$ it overestimates
it by more than 20\%.
NLO corrections reduce these differences at medium and large-$x$, but
for $x\simeq 10^{-3}$ the NLO structure functions still overshoot the NNLO result by 20\%.
Since PDF uncertainties in  $F_2$ are at the few percent level at most
(see below), at $Q=2$ GeV missing higher order uncertainties (MHOU)
are the main source of theory errors.
Concerning the parity-violating structure function $xF_3$, for neutrino beams
the LO calculation overestimates the NNLO result by 50\%
at $x=0.1$ and by a factor 2 for $x=10^{-2}$.
A similar pattern is observed
for antineutrinos, with now LO becoming  more negative at small-$x$
while NLO is similar to NNLO.
The longitudinal structure function $F_L$ vanishes at LO
and displays large NNLO corrections, up to $+40\%$
as compared to the NLO result.

Once we increase the scale to $Q=10$ GeV, the perturbative expansion exhibits
an improved convergence, and in particular differences between NLO and
NNLO structure functions (for a fixed common PDF set) are moderate
in all cases.
The only exception is $F_L$ at large-$x$, a region anyway not relevant
for phenomenology since $F_2$ is much larger there.
Nevertheless, neglecting NLO and NNLO coefficients functions still leads
to sizable differences, specially for $F_2$, of up to 10\% at small-$x$ and $-20\%$
at large-$x$.

\begin{figure}[!t]
 \centering
 \includegraphics[width=1.0\linewidth]{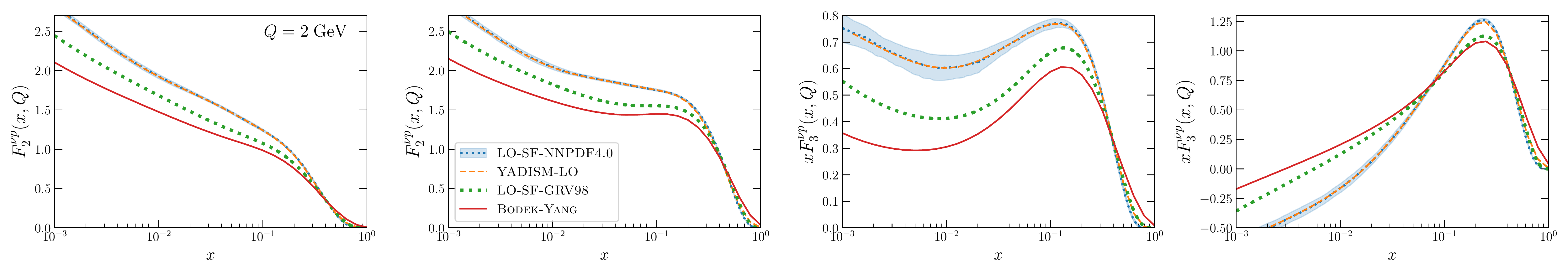}
 \includegraphics[width=1.0\linewidth]{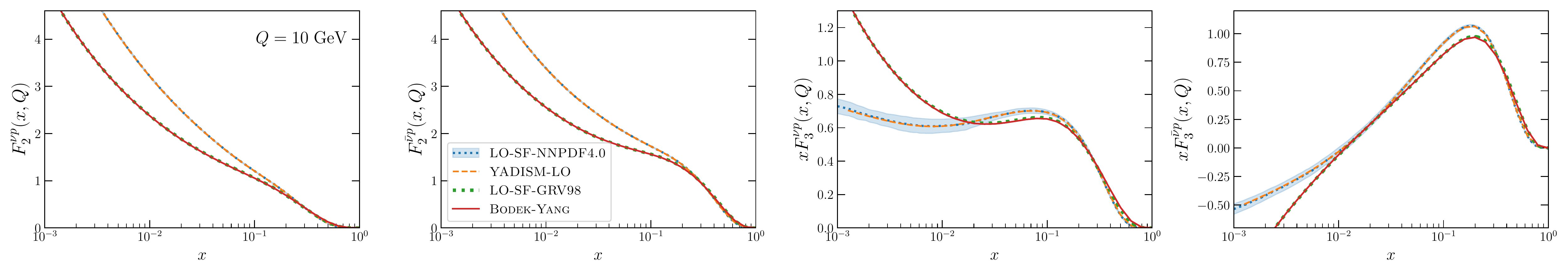}
 \includegraphics[width=1.0\linewidth]{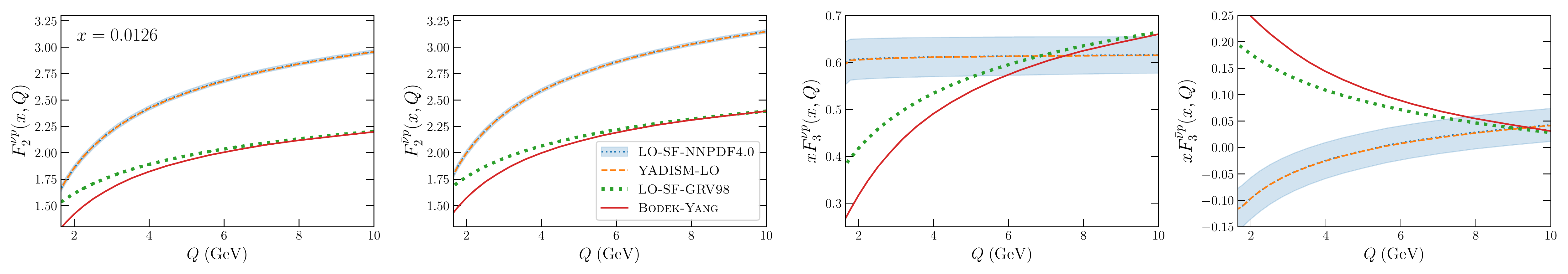}
 \includegraphics[width=1.0\linewidth]{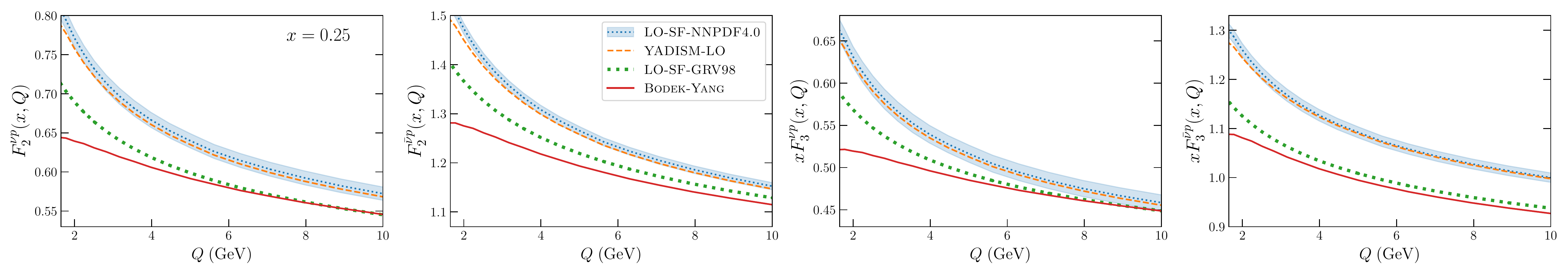}
 \caption{\small Neutrino structure functions on a proton target, comparing the
  {\sc\small YADISM}-LO, {\sc\small Bodek-Yang}, {\sc\small LO-SF-NNPDF4.0},
   and {\sc\small LO-SF-GRV98} predictions.
   We display, from top to bottom, $
   F_2^{\nu p}$, $F_2^{\bar{\nu} p}$, $xF_3^{\nu p}$, and $xF_3^{\bar{\nu} p}$
   first as a function of $x$ for $Q=2$ GeV and $Q=10$ GeV
   and then as a function of $Q$ for $x=0.0126$ and $x=0.25$.
   For {\sc\small LO-SF-NNPDF4.0} we also show the
    68\% CL PDF uncertainties.
 }    
 \label{fig:StructureFunction-Validation}
\end{figure}

\paragraph{Benchmarking.}
Fig.~\ref{fig:StructureFunction-Validation} compares
the neutrino structure functions on a proton target
in the  {\sc\small YADISM}-LO, {\sc\small Bodek-Yang}, {\sc\small LO-SF-NNPDF4.0},
and {\sc\small LO-SF-GRV98} calculations
as a function of $x$ for $Q=2$ GeV and $Q=10$ GeV
and then as a function of $Q$ (in the perturbative region) for $x=0.0126$ and $x=0.25$.
In the case of {\sc\small LO-SF-NNPDF4.0}, we  show the
68\% CL uncertainties,
evaluated over $N_{\rm rep}=100$ Monte Carlo replicas.

As expected, {\sc\small YADISM}-LO  coincides
with the central value of {\sc\small LO-SF-NNPDF4.0} for all values of $Q$.
Residual differences are found only at large-$x$ and small-$Q$
and are explained in terms of the target mass corrections (TMCs) accounted for in the {\sc\small YADISM}
calculation.
Likewise, {\sc\small Bodek-Yang} reduces to {\sc\small LO-SF-GRV98}
at large-$Q$, and in particular at $Q\simeq 10$ GeV the two calculations are
almost identical.
This agreement indicates that the phenomenological corrections to the GRV98 LO PDFs in the
Bodek-Yang model have a negligible effect for $Q\gsim 10$ GeV,
while they become instead important at low-$Q$.
For instance, at $Q=2$
GeV, the differences between {\sc\small Bodek-Yang} and {\sc\small LO-SF-GRV98}
range between 10\% and 30\% depending on the structure function
and the value of $x$.

The comparisons of Fig.~\ref{fig:StructureFunction-Validation} also
 highlight the typical behaviours
of the neutrino structure functions in different regions of the $(x,Q^2)$ plane.
Concerning the $x$ dependence, that of $F_2^{\nu p}$ and $F_2^{\bar{\nu} p}$ is similar
and displays the valence peak at $x \simeq 0.3$ followed by the rise at small-$x$ driven
by DGLAP evolution, which is more marked the higher the value of $Q$.
In the case of $xF_3$, being a non-singlet structure function, the valence peak is followed
by a small-$x$ behaviour that depends on the value of $Q$ and of whether the beam is
composed by neutrinos 
or anti-neutrinos, since in each case the quark flavour
combinations, Eq.~(\ref{eq:neutrinoSFs}), are different.
In terms of the $Q$ dependence, whether higher $Q$ values lead
to an increase or a decrease of the structure function depends on the value of $x$.
For $F_2^{\nu p}$ and $F_2^{\bar{\nu} p}$ the structure functions grow (decrease) with $Q$
for $x=0.0126$~($x=0.25$).
For $xF_3$ also for $x=0.25$ there is a similar decrease with $Q$,
while for $x=0.0126$ the $Q$ dependence varies strongly with the choice of input PDF set.

Given that in Fig.~\ref{fig:StructureFunction-Validation} all calculations shown
are based on LO coefficient functions,
the significant differences between  {\sc\small Bodek-Yang} and {\sc\small YADISM}-LO
as $Q$ is increased can only be attributed to those at
 the input PDFs level, GRV98LO and NNPDF4.0 NNLO respectively.
For instance, for $F_2^{\nu p}$ the GRV98LO calculation
undershoots the {\sc\small YADISM}-LO one based on NNPDF4.0 by 40\%
at $x\simeq 10^{-2}$ and $Q=10$ GeV and by 10\% at $x=0.25$ and $Q=6$ GeV.
In the case of $xF_3$, the small-$x$ behaviour is qualitatively different between
GRV98 and NNPDF4.0, for example at $Q=10$ GeV for $xF_3^{\nu p}$ the former predicts a steep
rise while a flat extrapolation is preferred by the latter.
This indicates that predictions based on the Bodek-Yang model,
and thus on the obsolete GRV98LO PDF set, will in general disagree with those based on modern PDF
determinations.

\begin{figure}[!t]
 \centering
 \includegraphics[width=0.95\linewidth]{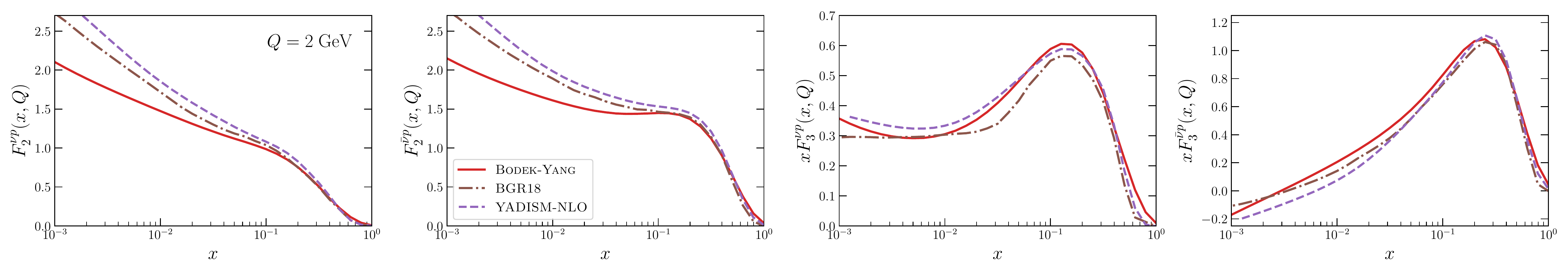}
 \includegraphics[width=0.95\linewidth]{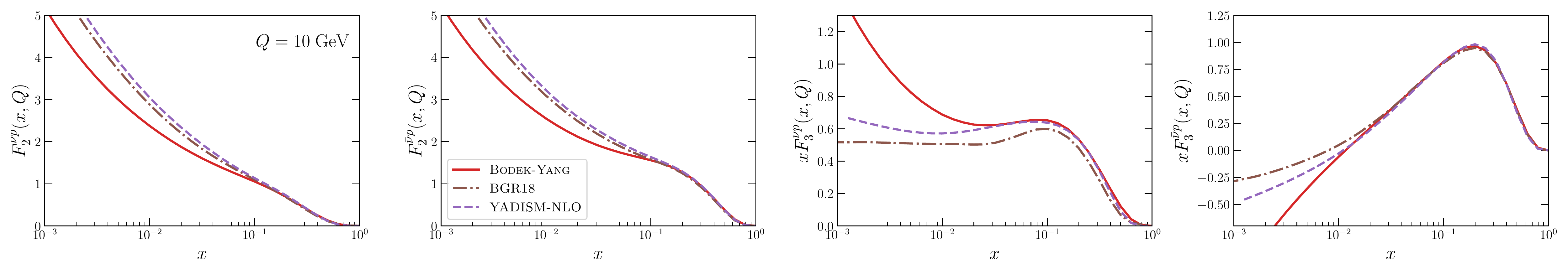}
 \includegraphics[width=0.95\linewidth]{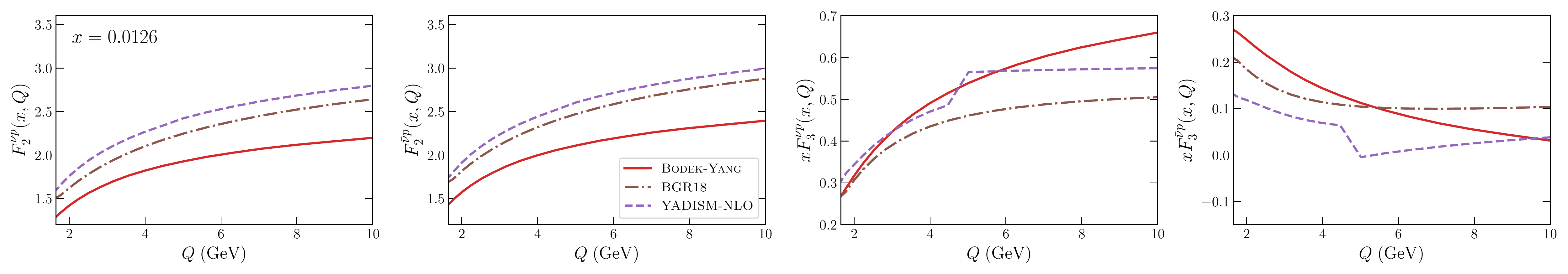}
 \includegraphics[width=0.95\linewidth]{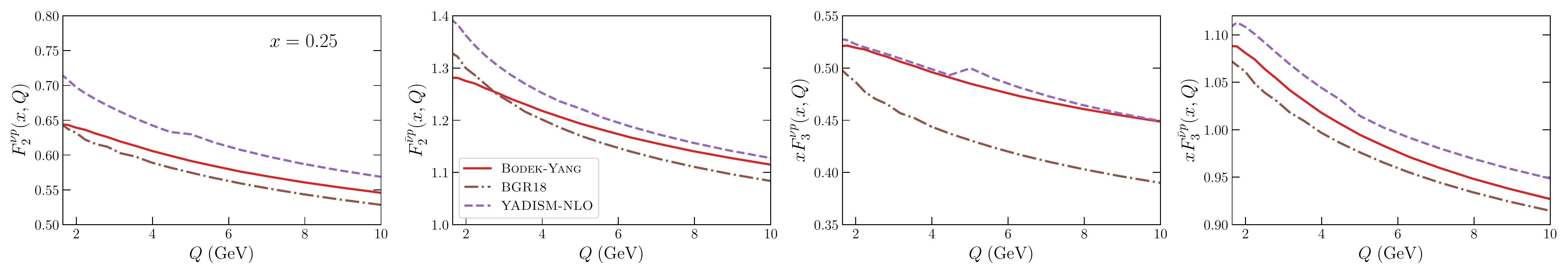}
 \caption{\small Same as Fig.~\ref{fig:StructureFunction-Validation} comparing
   the  {\sc\small Bodek-Yang},  {\sc\small BGR18}, and {\sc\small YADISM-NLO}
 structure functions.
 }    
 \label{fig:StructureFunction-ComparisonsPreFit}
\end{figure}

\paragraph{PDF dependence.}
We compare in Fig.~\ref{fig:StructureFunction-ComparisonsPreFit}
the {\sc\small YADISM}-NLO predictions with those from the  {\sc\small Bodek-Yang}
and {\sc\small BGR18} calculations.
The {\sc\small YADISM}-NLO
and {\sc\small BGR18} predictions are very similar, consistent with
the agreement within uncertainties of the underlying NNPDF4.0 and NNPDF3.1 PDF
fits respectively.
Differences between {\sc\small YADISM}-NNLO and {\sc\small Bodek-Yang} 
are significant, specially for $F_2$ in the $x\lsim 0.1$ region and for $xF_3$ at small-$x$.
These differences are explained by the reliance
of BY
on the obsolete GRV98LO PDF set (cfr Fig.~\ref{fig:StructureFunction-Validation})
and due to neglecting higher-order
QCD corrections (cfr Figs.~\ref{fig:StructureFunction-PerturbativeStab-xdep-allcomp_q2gev}
and~\ref{fig:StructureFunction-PerturbativeStab-xdep-allcomp_q10gev}).

\begin{figure}[!t]
 \centering
 \includegraphics[width=0.85\linewidth]{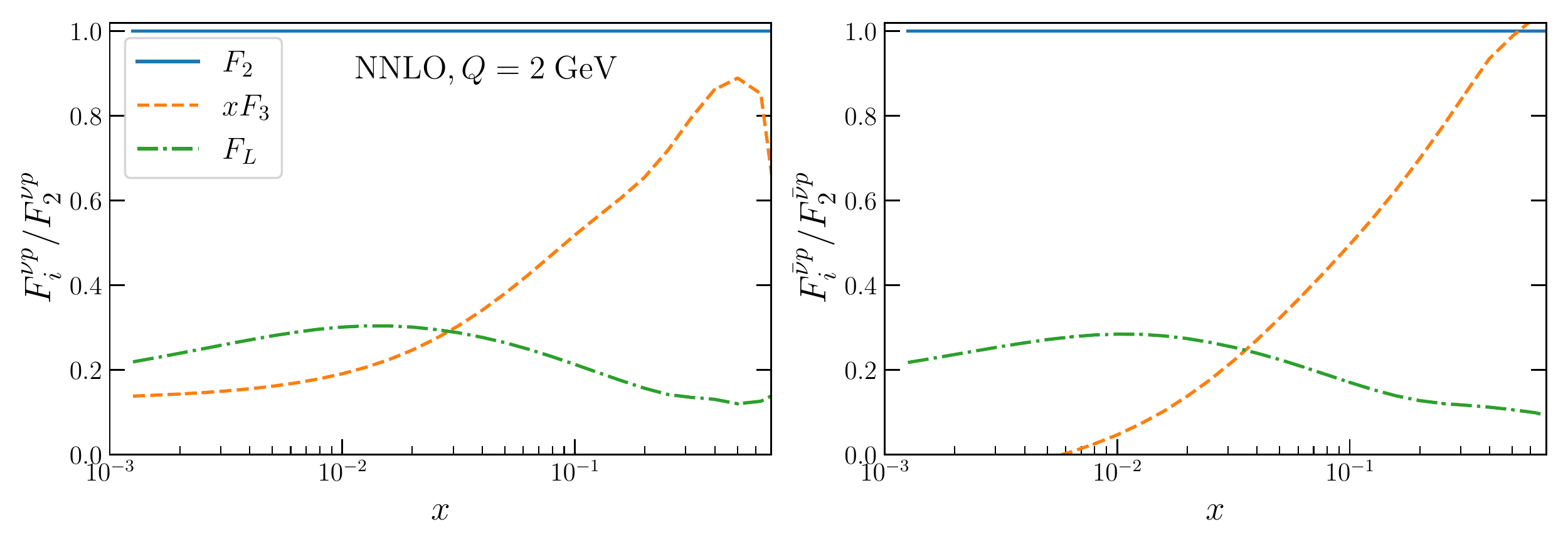}
 \includegraphics[width=0.85\linewidth]{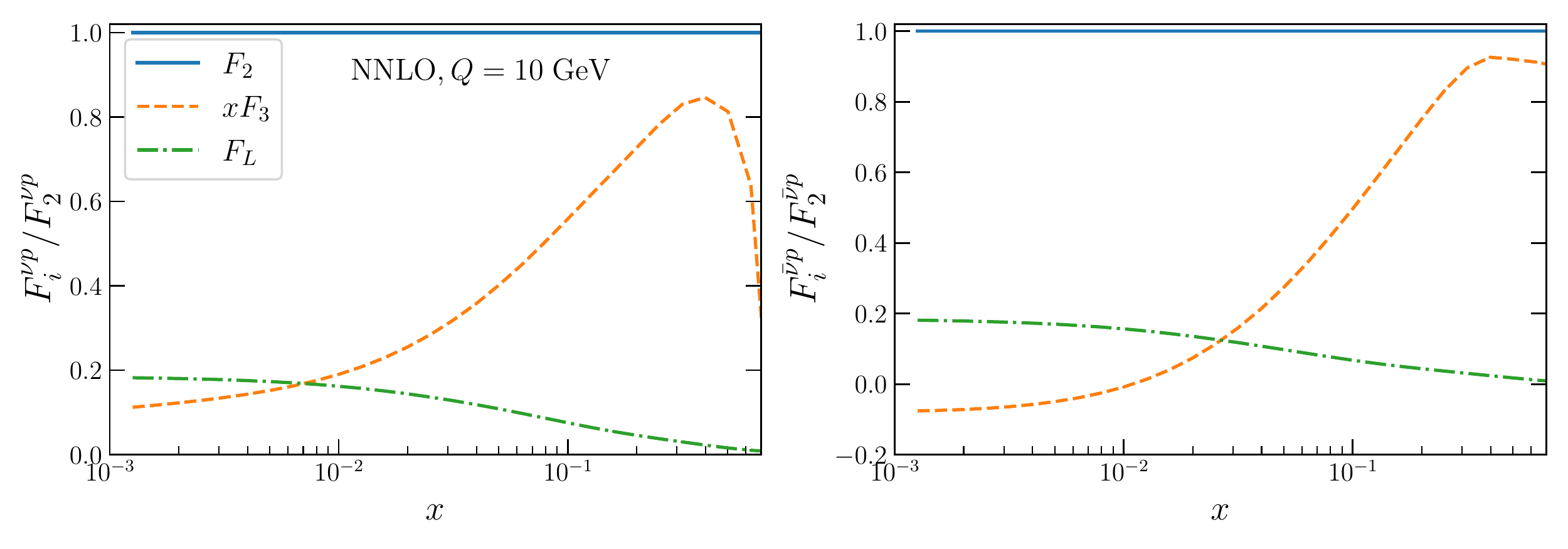}
 \caption{\small The ratio of the neutrino structure functions 
   to $F_2$ for neutrinos (left) and antineutrinos (right)
   for $Q=2$ GeV (top) and $Q=10$ GeV (bottom panels)
   in  {\sc\small YADISM}-NNLO.
 }    
 \label{fig:StructureFunction-RelSize-xdep-allcomp_q2gev}
\end{figure}

\paragraph{The longitudinal structure function.}
In Fig.~\ref{fig:StructureFunction-Validation} we display the
$F_2$ and $xF_3$ structure functions which provide the dominant
contribution to the double-differential cross section Eq.~(\ref{eq:neutrino_DIS_xsec_FL}).
While the longitudinal structure function $F_L$ vanishes at LO, it becomes
non-zero at NLO and in specific kinematic regions can lead
to non-negligible contributions to the scattering cross sections.
To illustrate this hierarchy in the relative magnitude
of the different structure functions,
Fig.~\ref{fig:StructureFunction-RelSize-xdep-allcomp_q2gev}
displays their ratio 
to $F_2$ for neutrinos  and antineutrinos 
for $Q=2$ GeV  and $Q=10$ GeV in the {\sc\small YADISM}-NNLO calculation.

From Fig.~\ref{fig:StructureFunction-RelSize-xdep-allcomp_q2gev} we observe how in the large-$x$
valence region the $F_2$ and $xF_3$
structure functions are of comparable magnitude, with $F_L$ being much smaller.
Since $xF_3$ is a valence structure function, it is suppressed as $x$ decreases and indeed
for $x= 10^{-2}$ it becomes at most 20\% of the value of the dominant $F_2$.
The relative contribution from $F_L$ is similar for neutrinos and neutrinos, 
since it is dominated by the gluon contribution, and becomes more important as both $x$
and $Q^2$ decrease.
In particular, for $x\lsim 10^{-2}$ the magnitude of $F_L$ becomes larger than that of $xF_3$.
At $Q=2$ GeV, $F_L$ can be up to 30\% the value of $F_2$, indicating a contribution
to the double differential cross section larger than the typical
experimental uncertainties and that hence must be accounted for.

Whenever possible, we fit data for the double-differential
cross section rather than for the individual $F_2$ and $xF_3$
structure functions separately, since the former provides also sensitivity
to $F_L$.

\section{The NNSF$\nu$ approach}
\label{sec:fitting}

Here  we describe the NNSF$\nu$ approach used
to determine neutrino-nucleon inelastic structure functions
and their associated uncertainties across the whole
range of $Q$ relevant for neutrino phenomenology.
We first describe the general strategy, based on the
combination of a machine-learning
parametrisation of experimental data with state-of-the-art QCD
calculations.
We then review the available measurements on neutrino structure functions
and cross sections used to constrain this parametrisation.
Subsequently, we discuss the neural network
parametrisation of neutrino structure functions,
how it is trained on both the data and the QCD predictions,
and the uncertainty estimate based on the Monte Carlo replica method.

\subsection{General strategy}
\label{subsec:general_strategy}

An schematic representation of the NNSF$\nu$ strategy to
determine neutrino structure functions is
displayed in Fig.~\ref{fig:general-strategy}
with the Bodek-Yang predictions at $x=0.0126$
for illustration.
The $(x,Q)$  plane is divided into three disjoint regions,
with  complementary methods to evaluate the structure functions in each of them:

\begin{figure}[!t]
 \centering
 \includegraphics[width=1.0\linewidth]{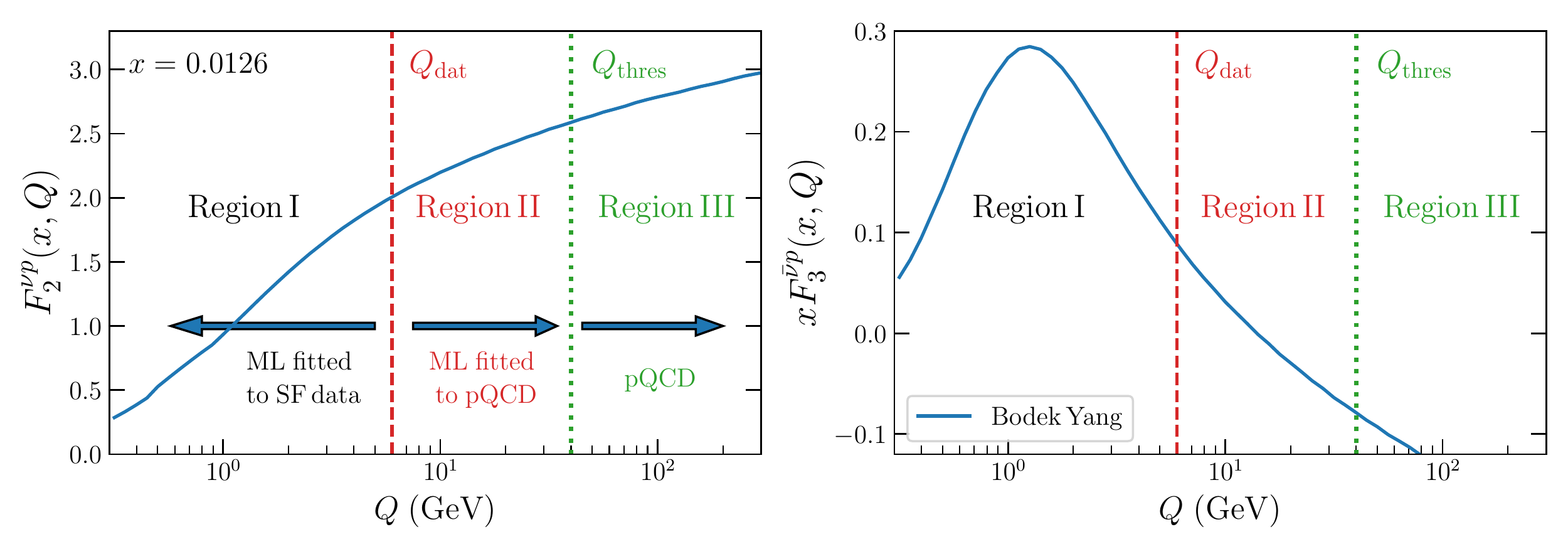}
 \caption{Schematic representation of the NNSF$\nu$ strategy.
   In the region with $Q \le Q_{\rm dat}$ (Region I), we adopt a data-driven approach where a neural
   network parametrisation
   is fitted to neutrino inelastic structure function data.
   In the intermediate region with $Q_{\rm dat}\le Q \le Q_{\rm thr}$ (Region II),
   the same parametrisation is fitted instead
   to the QCD calculations provided by {\sc\small YADISM} with nNNPDF3.0 as input.
   In the high-$Q$ region, $Q> Q_{\rm thr}$ (Region III),
   the neural
   network predictions are replaced by the outcome of the same {\sc\small YADISM}
   calculation.
   For small-$x$ values relevant for UHE neutrino scattering
   ($x \le 10^{-5}$), Region III is extended to cover 
   $Q_{\rm min} \le Q \le Q_{\rm thr}$ 
   with $Q_{\rm min}=2$ GeV, see App.~\ref{app:delivery} for more details
   and Fig.~\ref{fig:x_Q_Enu_coverage} in particular for the analogous
   division into regions at the level of the $(x,Q^2)$ plane.
 }
 \label{fig:general-strategy}
\end{figure}

\begin{itemize}

\item {\bf Region I.}
  At low momentum transfers $Q \lsim Q_{\rm dat}$, with $Q_{\rm dat} \simeq 5$ GeV,
  the perturbative calculation of neutrino  structure functions
  in Eq.~(\ref{eq:sfs_pqcd}) is either invalid or affected by significant
  theory uncertainties related to higher twists, missing higher perturbative orders, and
   large-$x$ resummation effects.

  In this region we parametrise
  the structure functions in terms of the information provided
  by the available experimental data on neutrino-nucleus
  inelastic scattering summarised in Sect.~\ref{subsec:expdata}.
  Following the
  NNPDF fitting methodology, this parametrisation
  combines neural networks as universal unbiased
  interpolants with the Monte Carlo replica method for the uncertainty estimate.

\item {\bf Region II.}
  The region of intermediate momentum transfers,
  $Q_{\rm dat} \lsim Q \lsim Q_{\rm thr}$ with $Q_{\rm thr}\simeq 25$ GeV,
  is well described by the perturbative QCD formalism.
  DIS structure functions are computed at NLO by 
  {\sc\small YADISM} with nNNPDF3.0 as  input for all targets.
  In this region the neural network parametrisation is
  fitted to these QCD predictions  rather than to the data as in Region I.
  The figure of merit is given in terms of the theory covariance matrix
  with PDF and the MHO uncertainties,
  see Sect.~\ref{subsec:error-prorpagation}.

 The nNNPDF3.0 determination already includes
  information from neutrino measurements, in particular from CHORUS (inclusive)
  and NuTeV (charm) structure functions,  and therefore in Region II no neutrino data needs to be
  explicitely used to further constrain
  the  parametrisation.

\item {\bf Region III.}
  For large momentum transfers,
  $Q \gsim Q_{\rm thr}$, the neural network predictions are replaced
  by the direct outcome of the same {\sc\small YADISM} calculation
  used to constrain the fit in  Region II.
  Hence, in Region III the central prediction and uncertainties
  of  NNSF$\nu$  coincide with the {\sc\small YADISM} ones,
  which  extend up to $Q=10$ TeV
  and down to $x=10^{-9}$ to cover the entire kinematic region
  relevant for neutrino phenomenology including the
  UHE scattering.

  Furthermore,  Region III is extended
  in the small-$x$ region with
   $x \le 10^{-5}$
  to cover also momentum
  transfers of $Q_{\rm min} \le Q \le Q_{\rm thr}$,
  with $Q_{\rm min}=2$ GeV.
  The reason for this choice is that for $x\le 10^{-5}$ the neural network extrapolation
  trained in Regions I and II exhibits large uncertainties and
  using the QCD calculation is preferred on theoretical grounds~\cite{Ball:2016spl}.
  
\end{itemize}

As we will show in the subsequent sections, such a strategy allows us to consistently
extend the state-of-the-art perturbative QCD computations into the non-perturbative
region and provide predictions that are valid across a wide range of energy relevant for
neutrino phenomenology.

We have verified that the NNSF$\nu$ determination is stable
with respect to  moderate variations of the
values of the $Q_{\rm dat}$ and $Q_{\rm thr}$ hyperparameters.
App.~\ref{app:delivery} provides additional details on
the implementation of the NNSF$\nu$  procedure including the prescriptions
to evaluate and match inelastic structure functions in the various
regions of the $(x,Q^2)$ kinematic plane.

\subsection{Experimental data}
\label{subsec:expdata}

The parametrisation of neutrino structure functions
applicable in Regions I and II defined in Fig.~\ref{fig:general-strategy} requires
two different inputs: experimental data in Region I
and the corresponding QCD calculations for Region II.
For the former, we consider
all available data on inelastic neutrino structure functions and double differential
cross sections.
We restrict our analysis to those measurements where the incoming neutrino energy $E_{\nu}$
is sufficiently large to ensure that the contribution
from the inelastic region dominates.
For this reason, we do not consider neutrino
measurements from experiments such as ArgoNeuT~\cite{ArgoNeuT:2011bms},
MicroBooNE~\cite{MicroBooNE:2019nio}, T2K~\cite{T2K:2017qxv},
or MINER$\nu$A~\cite{MINERvA:2016oql},  where $E_\nu$ is too low
to cleanly access inelastic scattering.

Two kinematic cuts are applied to the data used as input to the NNSF$\nu$ fit.
First, a cut in the invariant mass of the final hadronic state
$W^2  \ge 3.5 ~ \rm{GeV}^2$ filters away points in the quasi-elastic and resonant
scattering regions.
Second, data points with  $Q \ge Q_{\rm dat}$ 
are excluded according to the definition of Region I in Fig.~\ref{fig:general-strategy}.
This cut does not result on a net information loss in the fit,
since as mentioned above nNNPDF3.0 already includes the constraints
from neutrino DIS data present in the $ Q \ge Q_{\rm dat}$ region.

\begin{figure}[!t]
 \centering
 \includegraphics[width=0.80\linewidth]{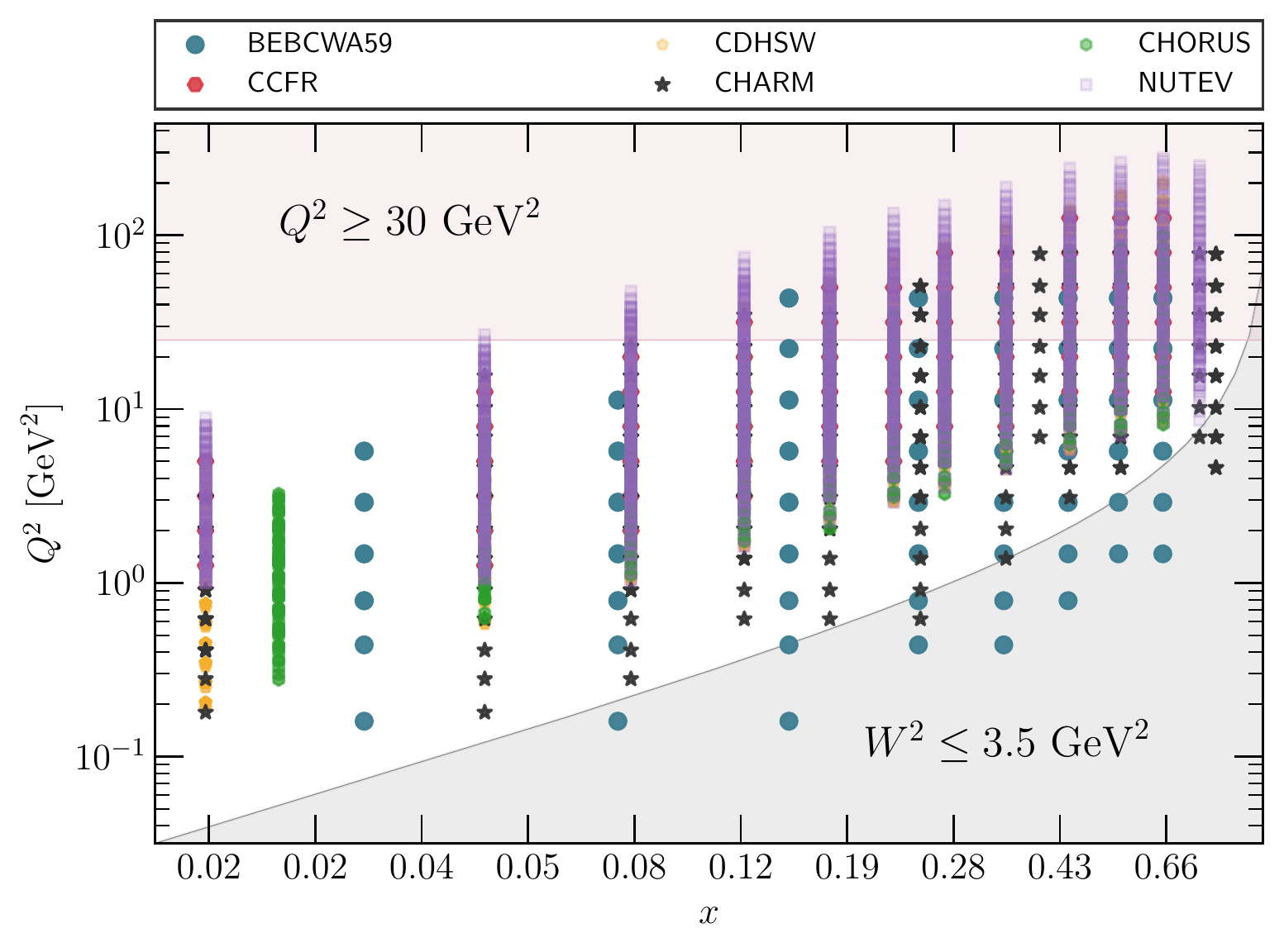}
 \caption{The kinematic coverage in the $(x,Q^2)$ plane
   of the neutrino scattering cross-section data listed in Table~\ref{tab:neutrino-DIS}.
   The region covered in grey is excluded from the fit
   from  $W^2 \ge 3.5~\rm{GeV}^2$ cut required to isolate
   inelastic scattering,
   while the one in light red is excluded from the $Q\le Q_{\rm dat}$
   condition that defines Region I in Fig.~\ref{fig:general-strategy}.
 }
 \label{fig:kin}
\end{figure}

Table~\ref{tab:neutrino-DIS}  lists
the datasets used to constrain neutrino
structure functions in Region I.
For each dataset, we indicate the publication reference, the range of $x$ and $Q^2$ covered,
the observables included, the scattering target, the final state
measured, and the number of available data points $n_{\rm dat}$ before and after applying
 kinematic cuts.
In total we have 6224 (4184) data points in the fit before (after) cuts.
  The corresponding kinematic coverage of these datasets in the $(x,Q^2)$ plane
 is displayed in Fig.~\ref{fig:kin}, which also indicates the regions
excluded by cuts.
Some of the datasets from Table~\ref{tab:neutrino-DIS} provide additional observables
on top of those indicated there, which however need to be excluded from the fit to prevent double counting.
In particular, the same
measurement is often presented in terms of  both the differential cross section $d^2\sigma/dxdQ^2$
and of the individual structure functions $F_2$ and $xF_3$.
We always select observables that are closer to the actual measurements,
in this case the double differential cross-sections as they also constrain
the longitudinal structure function $F_L$.

\begin{table}[t]
\centering
\footnotesize
\renewcommand{\arraystretch}{1.80}
\begin{tabularx}{\textwidth}{Xcccccccc}
\toprule
 Dataset & Ref. & $\lc x_{\text{min}},~x_{\text{max}}\rc $  & $\lc Q_{\text{min}}^2, Q^2_{\text{max}}\rc~(\rm{GeV}^2)$   &  Observables  
 & Target & Final state & $n_{\rm dat}\,\lp n_{\rm dat}^{(\rm fit)}\rp$ \\
\midrule
 BEBCWA59 & \cite{BEBCWA59:1987rcd} &  $[0.028, 0.65]$  & $[0.2, 44]$  &  $F_2, xF_3$ & Ne & $\nu+\bar{\nu}$  &   114 (71)  \\
 CCFR   & \cite{Oltman:1992pq}  &  $[0.015, 0.65]$  & $[1.3, 126]$  &  $F_2, xF_3$ & Fe & $\nu+\bar{\nu}$ &  256 (164) \\
 CHARM   &  \cite{CHARM:1982ods} &$[0.015, 0.80]$  & $[0.2, 78]$  &  $F_2, xF_3 $ & CaCO$_3$ & $\nu+\bar{\nu}$ &  320 (144) \\
 CHORUS & \cite{CHORUS:2005cpn}   & $[0.020, 0.65]$ & $[0.3, 101]$  & $d^2 \sigma/dxdQ^2$  & Pb & $\nu$, $\bar{\nu}$ &  1212 (966) \\
 CDHSW  & \cite{Berge:1989hr}   &$[0.015, 0.65]$  & $[0.2, 196]$  &  $d^2 \sigma/dxdQ^2$ & Fe & $\nu$, $\bar{\nu}$ &  1551 (1259) \\
 NuTeV  & \cite{NuTeV:2005wsg}  & $[0.015, 0.75]$  & $[1.1, 279]$  &  $d^2 \sigma/dxdQ^2$ & Fe & $\nu$, $\bar{\nu}$&  2874 (1580) \\
 \midrule
 {\bf Total}  & & & & & &  & {\bf  6224 (4184)} \\
 \bottomrule
\end{tabularx}
\vspace{0.2cm}
\caption{\small Datasets included in the NNSF$\nu$  fit for Region I of Fig.~\ref{fig:general-strategy}.
  We indicate the publication reference, the range of $x$ and $Q^2$,
  the observables included, the scattering target, the final state measured,
  and the number of data points before (after) applying
  kinematic cuts.
  The coverage of these datasets in the $(x,Q^2)$ plane
  is displayed in Fig.~\ref{fig:kin}.
  Some of experiments provide additional observables, which are excluded
 to prevent double counting.
}
\label{tab:neutrino-DIS}
\end{table}


Table~\ref{tab:neutrino-DIS} and Fig.~\ref{fig:kin} indicate
that the NNSF$\nu$ fit is sensitive to inelastic neutrino
structure functions for momentum transfers down to $Q\simeq 400$ MeV,
well in the non-perturbative region.
The range of $x$ covered reaches $x_{\rm min}= 0.015$, and as a consequence
of the  DIS kinematics the values of $Q$ being probed increase with $x$.
Measurements in the non-perturbative region with
$Q \lsim 1$ GeV are provided by several experiments
and cover momentum fractions up to $x\lsim 0.3$.
The datasets with the largest number of points (CHORUS, NuTeV, and CDHSW)
present their measurements in terms of the double differential
cross section $d^2\sigma/dxdy$, while BEBCWA59, CCFR, and CHARM
only provide data for separate structure functions $F_2$ and $xF_3$.

Concerning nuclear effects,  Table~\ref{tab:neutrino-DIS} shows that available data
have sensitivity to neutrino scattering on
 Ne ($A=20$), Fe $(A=56)$, and Pb ($A=208$) targets.
For CaCO3, the target used in the CHARM experiment, we assume $A=20$ as the average
atomic mass number of the nuclei that form this compound.
Neutrino structure function measurements are not available
on hydrogen or deuteron targets, and hence in Region I the low-$A$
behaviour is extrapolated from the measurements with $A \ge 20$.
The inter- and extrapolation to $A$ values not included in the fit is provided
by the smoothness of the neural network output, as we validate
in Sect.~\ref{sec:results}.

For momentum transfers $Q > Q_{\rm dat}$ (Regions II and III) the dependence with the atomic
mass number $A$ of NNSF$\nu$ follows that provided by nNNPDF3.0, and is hence
constrained  by
other types of processes beyond neutrino DIS, such as charged-lepton fixed-target
nuclear 
DIS and weak boson, dijet, and $D$-meson production cross sections in proton-lead
collisions at the LHC.

\subsection{Structure function parametrisation}
\label{subsec:fitting_nn}

The NNSF$\nu$ parametrisation of neutrino structure functions in Regions I and II
is obtained by training a  machine learning model to  experimental data
and to the QCD predictions, respectively.
It follows the  NNPDF fitting methodology based on the  combination of neural 
networks as universal unbiased interpolator with the Monte Carlo replica method 
for error estimate and propagation.
This methodology was originally developed for DIS neutral-current
structure functions~\cite{Forte:2002fg,DelDebbio:2004xtd} and subsequently
extended to proton
PDFs~\cite{DelDebbio:2007ee,Ball:2008by,Ball:2012cx,NNPDF:2014otw,NNPDF:2021uiq,NNPDF:2021njg}, helicity 
PDFs~\cite{Nocera:2014gqa,Ball:2013lla}, nuclear PDFs~\cite{AbdulKhalek:2019mzd,AbdulKhalek:2020yuc,AbdulKhalek:2022fyi}, and fragmentation
functions~\cite{Bertone:2018ecm,Bertone:2017tyb}.

Here we apply for the first time the NNPDF approach to {\it i)} the determination of neutrino 
(charged-current) structure functions and to {\it ii)} the parametrisation of a three-dimensional
function, with neural networks receiving $(x, Q^2, A)$ as inputs.
This is achieved by means
of a stand-alone open-source NNSF$\nu$ framework described in App.~\ref{app:delivery}.
This framework shares many  similarities with the  NNPDF4.0 codebase, 
in particular it is also built upon {\sc\small TensorFlow}~\cite{tensorflow2015-whitepaper}
and uses adaptative stochastic gradient descent (SGD) methods for the minimisation such as {\tt Adam}~\cite{adam_optimizer}.

The double-differential neutrino-nucleus cross sections,
Eqns.~(\ref{eq:neutrino_DIS_xsec}) and~(\ref{eq:antineutrino_DIS_xsec}),
are expressed in terms of three  independent structure functions and hence one needs to parametrise
six independent quantities which we choose to be
\bea
&&F_2^{\nu}(x,Q^2,A) \, , \quad
xF_3^{\nu}(x,Q^2,A) \, , \quad
F_L^{\nu}(x,Q^2,A) \, , \nonumber\\
&&F_2^{\bar{\nu}}(x,Q^2,A) \, , \quad
xF_3^{\bar{\nu}}(x,Q^2,A) \, , \quad
F_L^{\bar{\nu}}(x,Q^2,A) \, , \quad \label{eq:NNparam_SFs}
\eea
each of them function of the three inputs  $(x, Q^2, A)$.
In Eq.~(\ref{eq:NNparam_SFs}) the limit $A=1$ is to be understood
as that of the structure functions on an isoscalar, free-nucleon target, rather
than those for a proton target.
That is, $F_2^{\nu}(x,Q^2,A=1)$ coincides with $\lp F_2^{\nu p}(x,Q^2) + F_2^{\nu n}(x,Q^2) \rp/2 $,
and likewise for the other structure functions.

\begin{figure}[!t]
 \centering
 \includegraphics[width=1\linewidth]{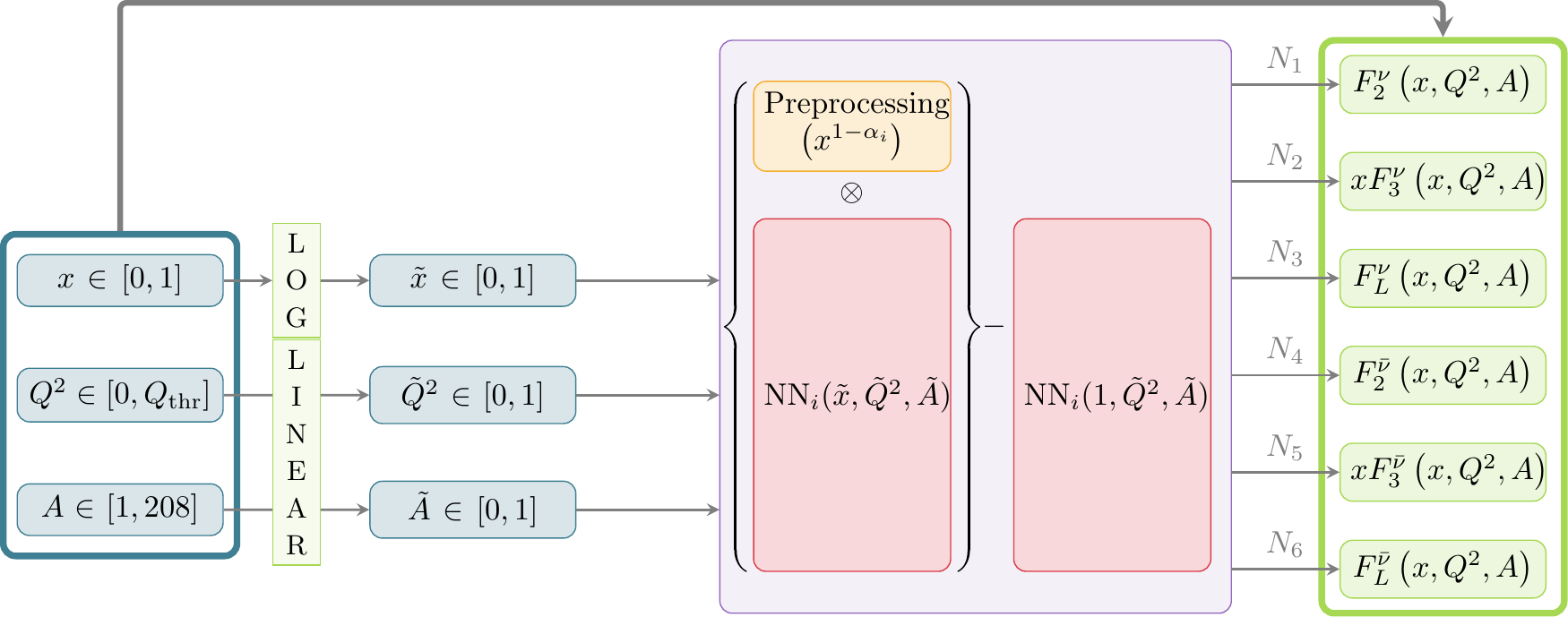}
 \caption{Flowchart summarizing the NNSF$\nu$ fitting framework.
   The three inputs are the momentum fraction 
   $x$, the momentum transfer squared $Q^2$, and the atomic mass number $A$,
   suitably preprocessed to lie in a common range.
   The network output is supplemented by a power-like term to facilitate the learning
   of the small-$x$ region and the endpoint behaviour at $x=1$ is subtracted
   to reproduce the elastic limit.
   The architecture of this neural network is 3-70-55-40-20-20-6 and hence composed
   by five hidden layers.
   The parameters of the neural network are optimised to reproduce the experimental data in Region I
 	and the QCD predictions in  Region II.
 }
 \label{fig:architecture}
\end{figure}

The mapping between the inputs $(x,Q^2,A)$ and the outputs $F_i^{\nu},F_i^{\bar{\nu}}$ 
with $i=2,3,L$ in Eq.~(\ref{eq:NNparam_SFs}) is
provided by a deep neural network as illustrated in Fig.~\ref{fig:architecture}.
The free parameters of this neural network, its weights and thresholds, are
determined by training the parametrisation to the experimental
data (in Region I) and to the QCD predictions (in Region II) for
neutrino structure functions and differential cross sections.
Hyperparameters like the network architecture are determined by means of a dedicated
optimisation procedure.
The best choice for the
architecture of this  network is found to be 3--70--55--40--20--20--6, hence
composed by five hidden layers with 70, 55, 40, 20, and 20 neurons in each
of them.

As indicated by Fig.~\ref{fig:architecture}, two corrections are applied
to the network output before it can be identified with the neutrino
structure functions.
First, we supplement the network output with a
preprocessing
factor $x^{1 - \alpha_i }$ that facilitates the learning and extrapolation
of structure functions in the small-$x$ region~\cite{Ball:2016spl}.
Second, we subtract the endpoint behaviour at $x=1$ 
to reproduce the elastic limit where structure functions vanish
due to kinematic constraints.
With these considerations, the relation between the network output
and the structure functions is given by
\bea
	F_i^{\nu}(x,Q^2,A) &=& x^{ 1 - \alpha_i } {\rm NN}_i(x,Q^2,A) -{\rm NN}_i(x=1,Q^2,A) \, , \label{eq:NN_output}\\
	F_i^{\bar{\nu}}(x,Q^2,A) &=&  x^{  1 - \bar{\alpha}_i } {\rm NN}_{3+i}(x,Q^2,A) -{\rm NN}_{3+i}(x=1,Q^2,A) \, , \nonumber
\eea
for $i =  1, 2, 3$ corresponding to $F_2, xF_3, F_L$,
and where ${\rm NN}_j$ indicates the activation state of the $j$-th neuron in the output 
layer of the network.

By construction, structure functions parametrised this way vanish in the elastic
limit $x=1$ for all values of $Q^2$ and $A$ without 
restricting the behaviour in the $x < 1$ region.
The small-$x$ preprocessing exponents $\lp \alpha_i, \alpha_i\rp$ in Eq.~(\ref{eq:NN_output})
are constrained from the data as part of the training procedure at the same
time as the neural network parameters.
Furthermore, the neural network inputs are rescaled to a common range,
logarithmically in $x$ and linearly in $Q^2$ and $A$
to ensure that no specific kinematic region is arbitrarily privileged by the training.
We do not enforce the positivity of the $F_2$ and $F_L$ structure functions, since
it is found that the  data together with the QCD constraints included are sufficient
to avoid the unphysical negative region.

\subsection{Fitting and error propagation}
\label{subsec:error-prorpagation}

The neural network parametrisation of neutrino structure functions
from Fig.~\ref{fig:architecture} is constrained
by the experimental data from Table~\ref{tab:neutrino-DIS} in Region I  
and by the QCD predictions in Region II.
Here we discuss the figures of merit used in the minimisation,
the error propagation strategy based on the Monte Carlo replica method,
and the settings and performance of the training procedure.

\begin{figure}[!t]
	\centering
	\includegraphics[width=0.95\linewidth]{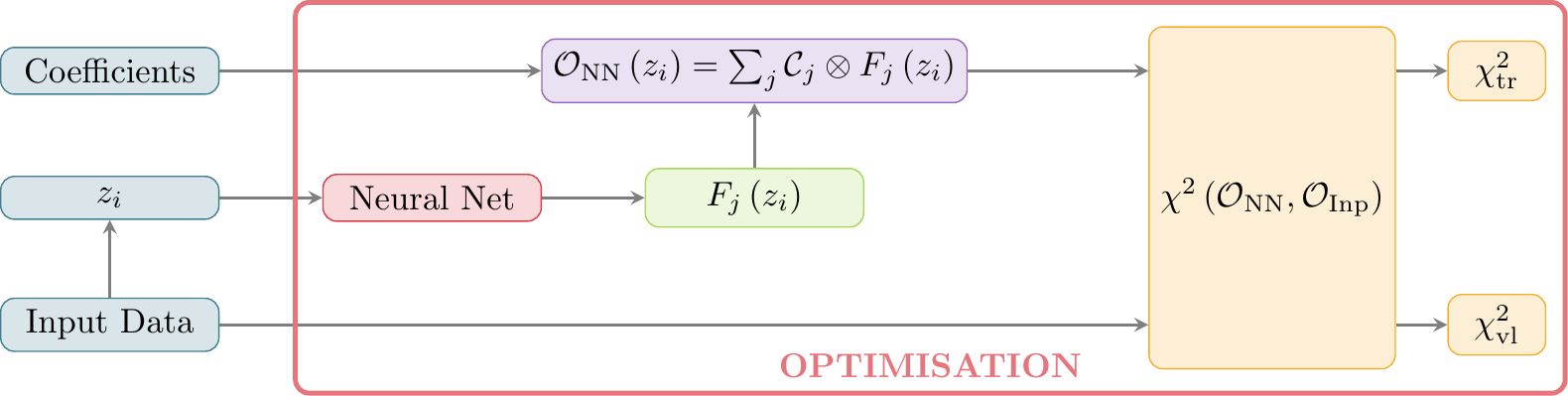}
	\caption{Diagrammatic representation of  the evaluation
          of the  figure of merit in the fit, $\chi^2$,  as a function of the kinematic inputs 
          ${\boldsymbol z}_i  = \left\{ x_i, Q^2_i, A_i \right\}$ with $i=1,\ldots, n_{\rm dat}^{(\rm fit)}$.
          The structure functions parametrised by neural networks
          according to Eq.~(\ref{eq:NN_output}) and
          Fig.~\ref{fig:architecture}
          are combined to construct the observables that input $(\mathcal{O}_{\rm Inp})$ the fit (and composed by either
          experimental data or QCD calculations)
          e.g. by means of Eqns.~(\ref{eq:neutrino_DIS_xsec_FL})
          and~(\ref{eq:antineutrino_DIS_xsec_FL}).
          The predicted observables are compared to the corresponding data points
          to evaluate the $\chi^2$ entering the optimisation process
          and the cross-validation stopping.
          $\chi^2_{\rm tr}$ and $\chi^2_{\rm vl}$ represent the training and validation
          losses, respectively.
	}
	\label{fig:pipeline_chi2}
\end{figure}

Fig.~\ref{fig:pipeline_chi2} provides a
diagrammatic representation illustrating the evaluation
of the  figure of merit in the fit, $\chi^2$,  as a function of the kinematic 
inputs ${\boldsymbol z}_i  = \left\{ x_i, Q^2_i, A_i \right\}$ with $i$ labelling 
the fitted data points.
The structure functions parametrised by neural networks
according to Eq.~(\ref{eq:NN_output}) and
Fig.~\ref{fig:architecture}
are combined to construct the observables that input the fit
e.g. by means of Eqns.~(\ref{eq:neutrino_DIS_xsec_FL})
and~(\ref{eq:antineutrino_DIS_xsec_FL}).
These fit inputs are classified into
experimental data and  QCD calculations depending on the range of $Q^2$ being
considered.
The predicted observables from the neural network
parametrisation are then compared to the corresponding input data points
to evaluate the $\chi^2$ entering both the optimisation process
and the cross-validation stopping.

\paragraph{Constraints from experimental data.}
In the Monte Carlo replica method one first generates a sample of 
$N_{\rm rep}$ artificial replicas of the experimental data as follows.
Given $n_{\rm dat}$ experimental measurements of neutrino structure functions
or differential cross sections characterised by central value $\mathcal{F}^{{\rm (dat)}}_i$,
 uncorrelated uncertainty $\sigma_{i}^{\rm (stat)}$,  $n_{\rm sys}$ correlated systematic uncertainties 
 $\sigma^{\rm (sys)}_{i,\alpha}$, and  $n_{\rm norm}$ normalisation uncertainties $\sigma^{\rm (norm)}_{i,n}$,
 artificial replicas of these measurements are generated by
\begin{equation}
	\label{eq:replicas}
	\mathcal{F}_{i}^{(\art)(k)} = \lc \prod_{n=1}^{n_{\rm norm}}\lp 1+r_{i,n}^{(k)}\sigma^{\rm (norm)}_{i,n}\rp \rc
	\mathcal{F}_{i}^{\rm (dat)}\lp 1 + r_{i}^{(k)}\sigma_{n}^{\rm (stat)} +
	\sum_{\alpha=1}^{n_{\rm sys}}r_{i,\alpha}^{(k)}\sigma^{\rm (sys)}_{i,\alpha}\rp
	\ , \quad k=1,\ldots,N_{\rep} \, ,
\end{equation}
for $i=1,\ldots,n_{\rm dat}$, where $r_{i}^{(k)}$, $r_{i,\alpha}^{(k)}$, and $r_{i,n}^{(k)}$
indicate univariate Gaussian random numbers generated such that
experimental correlations between systematic and normalisation errors are accounted for.
This procedure is formally equivalent to generating data replicas
according to the statistical model  provided by the experimental covariance matrix,
which is reproduced by averaging over  replicas,
\bea
	\la \mathcal{F}_{i}^{(\art)} \mathcal{F}_{j}^{(\art)}\ra_{\rm rep} - \la \mathcal{F}_{i}^{(\art)}\ra_{\rep}\la \mathcal{F}_{j}^{(\art)}\ra_{\rm rep} =\qquad\qquad \qquad\qquad\\
	\frac{1}{N_{\rm rep}} \sum_{k=1}^{N_{\rm rep}} \mathcal{F}^{{(\rm art)}(k)}_i\mathcal{F}^{{(\rm art)}(k)}_j
	- \frac{1}{N_{\rm rep}^2} \sum_{k=1}^{N_{\rm rep}} \mathcal{F}^{{\rm {(art)}}(k)}_i\sum_{k'=1}^{N_{\rm rep}} \mathcal{F}^{{\rm {(art)}}(k')}_j=  \nonumber
	\lp {\rm cov}_{\rm exp}\rp _{ij} \, ,
\eea
in the large replica limit, $N_{\rm rep}\to \infty$.
For the neutrino structure function data considered in NNSF$\nu$, ${\rm cov}_{\rm exp}$ contains only
additive systematic uncertainties and hence it is not affected by the D'Agostini
bias which would require introducing a $t_0$ covariance matrix~\cite{Ball:2009qv} based on
a previous iteration of the fit.

Subsequently, for each of the $N_{\rm rep}$ replicas generated according to Eq.~(\ref{eq:replicas}), a separate neural network with the structure of Fig.~\ref{fig:architecture} is trained by minimising
the error function defined as
\begin{equation}
 \label{eq:Et0}
E_{\rm exp}^{(k)}=\frac{1}{n_{\rm dat}}\sum_{i, j=1}^{n_{\mathrm{dat}}}\left(\mathcal{F}^{{\rm (net)}(k)}_i-\mathcal{F}^{{\rm (art)}(k)}_i \right)
\left(\operatorname{cov}_{\rm exp}^{-1}\right)_{i j}\left(\mathcal{F}^{{\rm (net)}(k)}_j-\mathcal{F}^{{\rm (art)}(k)}_j\right)\, , \quad k=1,\ldots,N_{\rep} \ ,
\end{equation}
in terms of the experimental covariance matrix.
For each replica, the training is stopped once the cross-validation stopping criterion
described below is satisfied.
The  overall goodness-of-fit between the model predictions and the experimental data is
then quantified by the $\chi^2$ defined
in a similar manner as Eq.~(\ref{eq:Et0}) now in terms of the average neural network prediction,
\begin{equation}
  \label{eq:chi2t0}
\chi^2_{\rm exp}=\frac{1}{n_{\rm dat}}\sum_{i, j=1}^{n_{\mathrm{dat}}}\left(\la \mathcal{F}^{\rm (net)}_i\ra_{\rm rep}-\mathcal{F}^{\rm (dat)}_i \right)
\left(\operatorname{cov}_{\rm exp}^{-1}\right)_{i j}\left(
\la \mathcal{F}^{\rm (net)}_j\ra_{\rm rep}-\mathcal{F}^{\rm (dat)}_j\right)\, ,
\end{equation}
with averages over replica sample are evaluated as
\be
\la \mathcal{F}^{\rm (net)}_i\ra_{\rm rep} = \frac{1}{N_{\rm rep}} \sum_{k=1}^{N_{\rm rep}}
\mathcal{F}^{{{\rm (net)}}(k)}_i \, .
\ee
The $N_{\rm rep}$ trained neutral network parametrisations provide a representation of the probability density
in the space of neutrino structure functions, from which expectation values and other statistical estimators
can be computed.
For instance, the $1\sigma$ uncertainty $\delta  \mathcal{F}$
in a structure function at generic $(x,Q^2,A)$ values
can be computed by evaluating
the standard deviation over the replica ensemble,
\be
\delta  \mathcal{F}(x,Q^2,A) = \lc   \frac{1}{N_{\rm rep}} \sum_{k=1}^{N_{\rm rep}} \lp \mathcal{F}^{{\rm {(net)}}(k)}(x,Q^2,A) \rp^2
- \lp \la \mathcal{F}^{\rm (net)}(x,Q^2,A)\ra_{\rm rep} \rp^2 \rc^{1/2} \, ,
\ee
where $\mathcal{F}$ represents any of the three structure functions $F_2, xF_3$, and $ F_L$.
Similar considerations apply to  other statistical estimators such as correlations coefficients,
higher moments, and confidence level intervals.

\begin{figure}[t!]
	\centering
	\includegraphics[width=0.495\linewidth]{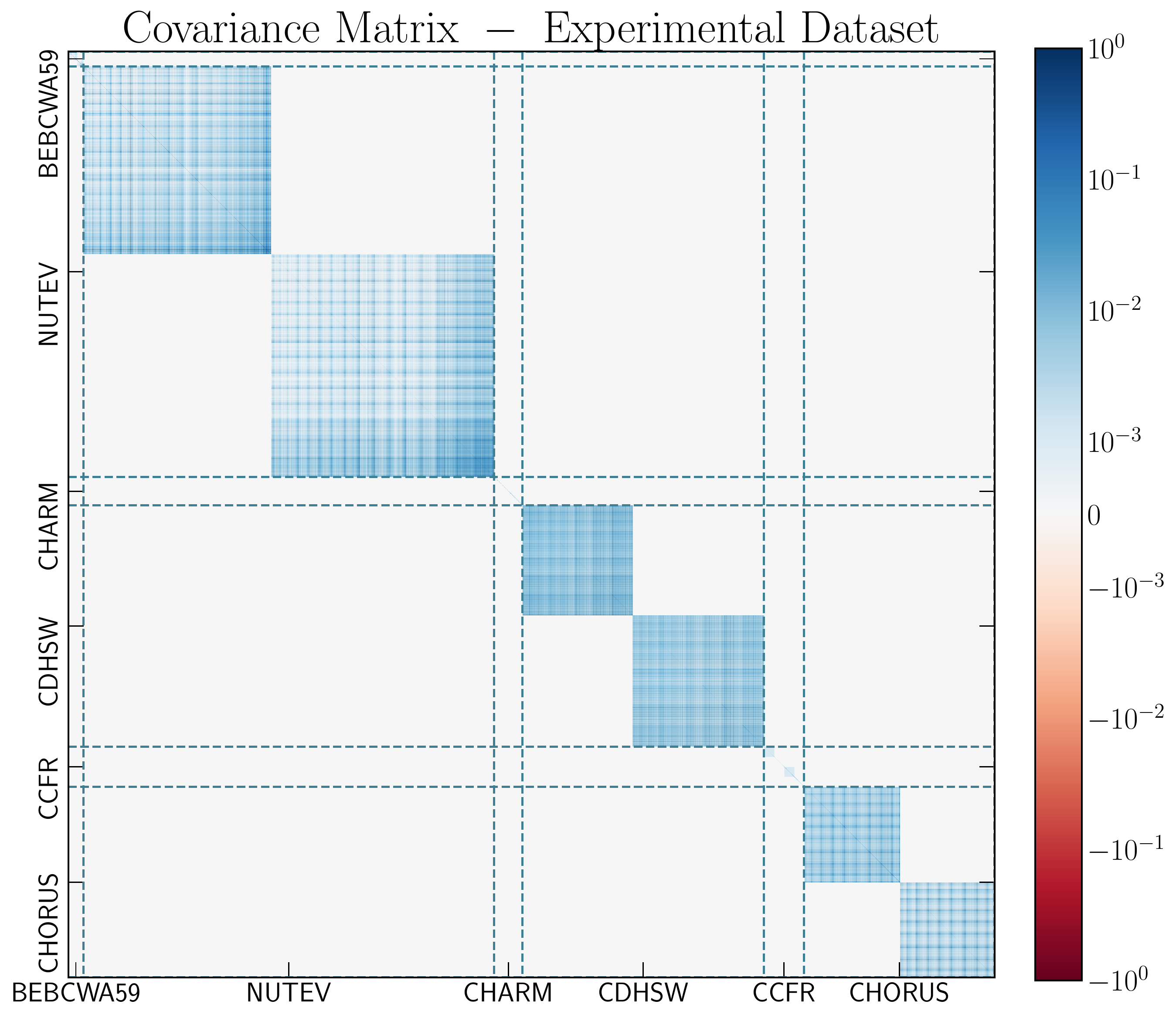}
	\includegraphics[width=0.495\linewidth]{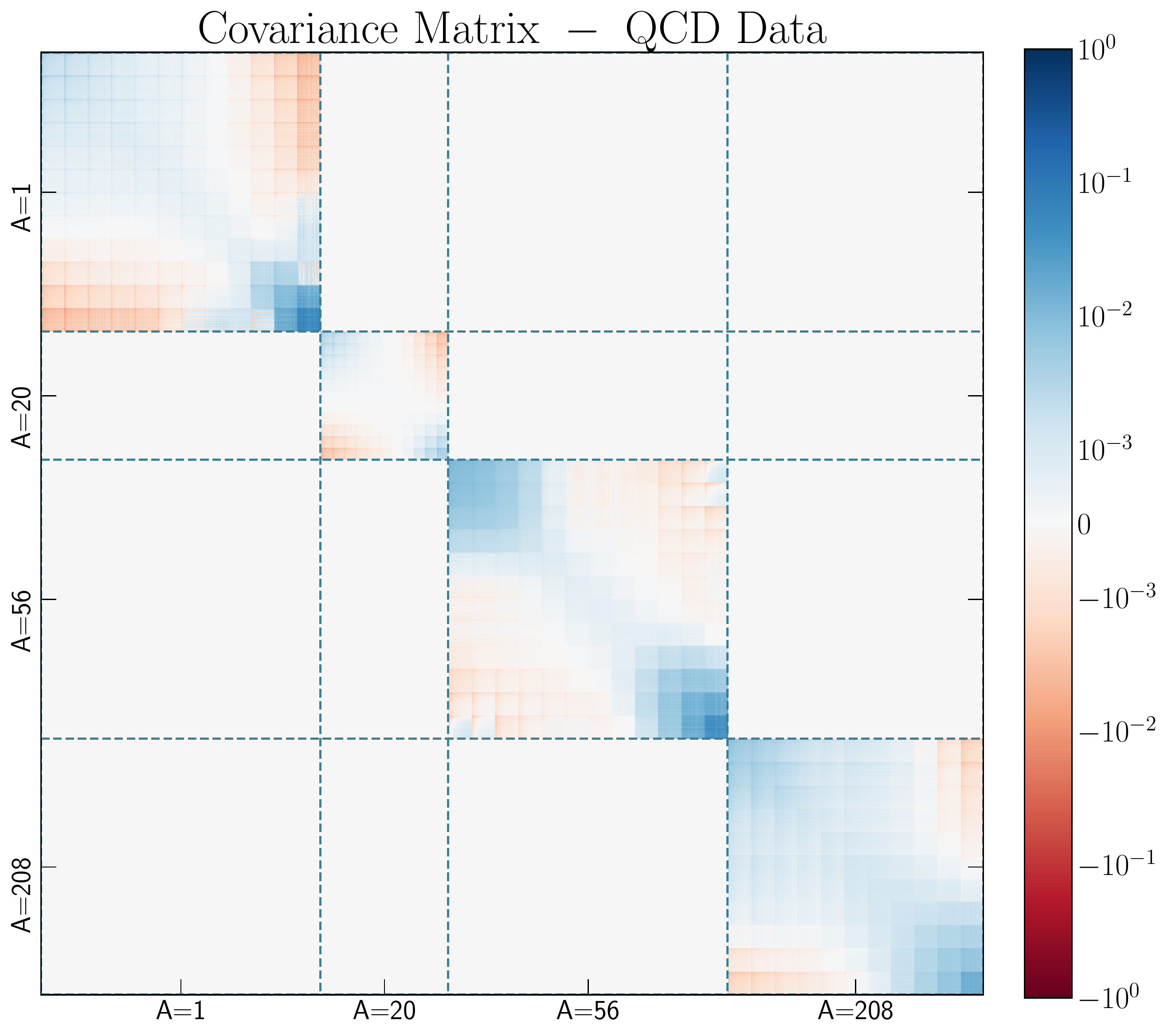}
	\includegraphics[width=0.495\linewidth]{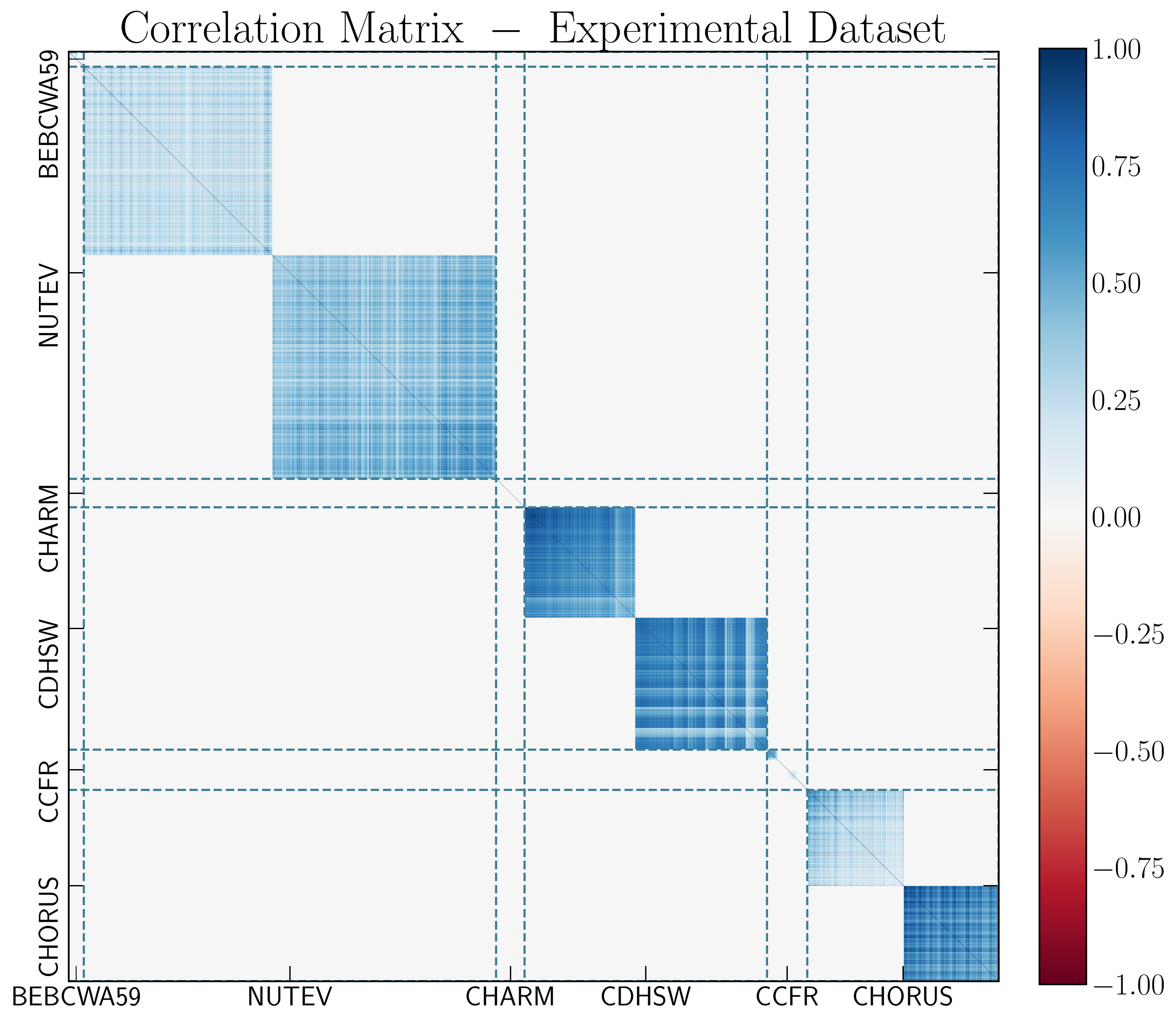}
	\includegraphics[width=0.495\linewidth]{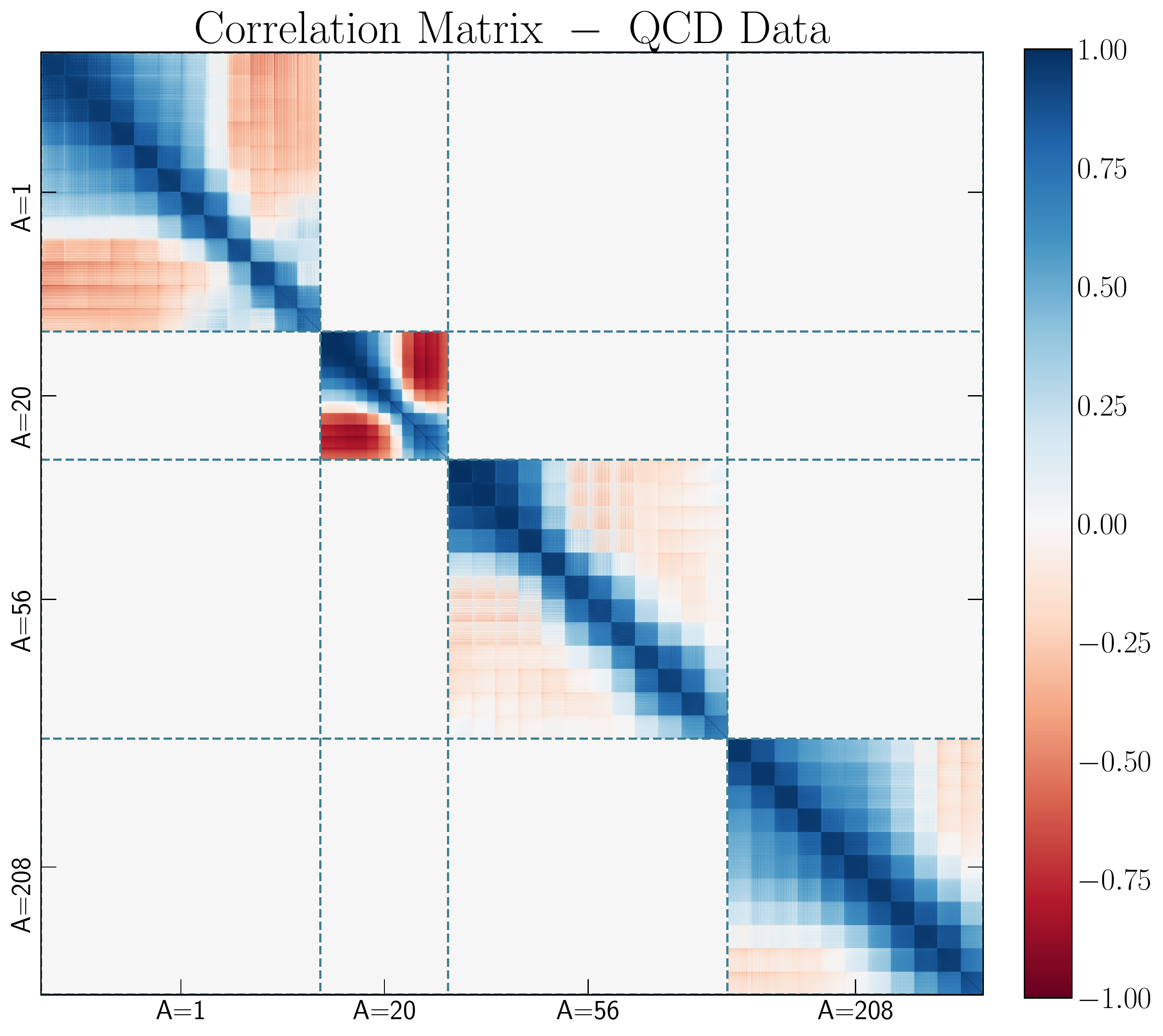}
	\caption{Left panels: the experimental covariance (top)
          and correlation (bottom) matrices relevant for the fit in Region I.
          Right panels: same for the theory
          covariance matrix associated to the QCD data entering the fit in Region II,
          with data points ordered in
          increasing order of $x$, $Q^2$, and $A$.
        There is no kinematic overlap between the
        entries of the experimental and theoretical matrices.
}
	\label{fig:covariance_matrix}
\end{figure}

Both the experimental  covariance matrix and the associated
correlation matrix given by
\be
 \label{eq:corr_mat_exp}
\lp \rho_{\mathrm{exp}}\rp_{ij}=\frac{\left(\operatorname{cov}_{\rm exp}\right)_{i j}}{
\sqrt{\left(\operatorname{cov}_{\rm exp}\right)_{i i}}\sqrt{\left(\operatorname{cov}_{\rm exp}\right)_{jj}}} \, ,
\ee
are displayed in the left panels of Fig.~\ref{fig:covariance_matrix}
for the data points listed in Table~\ref{tab:neutrino-DIS} after cuts.
The matrices are block-diagonal since the different experiments
are uncorrelated among them.
In most cases, neutrino inelastic scattering
experiments are limited by the correlated systematic uncertainties
rather than by the statistical errors.

\paragraph{QCD constraints.}
In Region II, the same neural network parameterisation
is trained now on QCD predictions based on {\sc\small YADISM} and
nNNPDF3.0 rather than to the experimental data.
Taking the central nNNPDF3.0 set as baseline,
we generate  $n_{\rm th}=100 \times n_{\rm sf}$ structure function QCD predictions distributed in 
$x$, $Q^2$, and $A$ covering Region II,
\be
\label{eq:qcd_pseudodata_region}
10^{-3} \le x \le 1 \, , \qquad  Q^2_{\rm dat}\le Q^2 \le Q^2_{\rm thr} \, , \qquad  1 \le A \le 208\, ,
\ee
with $n_{\rm sf}=6$ denoting the number of independent structure functions being parametrised.
As indicated in Eq.~(\ref{eq:qcd_pseudodata_region}) we include
a QCD boundary condition for $A=1$, understood as an isoscalar free-nucleon target.

By analogy with Eq.~(\ref{eq:replicas}), also in Region II one generates $N_{\rm rep}$
(independent) Monte Carlo replicas, this time starting from the
central QCD predictions rather than from the experimental data.
If we denote these central QCD structure functions by $\mathcal{F}_i^{(\rm qcd)(0)}$, we generate $N_{\rm rep}$ replicas as follows
\bea
\label{eq:qcd_replicas}
\mathcal{F}^{\rm (qcd)}_i &=& \mathcal{F}^{\rm (qcd)(0)}_i + \sum_{j=1}^{n_{\rm th}}
\left( \operatorname{cov}_{\rm th }^\star\right)_{i j} r_j^{{(\rm th)}} \,,  \quad i=1,\ldots,n_{\rm th} \, , \nonumber \\
\mathcal{F}^{{\rm (qcd)}(k)}_i &=& \mathcal{F}^{\rm (qcd)}_i + \sum_{j=1}^{n_{\rm th}}
\left( \operatorname{cov}_{\rm th }^\star\right)_{i j} r_j^{{(\rm th})(k)} \,, \quad k=1,\ldots,N_{\rm rep} \,, \quad i=1,\ldots,n_{\rm th} \, ,
\eea
where $\operatorname{cov}_{\rm th }^\star$ indicates the transpose of the Cholesky decomposition of
the theory covariance matrix and $r_j^{{(\rm th})}$, $r_j^{{(\rm th})(k)}$ represent stochastic noise sampled from a standard normal
distribution.
In order to treat the QCD data on the same footing
as the experimental data,
it is necessary to add two levels of stochastic noise to the
central predictions $\mathcal{F}_i^{(\rm qcd)(0)}$:
first to account the statistical fluctuations around the true underlying law,
and second to generate the Monte Carlo replicas themselves.
In the language of the closure test formalism~\cite{NNPDF:2014otw},
the first and second lines of Eq.~(\ref{eq:qcd_replicas}) correspond to level-1 and
level-2 pseudodata generation, while $\mathcal{F}_i^{(\rm qcd)(0)}$ is the level-0
underlying law.

The theory covariance matrix entering Eq.~(\ref{eq:qcd_replicas}) is constructed~\cite{NNPDF:2019ubu,NNPDF:2019vjt} as
the sum in quadrature of the contributions from the MHO and PDF uncertainties,
\begin{equation}
  \label{eq:cov_mat_def}
\left(\operatorname{cov}_{\rm th }\right)_{i j}=
\left(\operatorname{cov}_{\rm mho }\right)_{i j}+
\left(\operatorname{cov}_{\rm pdf }\right)_{i j} \, ,
\end{equation}
where the MHOU contribution
is evaluated using the NLO scheme-B with the 9-point scale variation prescription as implemented in {\sc\small YADISM}, 
and the PDF contribution is evaluated from the $\widetilde{N}_{\rm rep}=200$ replicas of the nNNPDF3.0 determination.
In constructing the theory covariance matrix Eq.~(\ref{eq:cov_mat_def}),
the correlations between different nuclear targets are neglected.

In Region II the figure of merit used for the neural network training should be, instead of Eq.~(\ref{eq:Et0_regionII}),
\begin{equation}
  \label{eq:Et0_regionII}
E_{\rm th}^{(k)}=\frac{1}{n_{\rm th}}\sum_{i, j=1}^{n_{\mathrm{th}}}\left(\mathcal{F}^{{\rm (net)}(k)}_i-\mathcal{F}^{{\rm (qcd)}(k)}_i \right)
\left(\operatorname{cov}_{\rm th}^{-1}\right)_{i j}\left(\mathcal{F}^{{\rm (net)}(k)}_j-\mathcal{F}^{{\rm (qcd)}(k)}_j\right)\, , \quad k=1,\ldots,N_{\rep} \ ,
\end{equation}
with goodness-of-fit after the training of all replicas quantified now by the counterpart of Eq.~(\ref{eq:chi2t0}),
\begin{equation}
 \label{eq:chi2t0_regionII}
\chi^2_{\rm th}=\frac{1}{n_{\rm th}}\sum_{i, j=1}^{n_{\mathrm{th}}}
\left(\la \mathcal{F}^{\rm (net)}_i\ra_{\rm rep}-\mathcal{F}^{\rm (qcd)}_i \right)\left(\operatorname{cov}_{\rm th}^{-1}\right)_{i j}\left(
\la \mathcal{F}^{\rm (net)}_j\ra_{\rm rep}-\mathcal{F}^{\rm (qcd)}_j\right)\, .
\end{equation}
Analogously to the experimental case, we can define the theoretical correlation matrix as
\be
 \label{eq:corr_mat_th}
\rho^{(\rm th)}_{ij}=\frac{\left(\operatorname{cov}_{\rm th}\right)_{i j}}{
\sqrt{\left(\operatorname{cov}_{\rm th}\right)_{i i}}\sqrt{\left(\operatorname{cov}_{\rm th }\right)_{jj}}}
\ee
which is displayed, together with the corresponding theory covariance matrix,
in the right panels of Fig.~\ref{fig:covariance_matrix}, where QCD data points
are sorted in increasing order of $x$, $Q^2$, and $A$.
Recall that there is no kinematic overlap between the entries of the experimental (Region I) and the theoretical (Region II) correlation matrices.
One difference between $\rho^{(\rm th)}$ and $\rho^{(\rm exp)}$
is that in the former case QCD induces correlations between all data points (for a given
value of $A$), arising
because the same underlying nPDF determination is used as well as due to the
correlation of factorisation scale variations entering DGLAP evolution.

\paragraph{Minimisation.}
Adding up the contributions from the experimental data in Region I and from
the QCD predictions in Region II,
the total figure of merit used for the training of the neural network parametrisation of neutrino structure functions is given by
\be
 \label{eq:E_total}
E_{\rm tot}^{(k)}=\frac{1}{\lp n_{\rm dat}+n_{\rm th}\rp } \lp n_{\rm dat} E_{\rm exp}^{(k)} + n_{\rm th} E_{\rm th}^{(k)}  \rp \, , \quad k=1,\ldots,N_{\rep} \, .
\ee
In minimising Eq.~(\ref{eq:E_total}), it is paramount to achieve a balanced description of the two regions, that is,
the average error function in both regions should be similar.
Furthermore,
 one expects $\la E_{\rm exp}\ra_{\rm rep}\sim \la E_{\rm th}\ra_{\rm rep}\sim 2$ 
in the absence of tensions or inconsistencies in the data~\cite{Forte:2002fg}.
A consequence of this requirement is that the fit quality to the data should not be distorted
by the inclusion of the QCD constraints,
and hence in a fit variant using  $E_{\rm tot}^{(k)}=E_{\rm exp}^{(k)}$ as figure or merit
the description of the experimental data should be comparable to that in the fits based
on Eq.~(\ref{eq:E_total}).
We will demonstrate this stability of the NNSF$\nu$ procedure in Sect.~\ref{sec:results}.

For the overall goodness-of-fit the total $\chi^2$ is evaluated as
\be
 \label{eq:chi2_total}
\chi^2_{\rm tot}=\frac{1}{\lp n_{\rm dat}+n_{\rm th}\rp } \lp n_{\rm dat} \chi^2_{\rm exp} + n_{\rm th} \chi^2_{\rm th} \rp \, ,
\ee
which again for a balanced fit should satisfy $ \chi^2_{\rm exp} \sim \chi^2_{\rm th}\sim 1$.
In the following we will only quote the values of the experimental contribution $ \chi^2_{\rm exp}$,
since we verify that a comparable
fit quality is obtained for the QCD component of the error function.

Eqns.~(\ref{eq:E_total}) and~(\ref{eq:chi2_total})
can also be understood as adding to the experimental $\chi^2$ an extra contribution
in the form of a Lagrange multiplier that enforces an external theory constraint, in this case
that the neural network extrapolation in $Q^2$ reproduces the QCD prediction.
Such Lagrange multiplier method is commonly used in the NNPDF framework to account
for theory constraints, such as positivity and integrability in NNPDF4.0
and the $A=1$ free-nucleon boundary condition in nNNPDF3.0.
One benefit of the approach adopted here is that the theory covariance matrix
Eq.~(\ref{eq:cov_mat_def})
provides automatically the appropriate normalisation for the Lagrange multiplier
contribution.

The minimisation of Eq.~(\ref{eq:E_total}) is carried out
 by means of the  adaptive SGD methods
available in {\sc\small TensorFlow}, specifically by {\tt Adam}.
The choice of optimisation algorithm, as well as that of other
hyperparameters defining the methodology, has been determined by
inspection of the fit results and performance.
Table~\ref{tab:hyperparameters} lists
the values of the hyperparameter configuration used
in the baseline NNSF$\nu$ determination, from the network
architecture to the learning rate and the settings
of the cross-validation stopping described below.
A gradient clipping procedure is used to normalize
gradient tensors such that their $L_2$-norm is less than or equal to the clipnorm
value.

\begin{table}[h]
  \centering
  \small
  \renewcommand{\arraystretch}{1.70}
  \begin{tabularx}{\textwidth}{XX}
    \toprule
      Hyperparameter  &Value         \\
    \hline
    Architecture            & 3-70-55-40-20-20-6    \\
    Activation function (hidden layers)    & hyperbolic tangent  \\
    Activation function (output layer)     & Scaled Exponential Linear Unit (SELU)  \\
    Optimizer               & {\tt Adam}          \\
    Clipnorm                & $10^{-5}$           \\
    Learning rate           & $10^{-3}$              \\
    Stopping patience       & $4.0 \cdot 10^4$     \\
    Maximum epochs          & $5.0 \cdot 10^5$              \\
    Training fraction       & 0.75                \\
    \hline
  \end{tabularx}
  \vspace{0.2cm}
  \caption{\small The values of the hyperparameter configuration used
    in the baseline NNSF$\nu$ determination.
   }
  \label{tab:hyperparameters}
\end{table}

Once all $N_{\rm rep}$ neural network replicas have been trained,
a post-fit selection procedure is carried out to filter out eventual
outliers associated to e.g. minimisation inefficiencies.
Specifically, we define (and remove) an outlier replica
as that exhibiting $E_{\rm tr}^{(k)}$ or $E_{\rm val}^{(k)}$ values
4$\sigma$ away from the mean of the associated distributions.
We also filter out replicas for which $E_{\rm val}^{(k)}$ never reaches
below a threshold of 3.

\paragraph{Stopping criterion.}
To avoid overfitting,  the same
early-stopping cross-validation algorithm as used in NNDPF4.0 determination is adopted.
For each dataset,
75\% of the datapoints are randomly sampled to generate a training dataset while
the other 25\% of datapoints constitute the validation set.
The SGD 
minimisation algorithm is trained of $E_{\rm tr}$   while simultaneously monitoring $E_{\rm val}$.
The optimal state of the neural network corresponds to the
training iteration at which the value of $E_{\rm val}$  has the lowest
value, and once training has ended the state of the model is reverted to this point
before producing the final outputs.
Training can end when one of two conditions is met, whichever comes earliest;
either the validation $\chi^2$  has not improved for a given number of steps (known
as stopping patience), or
a threshold number of training epochs is reached.
See Table~\ref{tab:hyperparameters} for the values of the associated
hyperparameters.

\paragraph{Performance.}
For the baseline hyperparameter configuration summarised in Table~\ref{tab:hyperparameters},
the training of a NNSF$\nu$ replica takes on average around 10 hours.
While the number of fitted data points $n_{\rm dat}$ is of the same order to that
of the NNPDF4.0 analysis, the fit takes a factor 25 more to converge,
see Table~3.4 of~\cite{NNPDF:2021njg}, despite the network output being directly
compared to the data without the intermediate requirement of the {\tt FK}-table convolution
present in PDF fits.
This behaviour has a two-fold explanation.
First of all, one is now exploring a three-dimensional parameter space
in $(x,Q^2,A)$, as compared to the one-dimensional space relevant for
a PDF determination where only the $x$ dependence is constrained by the data.
Second, the fit needs to reproduce not only the experimental data but also
the QCD constraints which impose boundary conditions also
in a three-dimensional  $(x,Q^2,A)$ space.

\section{NNSF$\nu$ structure functions}
\label{sec:results}

Here we present the main results of this work, the NNSF$\nu$
determination of inelastic neutrino structure functions valid for momentum
transfers $Q^2$ in Regions I and II as defined
in Fig.~\ref{fig:general-strategy}.
The corresponding implications for
 inclusive neutrino scattering cross-sections
are then presented in Sect.~\ref{sec:inclusive_xsec}, while the matching procedure of the NNSF$\nu$
outcome with the {\sc\small YADISM} QCD calculations appropriate
for Region III is described in App.~\ref{app:delivery}.

First of all we assess the quality of the fit to the experimental data on neutrino structure
functions  and compare NNSF$\nu$  with 
representative measurements.
Then we study the dependence with $x$ and $Q^2$ of the NNSF$\nu$ determination,
and in particular demonstrate that in Region II it correctly reproduces the QCD predictions
defining the theoretical boundary condition of the fit.
We also compare the NNSF$\nu$ determination
with the  Bodek-Yang and BGR18 calculations.
We present a number of alternative  NNSF$\nu$ fits with dataset or methodology variations
in order to assess the stability of our results.
Finally, we study the implications of our analysis for the Gross-Llewellyn Smith
sum rule
and verify the agreement with the perturbative QCD expectations.

\subsection{Fit quality and performance}

Here we assess the fit quality to the experimental data and the QCD boundary conditions,
quantify the fit performance including the small-$x$ preprocessing,
and provide representative comparisons between the
NNSF$\nu$ predictions and some of the fitted observables.

\begin{table}[t]
	\centering
	\small
	\renewcommand{\arraystretch}{1.60}
	\begin{tabularx}{\textwidth}{ c *{5}{Y} }
		\toprule
		Dataset & Target & Observable & $n_{\rm dat}$ (cuts) & $\chi^{2}_{\rm exp}$ (wo QCD) & $\chi^{2}_{\rm exp}$ (baseline) \\ 
		\hline
		\multirow{2}{*}{BEBCWA59} & \multirow{2}{*}{Ne} & $F_2$ & 57 (39) & 1.673 & 2.088 \\
		& & $x F_3$ & 57 (32) & 0.842 & 0.771 \\
		\hline
		\multirow{2}{*}{CCFR} & \multirow{2}{*}{Fe} & $F_2$ & 128 (82) & 1.902 & 2.292 \\
		& & $x F_3$ & 128 (82) & 0.857 & 0.946 \\
		\hline
		\multirow{5}{*}{CDHSW} & \multirow{5}{*}{Fe} & $\left[ F_2 \right]$ & 143 (92) & [6.17] & [5.32] \\
		& & $\left[ xF_3 \right]$ & 143 (100) & [22.9] & [11.7] \\
		& & $\left[ F_W \right]$ & 130 (95) & [15.9] & [16.4] \\
		& & $d\sigma^\nu / dxdQ^2$ & 847 (676) & 1.298 & 1.351 \\
		& & $d\sigma^{\bar{\nu}} / dxdQ^2$ & 704 (583) & 1.139 & 1.237 \\
		\hline
		\multirow{2}{*}{CHARM} & \multirow{2}{*}{CaCO$_3$} & $F_2$ & 160 (83) & 1.368 & 1.324 \\
		& & $x F_3$ & 160 (61) & 0.721 & 0.850 \\
		\hline
		\multirow{4}{*}{CHORUS} & \multirow{4}{*}{Pb} & $\left[ F_2 \right]$ & 67 (53) & [63.8] & [38.3] \\
		& & $\left[ x F_3 \right]$ &67 (53) & [6.881] & [2.904] \\
		& & $d\sigma^\nu / dxdQ^2$ & 606 (483) & 0.986 & 1.185 \\
		& & $d\sigma^{\bar{\nu}} / dxdQ^2$ & 606 (483) & 0.709 & 0.797 \\
		\hline
		\multirow{4}{*}{NuTeV} & \multirow{4}{*}{Fe} & $\left[ F_2 \right]$ & 78 (50) & [9.854] & [10.41] \\
		& & $\left[ x F_3 \right]$ & 75 (47) & [6.24] & [3.810] \\
		& & $d\sigma^\nu / dxdQ^2$ & 1530 (805) & 1.436 & 1.542 \\
		& & $d\sigma^{\bar{\nu}} / dxdQ^2$ & 1344 (775) & 1.254 & 1.311 \\
		\hline
		\textbf{Total} & & & {\bf 6197 (4089)} & {\bf 1.187} & {\bf 1.287}  \\
		\hline
	\end{tabularx}
	\vspace{0.2cm}
	\caption{\small The values of the experimental $\chi^2_{\rm exp}$ per data point, Eq.~(\ref{eq:chi2t0}),
          for the individual datasets entering the NNSF$\nu$ determination in Region I as well as for
          the total dataset.
          The datasets in brackets are not included  in the baseline to avoid double counting,
          and their $\chi^2$ values are evaluated a posteriori using the outcome of the
          NNSF$\nu$ fit.
          For each dataset we indicate the nuclear target, type of observable, number
          of data points before and after kinematic cuts, and the resulting values of $ \chi^2_{\rm exp}$.
          The latter are also provided by the fit variant where only the experimental
          data (Region I), but not the QCD predictions (Region II), are included in the fit
          and labelled as ``wo QCD''.
	}
	\label{tab:chi2-baseline}
\end{table}


\paragraph{Fit quality.}
Table~\ref{tab:chi2-baseline} we display the values of the experimental $\chi^2_{\rm exp}$ per data 
point, defined by Eq.~(\ref{eq:chi2t0}), for the individual datasets entering the NNSF$\nu$ 
determination as well  as for the total dataset.
For each dataset we indicate the nuclear target, the fitted observables, the number
of data points before and after kinematic cuts, and the values of the $ \chi^2_{\rm exp}$.
The latter are also provided by a fit variant (labelled as ``wo QCD'') where only the experimental data (in Region I), but not the
QCD predictions (in Region II), are included in the fitted error function.
The datasets in brackets are not included in the baseline fit to avoid double counting.
The quoted $\chi^2_{\rm exp}$
values are instead evaluated a posteriori using the outcome of the baseline NNSF$\nu$ fit.

From Table~\ref{tab:chi2-baseline} one finds that in the baseline fit a good description 
of the experimental data is obtained, with a total of
$\chi^2_{\rm exp}=1.287$ per data point for the 4089 points of 
considered.
The fit quality is in general similar among the input datasets, without major outliers.
This feature holds specially true for the three datasets that contribute the most in terms of statistical 
weight in the fit, namely CDHSW, CHORUS, and NuTeV, which are also the ones provided in terms
of the cleaner double-differential cross-sections.
In addition, a balanced descriptions of  neutrino and antineutrino data is obtained whenever
measurements for the two initial states are provided.
A somewhat worse fit quality is obtained for the $F_2$ data from BEBCWA59
and CCFR, which in any case carry less weight in the fit and are potentially affected by
the model-dependent separation from the measured cross-section.
For the CHORUS observables, we find $\chi^2_{\rm exp}=1.185$ and $0.797$ per data point
for the neutrino and antineutrino data respectively in the baseline fit.
These values can be compared with the corresponding ones
of 1.12 and 1.06 obtained in the NNPDF4.0 NNLO fit, where more stringent kinematic
cuts are applied resulting in $n_{\rm dat}=832$ data points as compared to 1580 in
the NNSF$\nu$ analysis.

As discussed in Sect.~\ref{sec:fitting}, in the context
of a matched analysis such as NNSF$\nu$ it is crucial to achieve a balanced
description of the experimental data (in Region I) and the QCD predictions (Region II)
during the fit.
In this respect,
we have verified that in the baseline fit the contribution to the total $\chi^2_{\rm tot}$
arising from the QCD constraints, $\chi^2_{\rm th}$ in Eq.~(\ref{eq:chi2_total}), is
of similar size as that for the experimental component $\chi^2_{\rm exp}$ as expected
for a balanced training.

Furthermore, by comparing the last two columns of Table~\ref{tab:chi2-baseline}, one can
assess the impact of accounting for the QCD structure function constraints 
in the fit quality to the experimental data.
As expected, the total fit quality is improved, $\chi^2_{\rm exp}=1.187$
down from $\chi^2_{\rm exp}=1.287$ in the baseline,
in the fit where only the experimental data enters the figure of merit, $E_{\rm tot}=E_{\rm th}$
instead of Eq.~(\ref{eq:E_total}).
Indeed, a better (or comparable) description of the data is generically expected 
once theoretical constraints are removed from the figure of merit,  given that the functional 
form space available for $F_i(x,Q^2,A)$ becomes less restricted.
Nevertheless, this improvement remains moderate confirming that the addition
of the QCD constraints in Region II does not significantly distort the description of the experimental 
data in Region I.
Furthermore, this improvement in the $\chi^2$ is homogeneously spread
among the input datasets, rather than being associated to specific ones.
We hereby conclude that the NNSF$\nu$ fit is dominated in Region I by
the experimental data constraints, with QCD boundary conditions  providing
a smooth transition to Region  II.

Table~\ref{tab:chi2-baseline} also indicates the values
of the $\chi^2_{\rm exp}$ corresponding to datasets not included in the baseline fit,
but rather evaluated a posteriori using the NNSF$\nu$ predictions.
Specifically, we list the values of the  $\chi^2_{\rm exp}$  for
the individual structure functions from CDHSW, CHORUS, NuTeV experiments.
These structure function datasets
 are not considered in the baseline fit since they would overlap
 with the corresponding reduced cross-sections.
 The NNSF$\nu$ predictions for the separate $F_2$ and $xF_3$ data excluded from the fit
lead to a poor $\chi^2$, indicating a potential internal inconsistency between the reduced cross-section
and separate structure function data.
In Sect.~\ref{sec:dataset-variations} we investigate this issue
by assessing the stability of
the NNSF$\nu$ fit results when the differential  cross-section
data is actually replaced by these separate structure function measurements.

\begin{figure}[t!]
	\centering
	\includegraphics[width=0.495\linewidth]{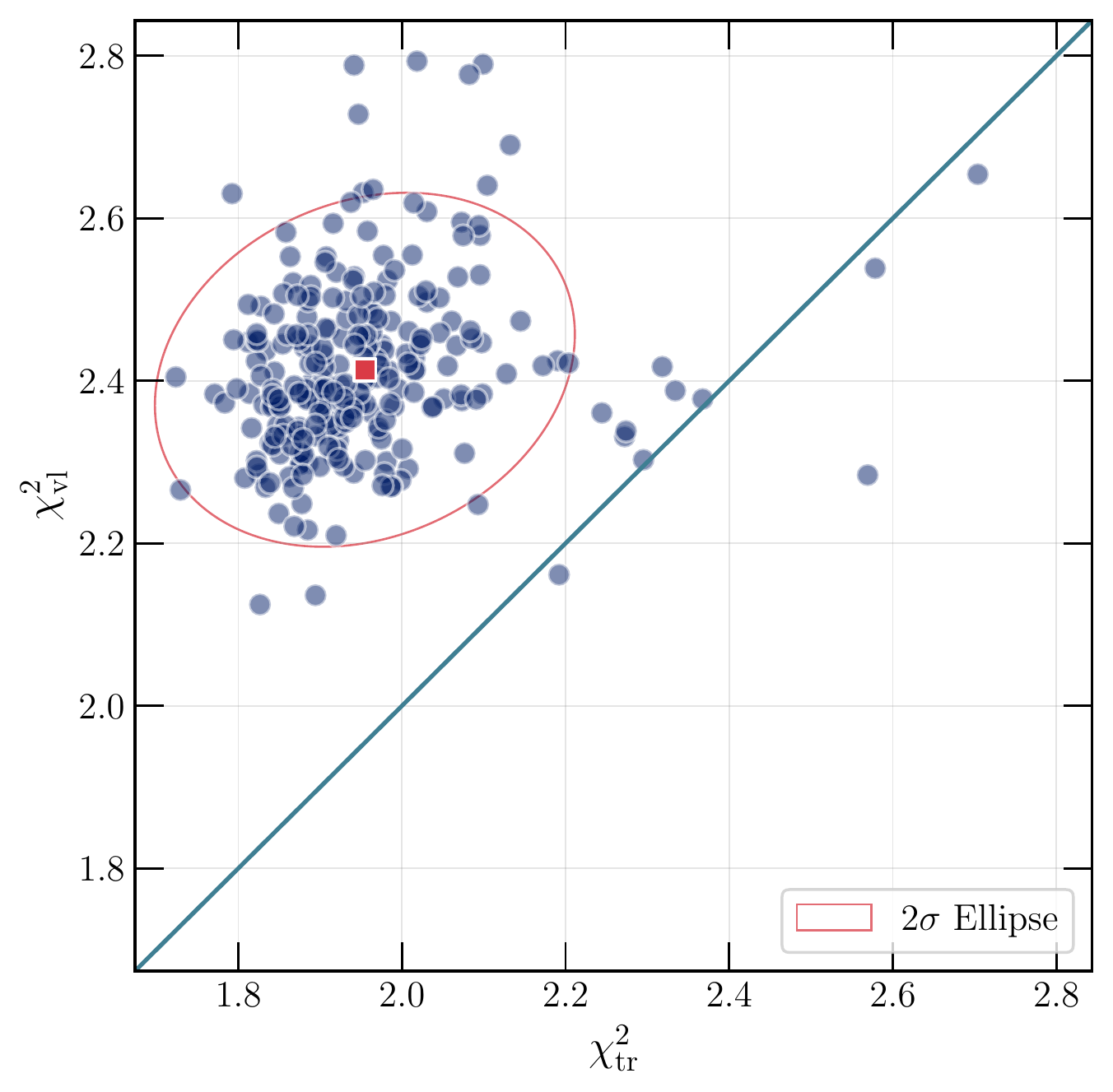}
	\includegraphics[width=0.495\linewidth]{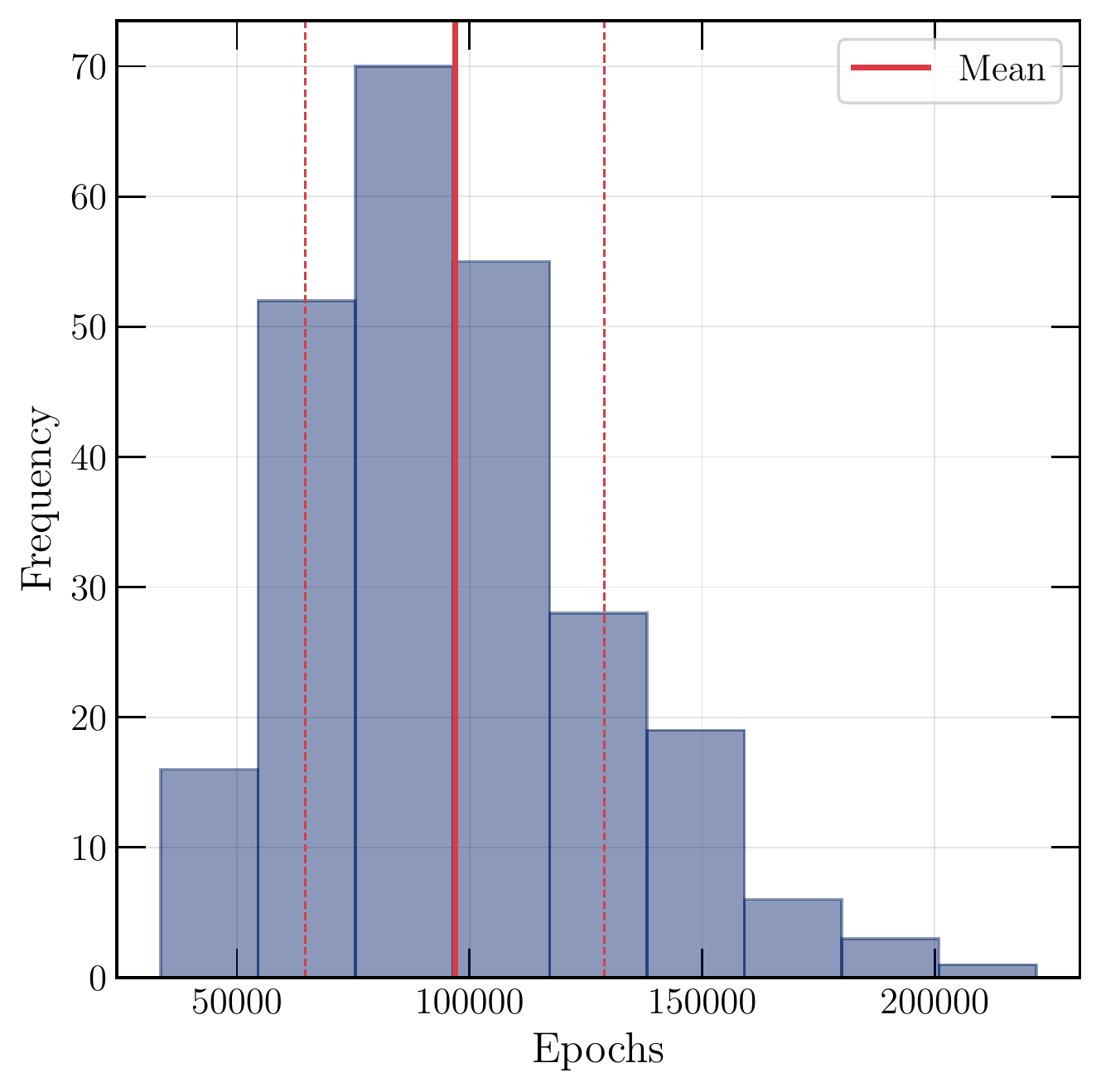}
	\caption{Left:  distribution of the training $(\chi^2_{\rm tr})$ and validation $(\chi^2_{\rm vl})$
	  	values for the fit error function $E_{\rm exp}^{(k)}$ over the $N_{\rm rep}=200$ replicas entering
          the NNSF$\nu$ data-driven parametrisation.
          We consider only the contribution from the experimental data in this comparison.
          The red square indicates the mean value over the replicas, while the red ellipse represents the 
          $2\sigma$-contour.
          Right: the corresponding distribution of training lengths. The solid red line indicates the
		mean value while the dashed red lines indicate the $1\sigma$ standard deviation.}
	\label{fig:fit_quality}
\end{figure}

\paragraph{Fit performance.}
Fig.~\ref{fig:fit_quality} displays the distribution of
the experimental training and validation error functions,
$E^{(k)}_{\rm tr}$ and $E^{(k)}_{\rm val}$, evaluated over
the $N_{\rm rep} = 200$ Monte Carlo replicas used in the fit.
The red square indicates the mean value over the replicas, while the red ellipse represents the
$2\sigma$-contour.
Since the validation datasets are not used for the optimisation, in general
one expects the 
values of $E^{(k)}_{\rm val}$ to be somewhat higher than those
of $E^{(k)}_{\rm tr}$, and indeed  $\la E_{\rm tr}\ra \sim 2.0 $
while $\la E_{\rm val}\ra \sim 2.4 $.
The replicas are clustered around the mean value and only a few outliers
are present.
Specifically, around 18 replicas fall outside the $2\sigma$-ellipse,
to be compared with the 10 that would be expected to lie outside the 95\% CL interval of
a purely Gaussian
distribution.
As explained in Sect.~\ref{subsec:error-prorpagation},
the post-fit procedure removes outlier replicas
exhibiting $E_{\rm tr}^{(k)}$ or $E_{\rm val}^{(k)}$ values
4$\sigma$ away from the corresponding mean value.

The right panel of Fig.~\ref{fig:fit_quality} shows the distribution of training lengths,
defined as the number of epochs at which the optimal stopping conditions are reached,
over the $N_{\rm rep} = 200$  replicas.
None of the replicas reach
the maximum number of iterations (see Table~\ref{tab:hyperparameters}), demonstrating
that in all cases  convergence is reached in a way that satisfies
the cross-validation stopping 
criterion.
Furthermore, the distribution is approximately Gaussian, with a mean of $N_{\rm ep} \sim 10^5$ epochs,
and does not present any significant long tails.
These two considerations point to a stable training which is evenly distributed among the
replica distribution.

\begin{figure}[t!]
	\centering
	\includegraphics[width=0.99\linewidth]{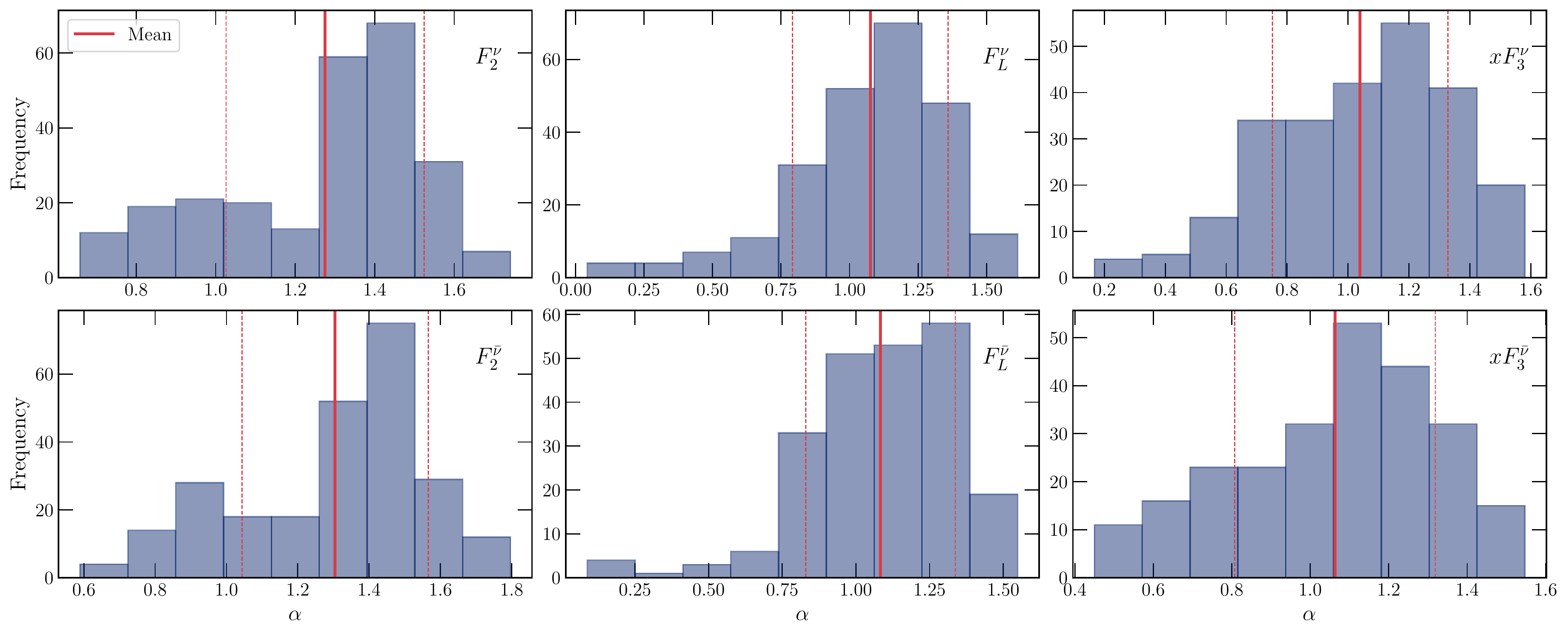}
	\caption{Posterior probability distributions for 
          the preprocessing exponents $\alpha_i,\bar{\alpha}_i$ as defined
	  in Eq.~(\ref{eq:NN_output}) for each of the
          fitted neutrino and antineutrino structure functions.
          As in Fig.~\ref{fig:fit_quality}, the distribution is sampled
          from the  $N_{\rm rep}=200$ replicas entering NNSF$\nu$.
          The vertical solid (dashed) lines indicate the mean values ($1\sigma$ ranges).
        }
	\label{fig:smallx_exponent}
\end{figure}

Fig.~\ref{fig:smallx_exponent} displays the
posterior probability distributions associated to
the preprocessing exponents $\alpha_i,\bar{\alpha}_i$ as defined
in Eq.~(\ref{eq:NN_output}) for each of the
fitted neutrino and antineutrino structure functions.
Recall that these exponents are part of the fitted parameters and their
values are restricted to the interval $\alpha_i \in \left[ 0, 2 \right]$. 
As in Fig.~\ref{fig:fit_quality}, the distribution is sampled
from the  $N_{\rm rep}=200$ replicas entering NNSF$\nu$.
For each structure function, the distributions
for the small-$x$ neutrino and antineutrino exponents, $\alpha_i$ and $\bar{\alpha}_i$
respectively, turn out to be  similar, indicating that the small-$x$ behaviour
of NNSF$\nu$ depends only mildly on the neutrino flavour.
The fact that despite $\alpha_i$ and $\bar{\alpha}_i$ being fitted separately,
the same distributions are obtained, is another indication of the fit stability,
given that QCD predicts that asymmetries between neutrino and antineutrino
structure functions are washed out in the small-$x$ region.
The three structure functions also prefer similar small-$x$ preprocessing
exponents, with median values of $\alpha_1 \sim 1.25$, $\alpha_2 \sim 1.1$, and
$\alpha_3 \sim 1.05$ for $F_2$, $xF_3$, and $F_L$ respectively,
and in agreement within uncertainties.
The distributions of $\alpha_i,\bar{\alpha}_i$ are non Gaussian and exhibit
a skewed tail towards smaller values of the exponents.

\subsection{Comparison with data and previous calculations}

Here we compare the NNSF$\nu$ results with other calculations
of neutrino structure functions, in particular
with BGR18 and Bodek-Yang, as well as with the {\sc\small YADISM} 
predictions based on nNNPDF3.0 which enter the fit as the QCD boundary
condition.
We then show the agreement between  NNSF$\nu$ and representative
datasets used in the fit.
Subsequently, we study the NNSF$\nu$ uncertainties and their
dependence with respect to variations in the
$x$, $Q^2$ and $A$ inputs of the parametrisation.

\paragraph{Comparison with Bodek-Yang and BGR18.}
First, we compare the NNSF$\nu$ predictions with those
from the  Bodek-Yang  and BGR18
calculations described in Sect.~\ref{sec:dis-sf}.
In both cases, we access their predictions
by means of their  implementation in the {\sc\small GENIE}
event generator.
For BGR18, we consider the variant  based on NLO coefficient functions and NNPDF3.1 NLO 
as input PDF.
We also display the {\sc\small YADISM} predictions based on nNNPDF3.0
which enter the fit as the QCD boundary condition in Region II.
The corresponding comparisons at the  inclusive neutrino 
cross-section level will be presented in Sect.~\ref{sec:inclusive_xsec}.
We consider predictions for an isoscalar free-nucleon
target, and also in Sect.~\ref{sec:inclusive_xsec} we will compare
results for other nuclear targets for inclusive cross-sections.

\begin{figure}[!t]
	\centering
	\includegraphics[width=\linewidth]{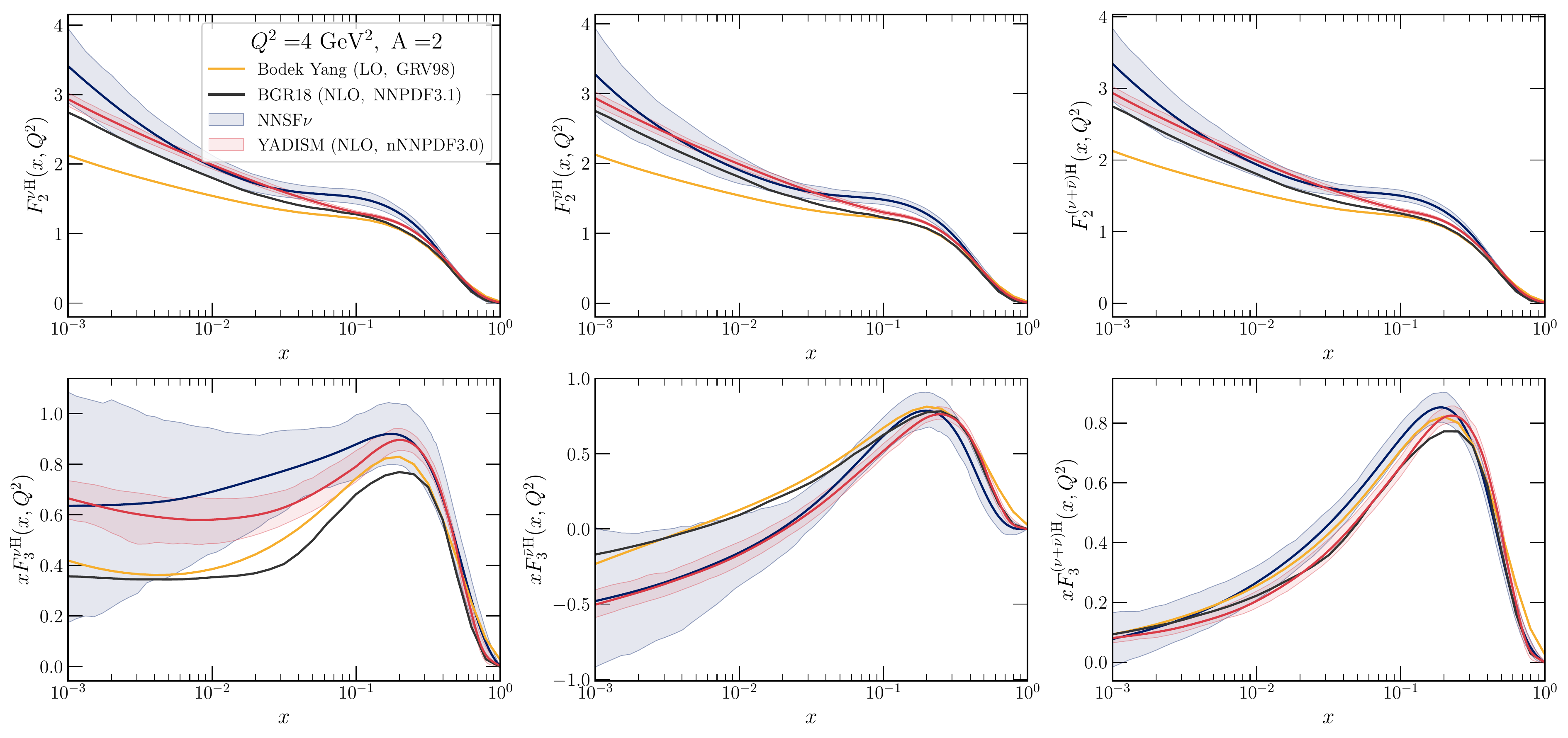}
	\includegraphics[width=\linewidth]{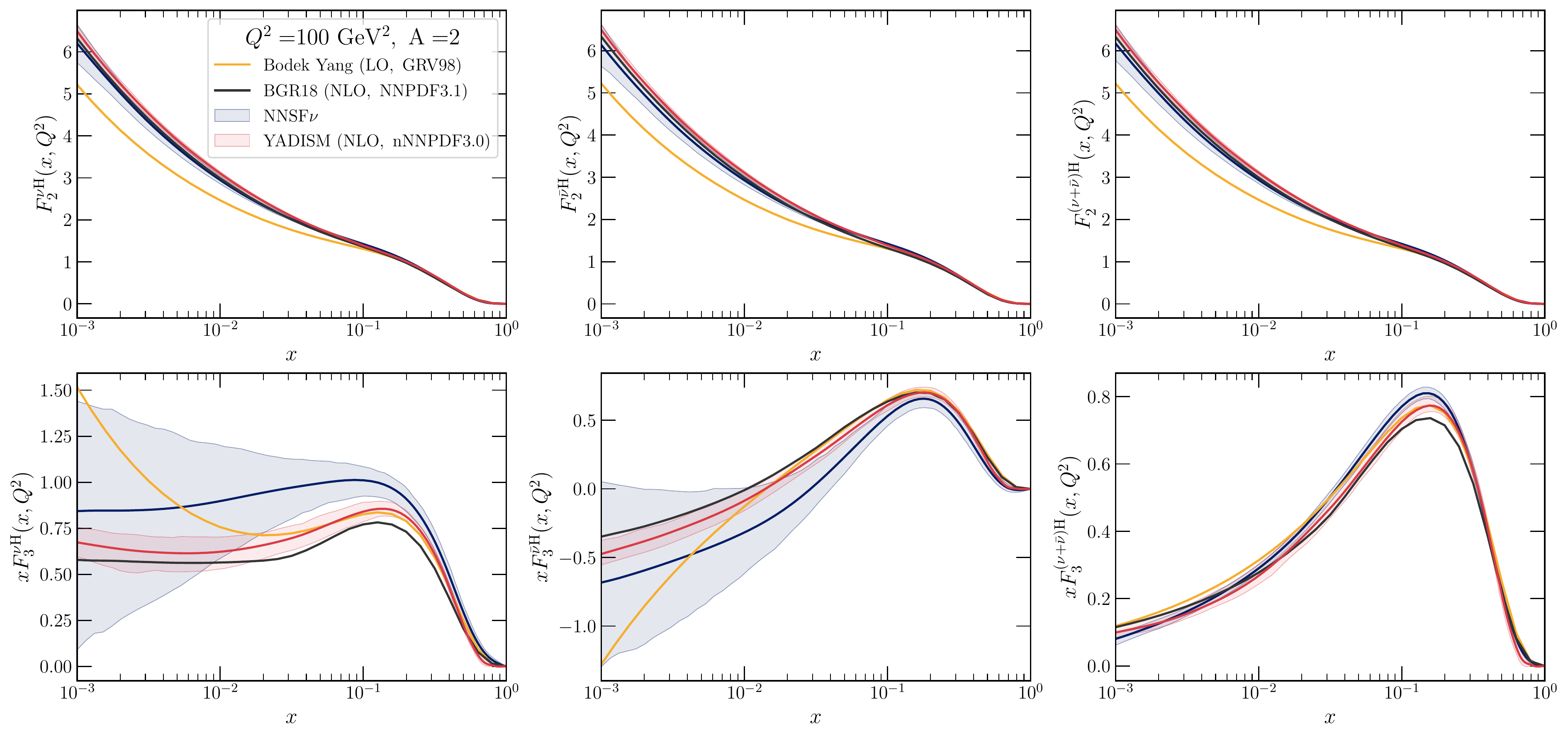}
	\caption{The NNSF$\nu$ predictions, together
          with the corresponding 68\% CL uncertainties,
          as function of $x$ at $Q^2=4~\rm{GeV^2}$ 
	  (Region I, top panels) and $Q^2=100^2~\rm{GeV^2}$ (Region II, bottom panels).
	  We display the $F_2$ and $xF_3$ structure functions for neutrinos, antineutrinos, and
	  their sum on an isoscalar free-nucleon target ($^2$H).
          The NNSF$\nu$ results are compared with the central values
          of the Bodek-Yang and BGR18 calculations, as well as with the {\sc\small YADISM}
          calculation based on nNNPDF3.0 (with the red band indicating the
          PDF uncertainties).
	}    
	\label{fig:StructureFunction-Comparisons-xdep}
\end{figure}

\begin{figure}[!t]
	\centering
	\includegraphics[width=\linewidth]{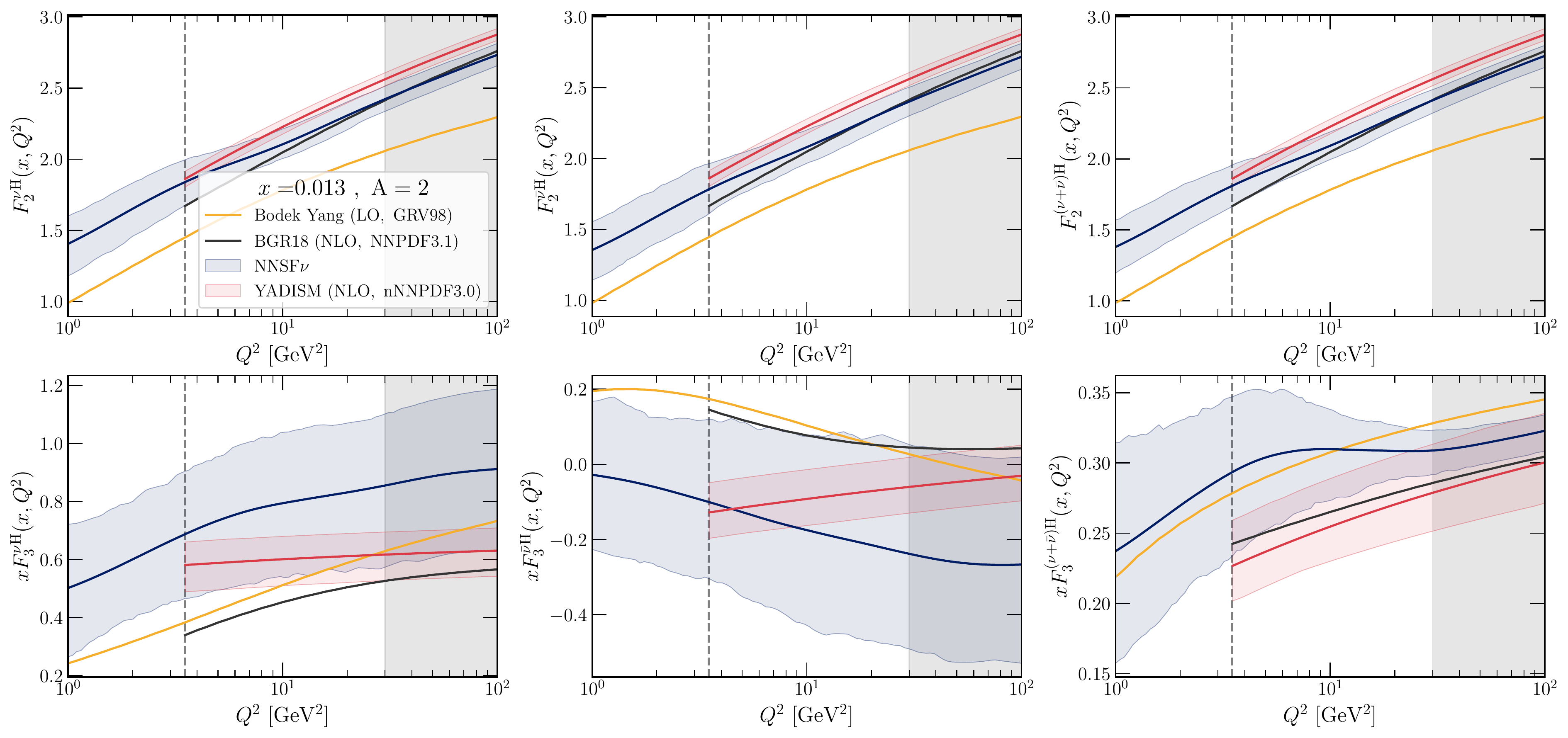}
	\includegraphics[width=\linewidth]{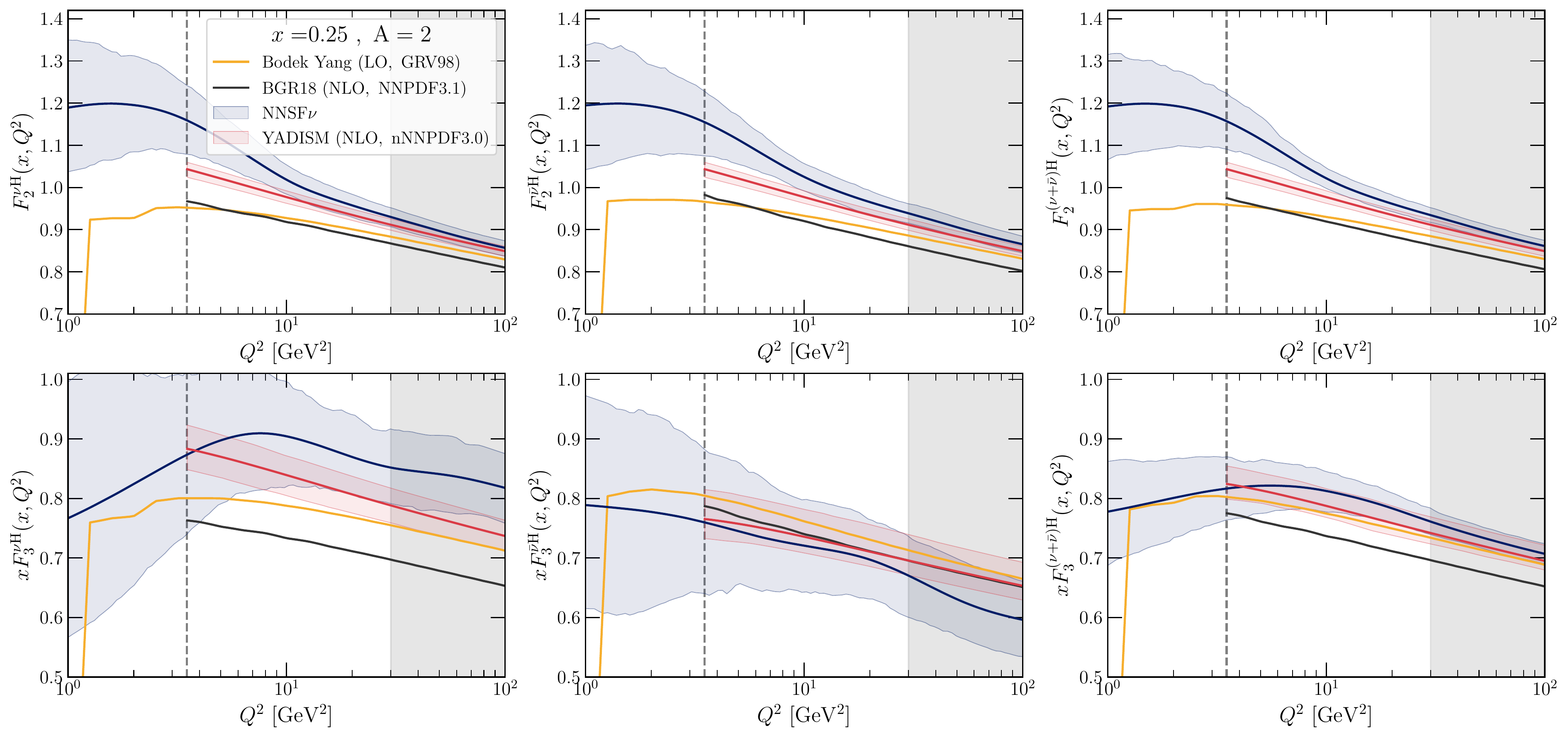}
	\caption{Same as Fig.~\ref{fig:StructureFunction-Comparisons-xdep}
          now
          as a function of 
	  $Q^2$ for $x=0.0125$ (upper) and $x=0.25$ (lower panels).
          The BGR18 and 
	  {\sc\small YADISM} calculations are restricted to the region $Q^2 \ge 3.5~\rm{GeV^2}$
          (indicated by a dashed grey vertical line) 
	to ensure the validity of perturbative QCD.
        The area covered in light gray indicates Region II,
        where the NNSF$\nu$ parametrisation is constrained 
	by the {\sc\small YADISM}  boundary condition rather than by the experimental data.
        }    
	\label{fig:StructureFunction-Comparisons-qdep}
\end{figure}

Figs.~\ref{fig:StructureFunction-Comparisons-xdep}
and~\ref{fig:StructureFunction-Comparisons-qdep} display the NNSF$\nu$ predictions as 
function of $x$ for $Q^2=4~\rm{GeV^2}$ (Region I) and $Q^2=100~\rm{GeV^2}$ (Region II)
and then as a function of  $Q$ for $x=0.0126$ and $x=0.25$, respectively.
We display the $F_2$ and $xF_3$ structure functions for neutrinos, antineutrinos, and
for their sum for an isoscalar free-nucleon target ($^2$H).
The error band on the  NNSF$\nu$ predictions indicates the  68\% confidence
level intervals evaluated over the $N_{\rm rep}=200$ Monte Carlo replicas.
We also consider the
central values of the Bodek-Yang and BGR18 calculations, as well
as the {\sc\small YADISM} prediction including PDF uncertainties.
For the  BGR18 and {\sc\small YADISM} calculations we only display results corresponding 
to the perturbative region with $Q^2 > 3.5~\rm{GeV^2}$.
In Fig.~\ref{fig:StructureFunction-Comparisons-qdep},
 the area covered in light gray indicates the $Q^2$ coverage of Region II, where 
 the NNSF$\nu$ parametrisation is constrained to reproduce the {\sc\small YADISM}
 QCD boundary condition.

From the comparisons in Figs.~\ref{fig:StructureFunction-Comparisons-xdep}
and~\ref{fig:StructureFunction-Comparisons-qdep} one can observe how the NNSF$\nu$ 
predictions reproduce the {\sc\small YADISM} boundary conditions at $Q^2=100~\rm{GeV^2}$
in the relevant region of $x$.
We verify that within the whole Region II there is agreement within uncertainties between NNSF$\nu$
and {\sc\small YADISM}, demonstrating that as required the QCD boundary
condition is being reproduced by the structure function parametrisation.
At medium and small-$x$, there is a good agreement between the BGR18 calculation and the NNSF$\nu$ 
predictions in the region of validity of the former (for $Q^2 \ge 4~\rm{GeV^2}$),
with some differences in the large-$x$ region.
It is  interesting to note that the agreement found between NNSF$\nu$ and {\sc\small YADISM}
in Region II is not automatic: for instance at $Q=2$ GeV (Region I), we find that for $x \in \lc 0.05, 0.3 \rc $
the two results disagree within uncertainties, showing that the experimental neutrino
data (rather than the QCD boundary condition) is driving the fit results there.

Furthermore, one observes from these comparisons how the Bodek-Yang calculation falls outside the $1\sigma$ error 
band of the NNSF$\nu$ prediction for a significant region of the relevant $x$ and $Q^2$ values,
in particular for $F_2$ at small- and large-$x$. 
For instance, for $x \simeq 0.25$ in the low-$Q$ region, the Bodek-Yang prediction is around 25\%
smaller than the NNSF$\nu$ one.
The agreement between the NNSF$\nu$ and Bodek-Yang structure functions improves for heavier
nuclei, as we will demonstrate in Sect.~\ref{sec:inclusive_xsec} when evaluating
the inclusive neutrino cross-sections on iron, tungsten, and lead targets.

Another interesting feature
of Figs.~\ref{fig:StructureFunction-Comparisons-xdep}
and~\ref{fig:StructureFunction-Comparisons-qdep} is the behaviour of NNSF$\nu$ in the extrapolation regions.
Concerning the $x$ dependence, the  NNSF$\nu$  uncertainties increase at small-$x$ specially at low-$Q$
due to the lack of direct experimental data, while in the same $x$ region at higher values of $Q$
these uncertainties are reduced due to the information provided by the QCD boundary condition.
Note that in the case of the $xF_3$ structure function,
the small-$x$ behaviour is fixed in the case of the $\nu+\bar{\nu}$ combination,
rather than for the individual $\nu$ and $\bar{\nu}$ structure functions.
Concerning the extrapolation in $Q^2$, the NNSF$\nu$  uncertainties increase as $Q^2$ decreases due to the lack
of data, and decrease as $Q^2$ increases as a consequence of the constraints from the QCD boundary condition. 

\begin{figure}[t!]
	\centering
	\includegraphics[width=0.325\linewidth]{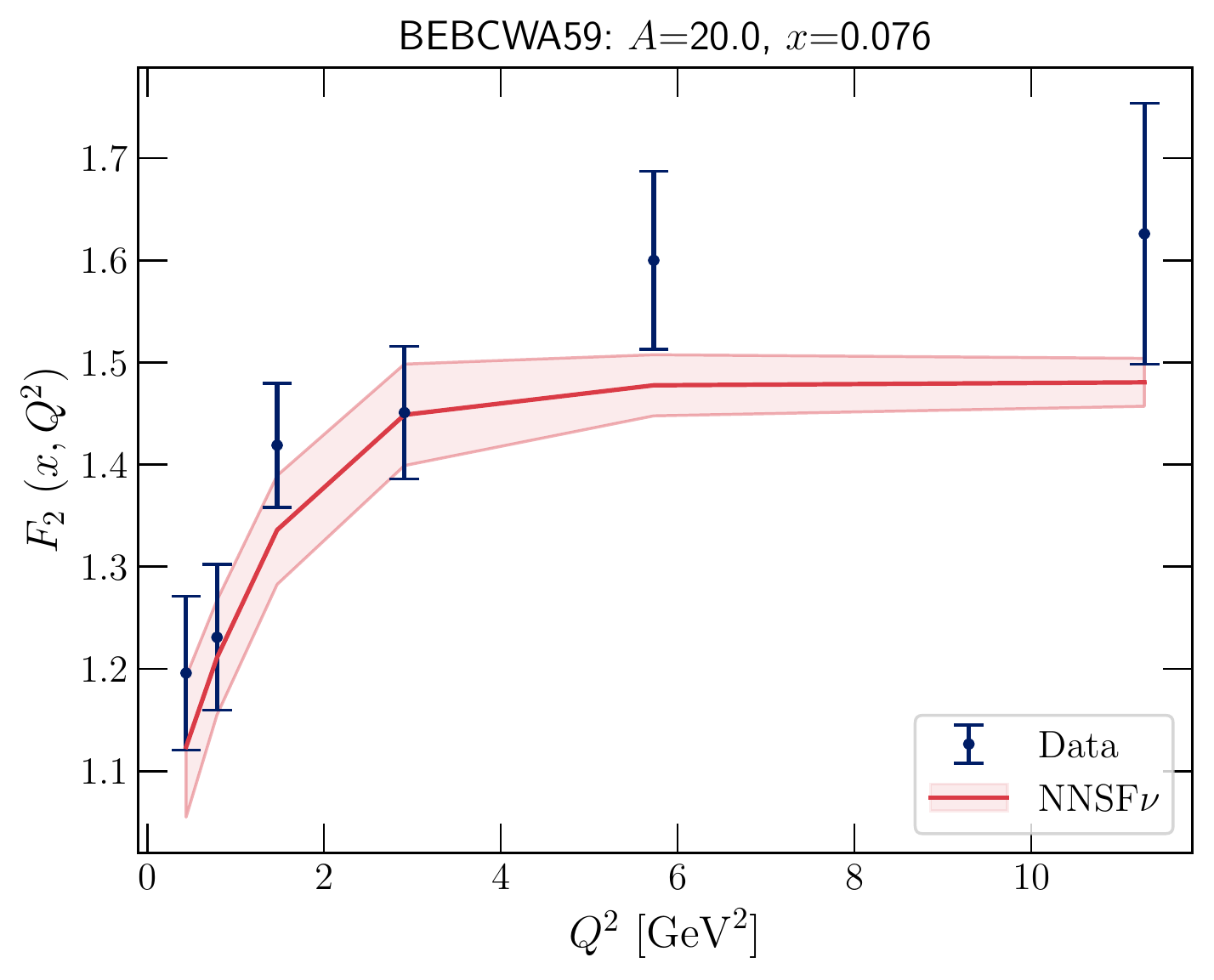}
	\includegraphics[width=0.325\linewidth]{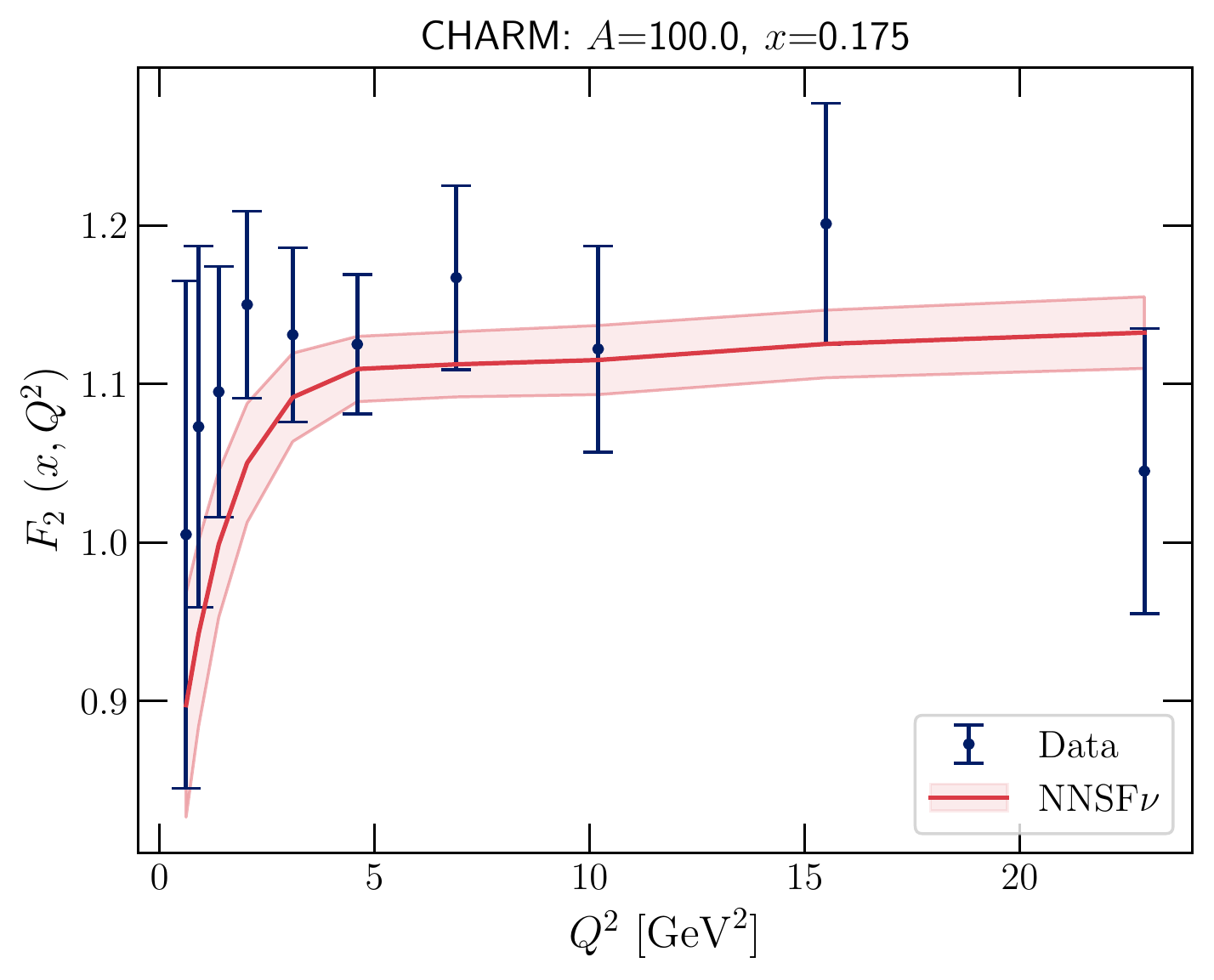}
	\includegraphics[width=0.325\linewidth]{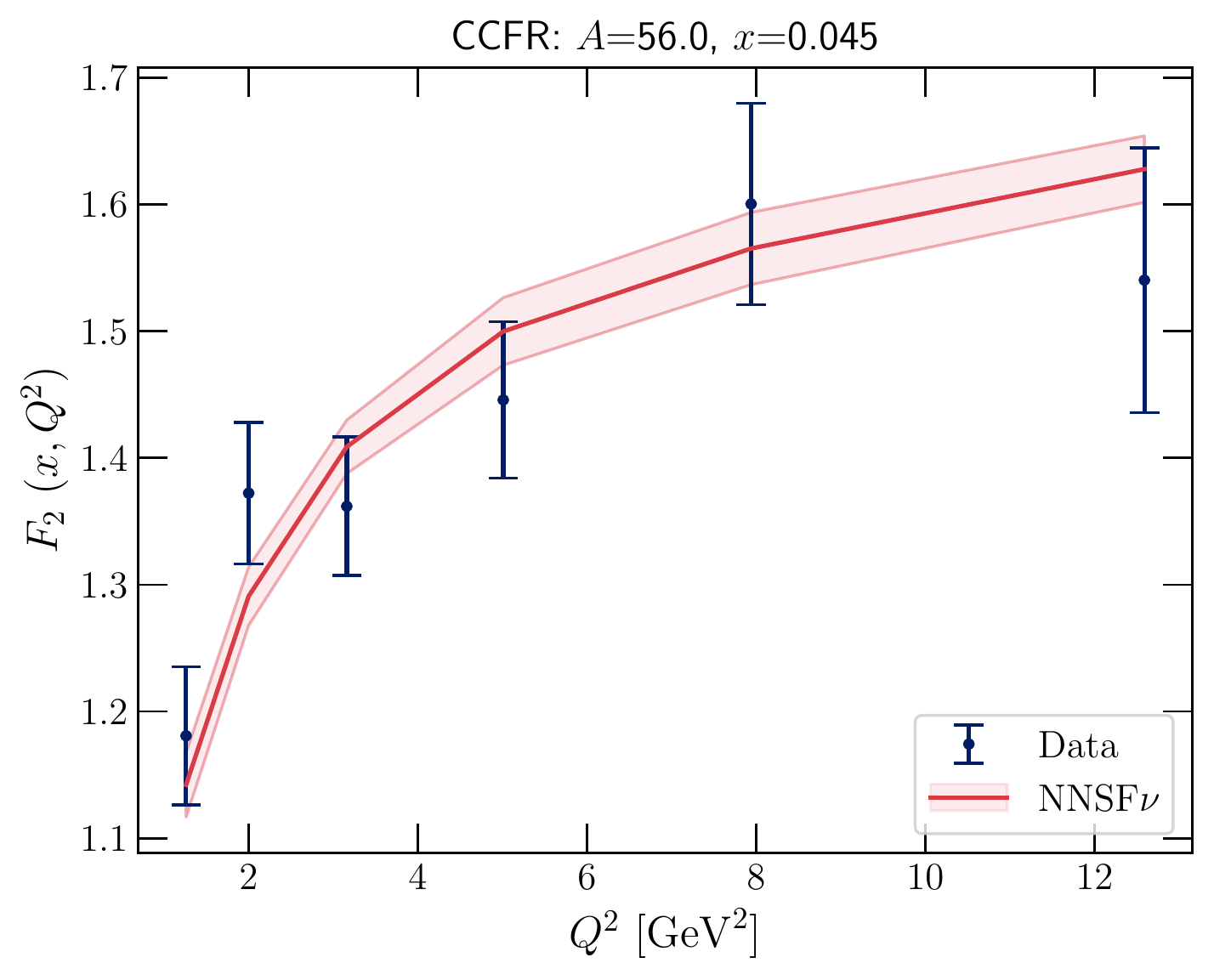}
	\includegraphics[width=0.325\linewidth]{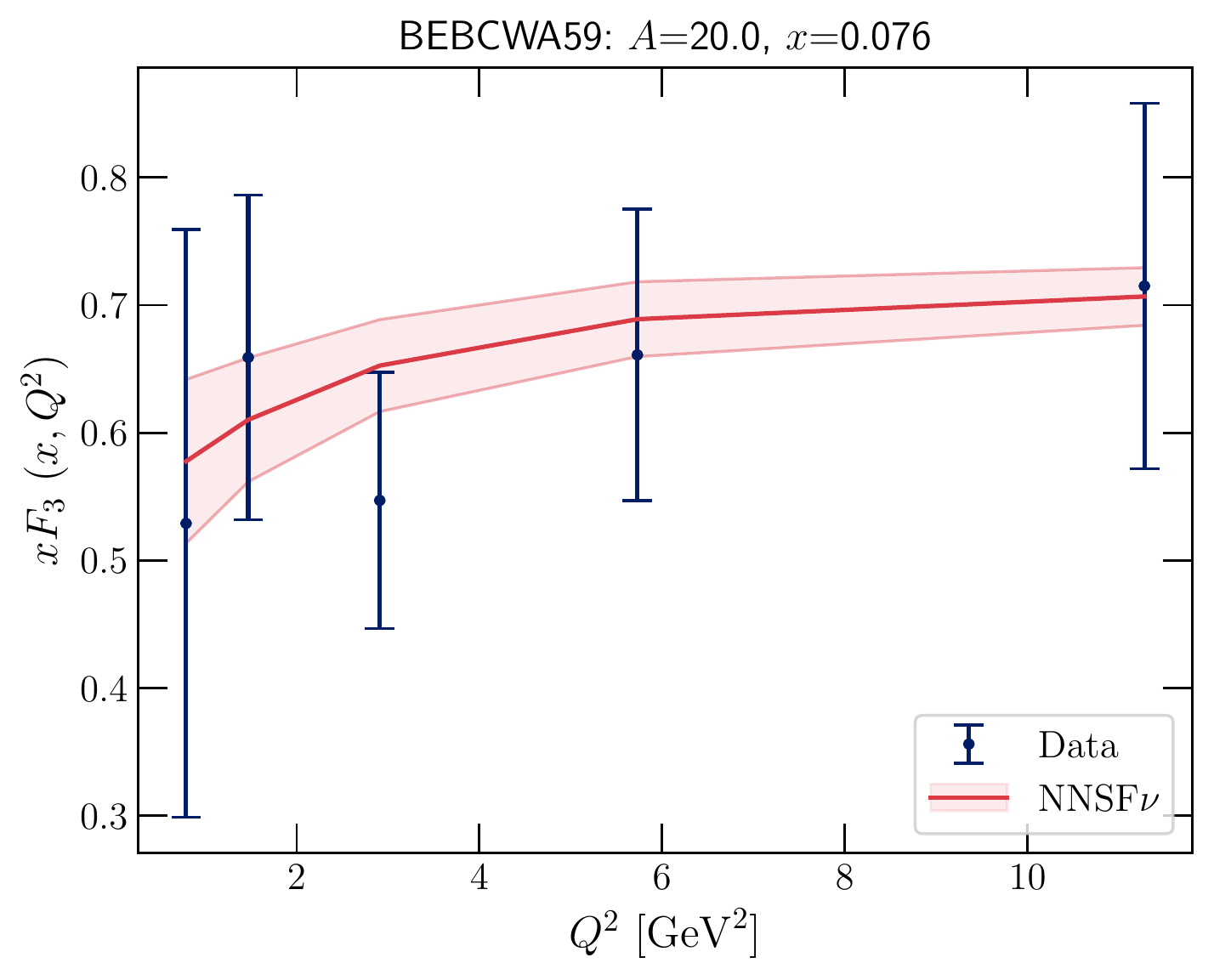}
	\includegraphics[width=0.325\linewidth]{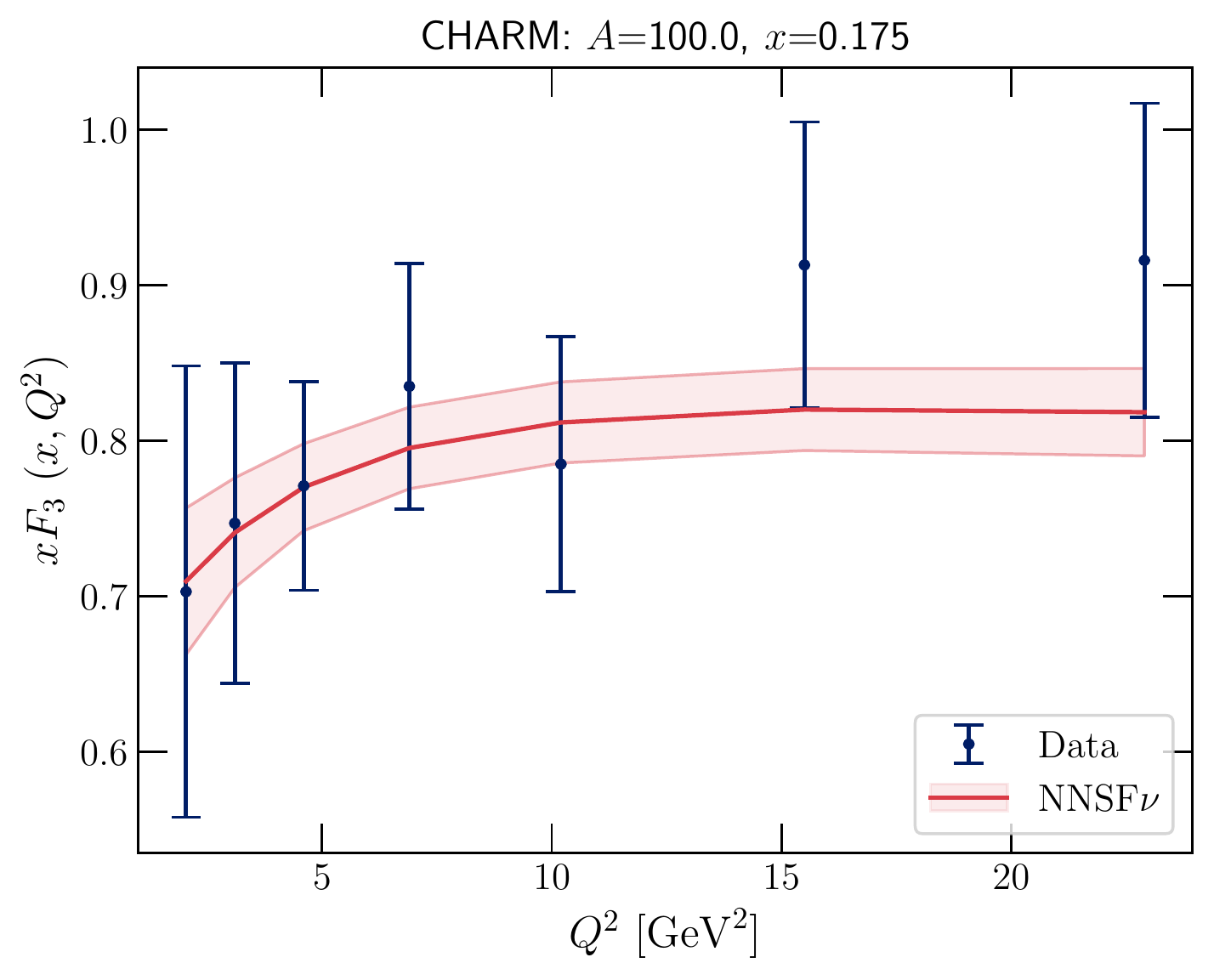}
	\includegraphics[width=0.325\linewidth]{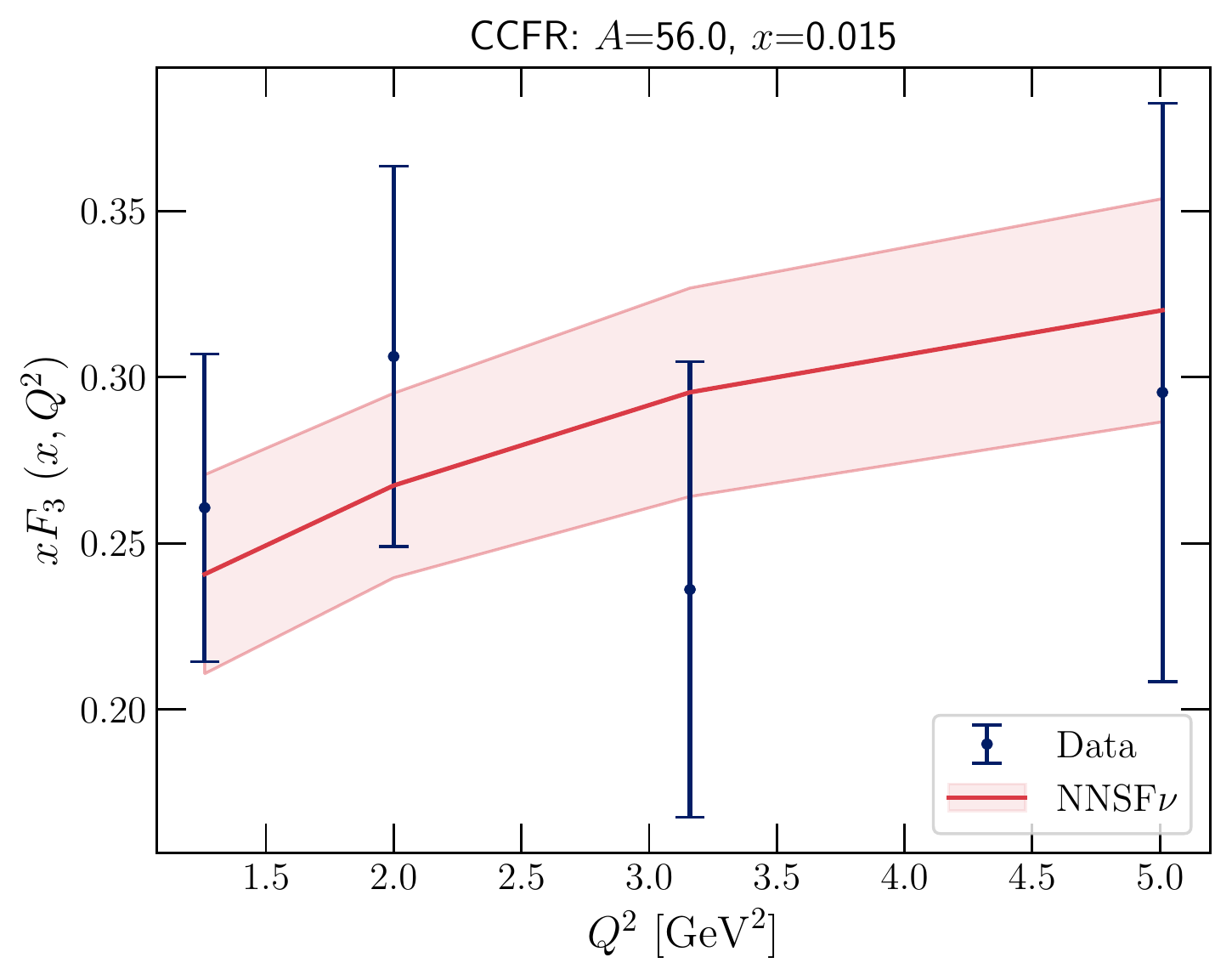}
	\caption{Comparison between a representative subset of the experimental
		data entering the fit in Region I and the corresponding NNSF$\nu$ predictions as a function
		of $Q^2$.
                For each dataset we also indicate the values of $x$ and $A$
                corresponding to the bin shown.
                The uncertainty band  in  NNSF$\nu$ is the standard deviation
                over the $N_{\rm rep}=200$ Monte Carlo replicas.
                For the experimental data, the error band corresponds to the
                diagonal entry of the covariance matrix. From top to bottom we show the 
		 $F_2$ and $x F_3$  structure functions (averaged over $\nu$ and $\bar{\nu}$)
               for  the BEBCWA59, CHARM, and CCFR experiments.
	}    
	\label{fig:prediction_data_comparison}
\end{figure}

\paragraph{Comparisons with experimental data.}
The values of the $\chi^2_{\rm exp}$ reported in Table~\ref{tab:chi2-baseline}
indicate good agreement between the experimental data and the NNSF$\nu$
parametrisation.
This agreement can be further illustrated by comparing the  NNSF$\nu$ predictions
with  representative datasets entering the fit in Region I
in selected kinematic regions as a function of $Q^2$,
as done in Fig.~\ref{fig:prediction_data_comparison}.
For each dataset we also indicate the values of $x$ and $A$
for the bin shown.
For the experimental data points, the error band corresponds to the
diagonal entry of the associated covariance matrix.
Specifically, we show the 
$F_2$ and $x F_3$  structure functions (averaged over $\nu$ and $\bar{\nu}$)
for  the BEBCWA59, CHARM, and CCFR experiments.
A similar level of agreement is obtained for the various regions of $x$, $Q^2$, or $A$ considered
in this analysis, again indicating a well-balanced fit where the different kinematic regions
are satisfactorily described.

\begin{figure}[t!]
	\centering
	\includegraphics[width=1\linewidth]{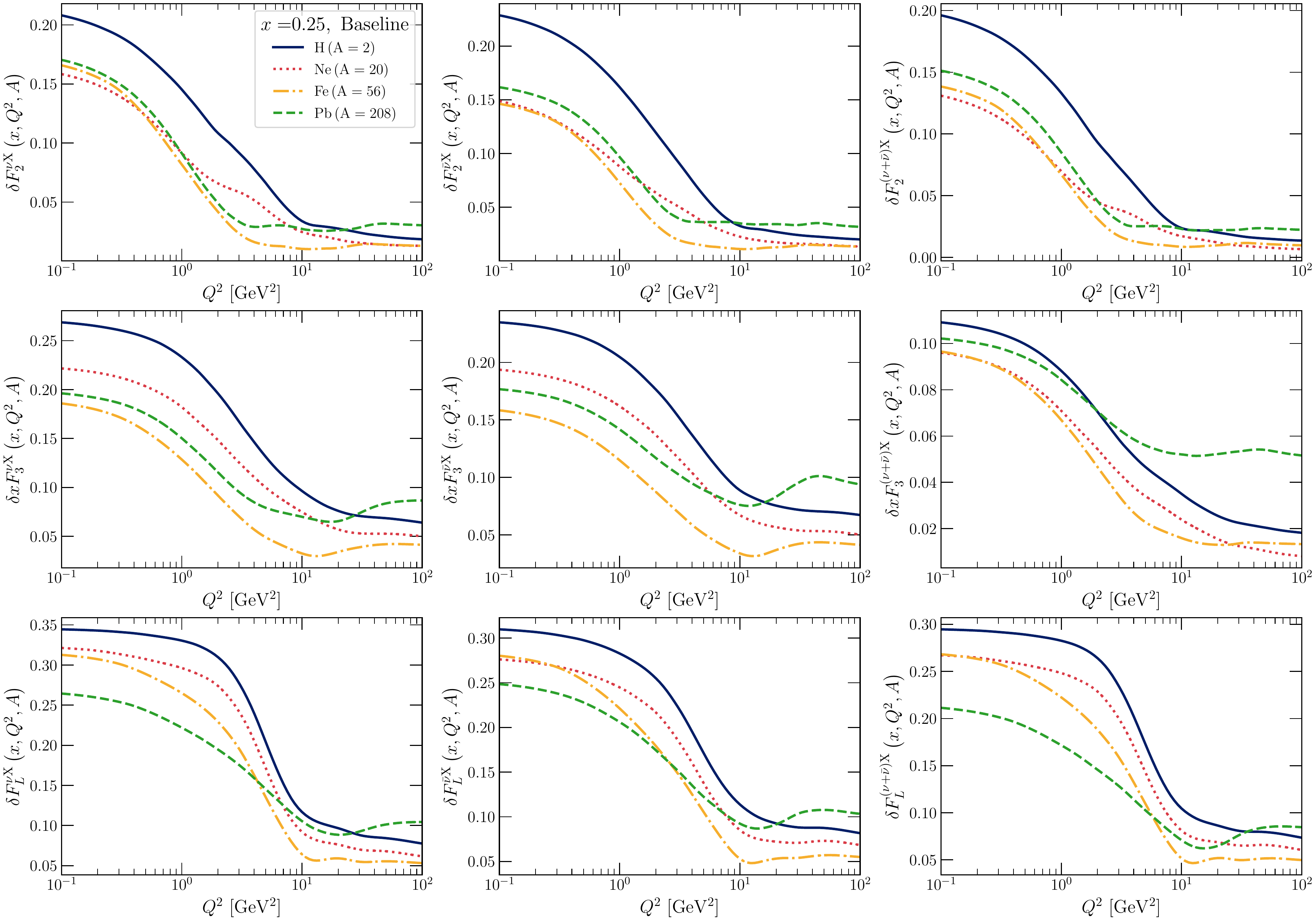}
	\caption{The 68\% CL (absolute) uncertainties in the NNSF$\nu$
          baseline fit as a function of $Q^2$ in the region between 1 GeV$^2$
          and 100 GeV$^2$ for $x=0.25$.
          We compare the results of an isoscalar nucleus ${ }^{2} \mathrm{H}$
          with those corresponding to various  nuclear
	targets entering the fit, namely ${ }^{20} \mathrm{Ne}$,  ${ }^{56} \mathrm{Fe}$, and 
	${ }^{208} \mathrm{Pb}$.
        We display results for the three structure functions $F_2, xF_3,$ and $F_L$, separately
        for the $\nu$, $\bar{\nu}$ and $\nu+\bar{\nu}$ initial states.
        }
	\label{fig:relative_errors_q2dep}
\end{figure}

\begin{figure}[t!]
	\centering
	\includegraphics[width=1\linewidth]{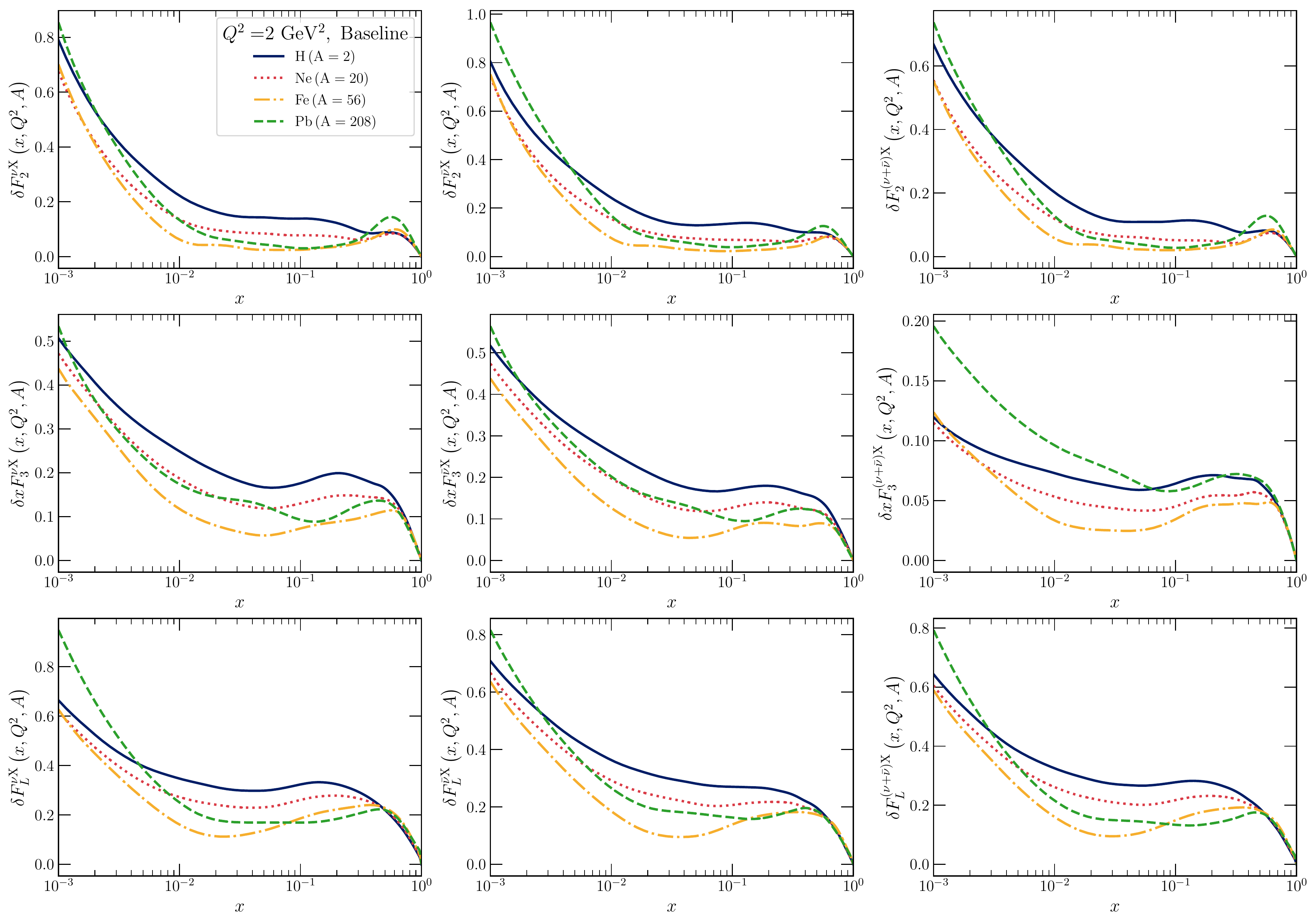}
	\caption{Same as Fig.~\ref{fig:relative_errors_q2dep} now as a function
          $x$ for $Q^2 = 2~\rm{GeV^2}$}    
	\label{fig:relative_errors_xdep}
\end{figure}

\paragraph{Uncertainty estimate and kinematic dependence.}
Concerning the uncertainty estimate of the NNSF$\nu$ determination,
in general its $1\sigma$ errors are found to be rather smaller as compared those
of the corresponding experimental measurements, see also Fig.~\ref{fig:prediction_data_comparison}.
This behaviour is expected, 
since effectively the neural network parametrisation is averaging over
the input data~\cite{Forte:2002fg}
which displays partly overlapping kinematic coverage.
In addition, one has to account for the effects of the QCD boundary conditions,
and indeed one can verify that as the lower boundary of Region II
($Q_{\rm dat}$) is approached,  the NNSF$\nu$ uncertainties
decrease as a consequence of these constraints.
These effect are illustrated in Figs.~\ref{fig:relative_errors_q2dep}
and~\ref{fig:relative_errors_xdep}, which display the
absolute 68\% CL relative uncertainties in the NNSF$\nu$ structure functions as $Q^2$
is varied  for $x = 0.25$ and as $x$ is varied for $Q^2=2$ GeV$^2$, respectively.
We compare the uncertainties for the isoscalar free nucleon $^2$H target
with those for  Ne, Fe, and Pb targets, for the three structure functions
and different initial states.


The uncertainties of the NNSF$\nu$ determination stabilize in Region II, where
they approach those of the QCD boundary condition based on {\sc\small YADISM} and nNNPDF3.0.
In Region II, the $Q^2$ dependence of the absolute uncertainties is moderate
and arising from scaling violations.
In the low-$Q^2$ extrapolation region, the NNSF$\nu$ uncertainties increase
as a consequence of the limited experimental constraints, as also mentioned above.
Uncertainties also increase as $x$ decreases, both for $F_2$ (which rises at small-$x$)
as for $xF_3$ (which being a non-singlet does not).
Lastly, the NNSF$\nu$ uncertainties become  small in the large-$x$ region where structure
functions vanish due to the elastic limit.

Another noticeable feature from Figs.~\ref{fig:relative_errors_q2dep}
and~\ref{fig:relative_errors_xdep} is the dependence of the
NNSF$\nu$ uncertainties with respect to the atomic mass number $A$.
At low $Q$ values (Region I), uncertainties are the largest for $^2$H,
consistent with the fact that there is no experimental information
in this region.
As $Q$ is increased, the NNSF$\nu$ uncertainties for all nuclei become
similar, specially for the $F_2$ and $F_L$ structure functions.
In general, the most precise NNSF$\nu$ prediction is obtained
for an iron target, at least in the region of $x$ and $Q^2$ being shown,
which is consistent with the fact that Fe is the most abundant target
in the input dataset.
However, we point out that this is not the case in the small-$x$ region with $x\lsim 10^{-5}$,
since there are no constraints from $D$-meson production on an Fe target,
and hence for the UHE neutrino cross-sections the uncertainties on a Fe target
are higher than those of either a $^2$H or a Pb target as shown in Sect.~\ref{sec:impact_nuclear}.

\subsection{Stability and validation}
\label{sec:stability}
\label{sec:matching}
\label{sec:dataset-variations}
\label{sec:a_dependence}

We now study the stability of the  NNSF$\nu$ determination
by comparing the baseline fit with variants where
either the input dataset or some aspect of the fitting methodology
are modified.
In particular, we assess the
dependence of the fit on the value of the threshold
scale $Q^2_{\rm dat}$ separating
Regions I and II;
quantify its stability when double-differential cross-section data is replaced
with their structure functions $(F_2, xF_3)$ counterparts;
and assess the quality of the NNSF$\nu$ interpolation to values
of the atomic mass number $A$ not considered in the fit.

\begin{figure}[htbp]
	\centering
	\includegraphics[width=0.96\linewidth]{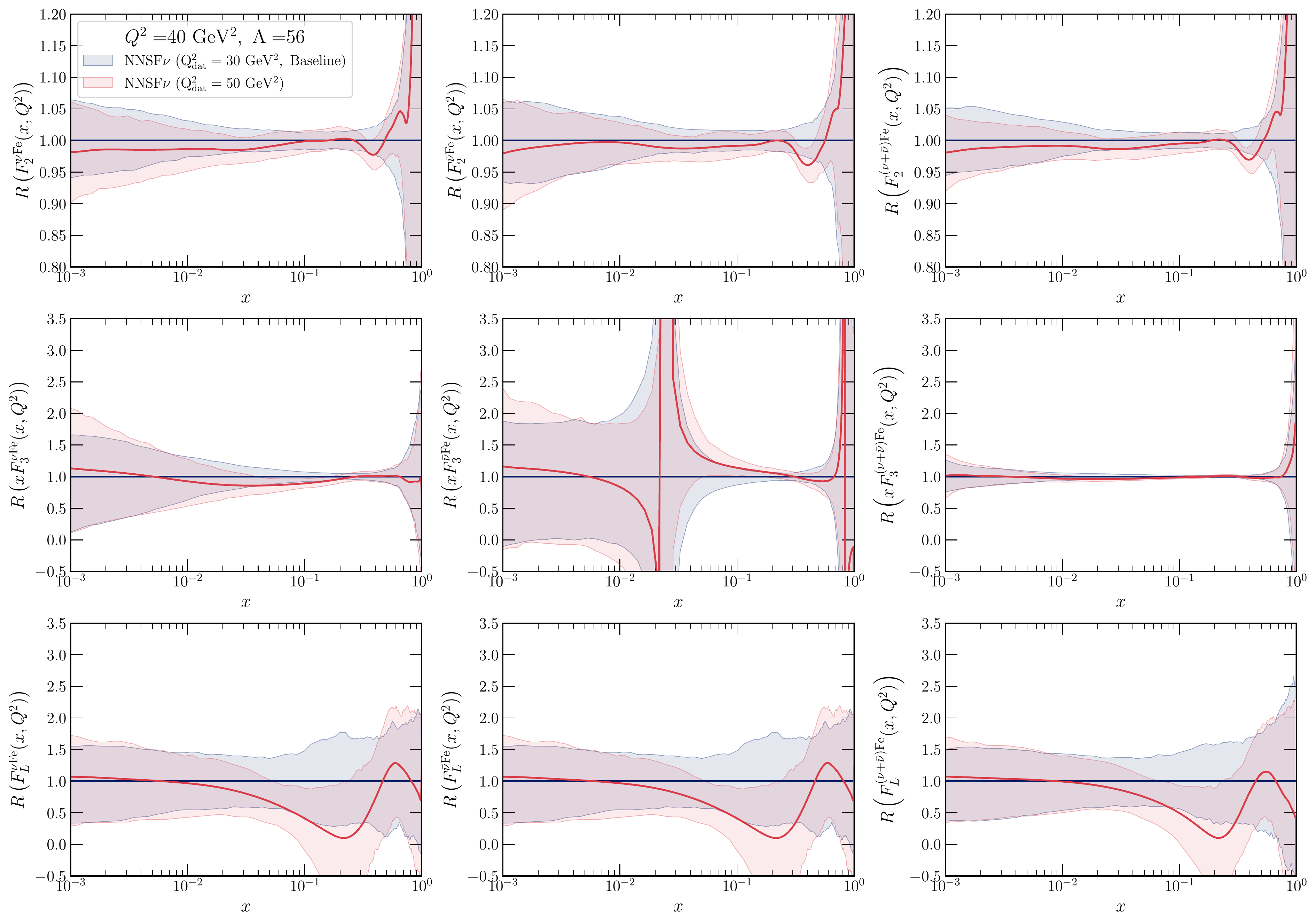}
	\includegraphics[width=0.96\linewidth]{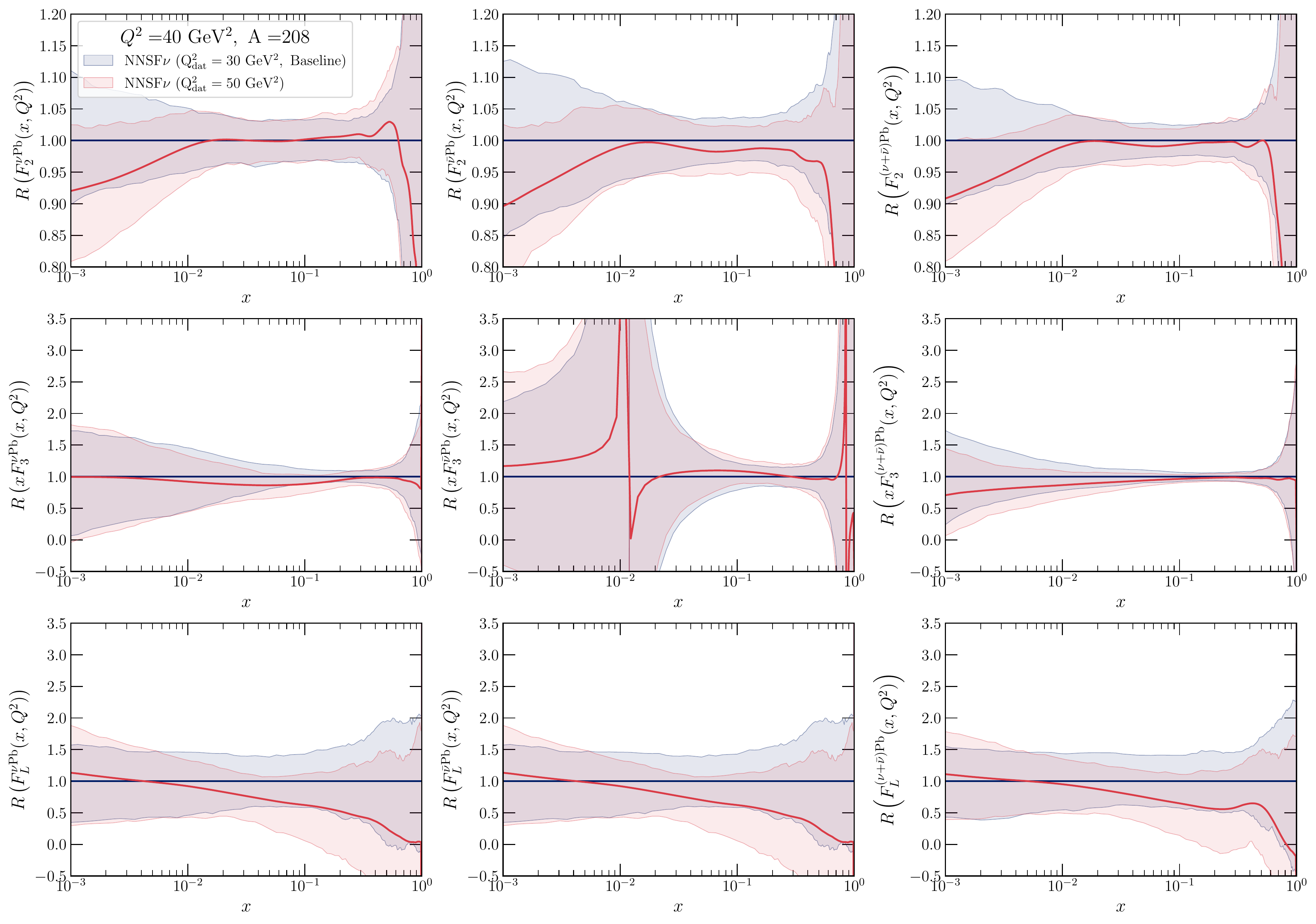}
	\caption{Comparison of the neutrino structure functions at $Q^2=40$ GeV$^2$
          as a function of $x$ in the NNSF$\nu$ baseline fit, where
          $Q^2_{\rm dat} = 30~\rm{GeV^2}$, with a variant in which
          Region I is extended up to $Q^2_{\rm dat} = 50~\rm{GeV^2}$,
          normalised to the central value of the baseline.
          We show results for Fe (top) and Pb (bottom panels) targets
          for the three structure functions $F_2, xF_3,$ and $F_L$ separately
        for the $\nu$, $\bar{\nu}$ and $\nu+\bar{\nu}$ initial states.}    
	\label{fig:qdat_dependence}
\end{figure}

\paragraph{Dependence on the matching scale.}
Fig.~\ref{fig:qdat_dependence} compares the NNSF$\nu$ baseline
results with a fit variant in which the matching scale between
Regions I and II has been increased from $Q^2_{\rm dat} = 30~\rm{GeV^2}$
(just above the bottom mass) to $Q^2_{\rm dat} = 50~\rm{GeV^2}$ .
Results are shown at $Q^2=40$ GeV$^2$ as a function of $x$
for both Fe and Pb targets, and
are normalised to the central NNSF$\nu$ baseline.
We display results for the three structure functions separately for neutrinos,
antineutrinos, and for their sum.
One finds that, in all cases considered, this NNSF$\nu$ variant is in agreement
with the baseline fit within uncertainties,
demonstrating the fit stability with respect to variations
of the  $Q^2_{\rm dat} $ threshold value.

This stability is particularly visible for $F_2$ and $xF_3$, while
somewhat larger effects are observed for $F_L$ in the large-$x$ region.
The likely explanation of this effect is that
experimental constraints on $F_L$
are limited and hence  this structure function is more sensitive to the settings
of the matching to the QCD predictions.
Some differences are also observed for the lead structure function
$F_2$ for $x\lsim 0.01$.
In the case of this target the direct experimental
constraints end at $x \approx 0.025$, so again the matching scale has
some influence on the results.
Nevertheless, agreement within uncertainties is preserved in all cases,
demonstrating that the NNSF$\nu$ analysis is robust with respect
to moderate variations  of the  hyperparameter $Q^2_{\rm dat}$.
This said, as discussed in Sect.~\ref{subsec:general_strategy}, $Q^2_{\rm dat}$ can
neither
be arbitrarily reduced, which would cut away most of the neutrino data,
nor increased, since a gap would arise between the data region
and the region where QCD boundary conditions are imposed.

\begin{figure}[htbp]
	\centering
	\includegraphics[width=0.96\linewidth]{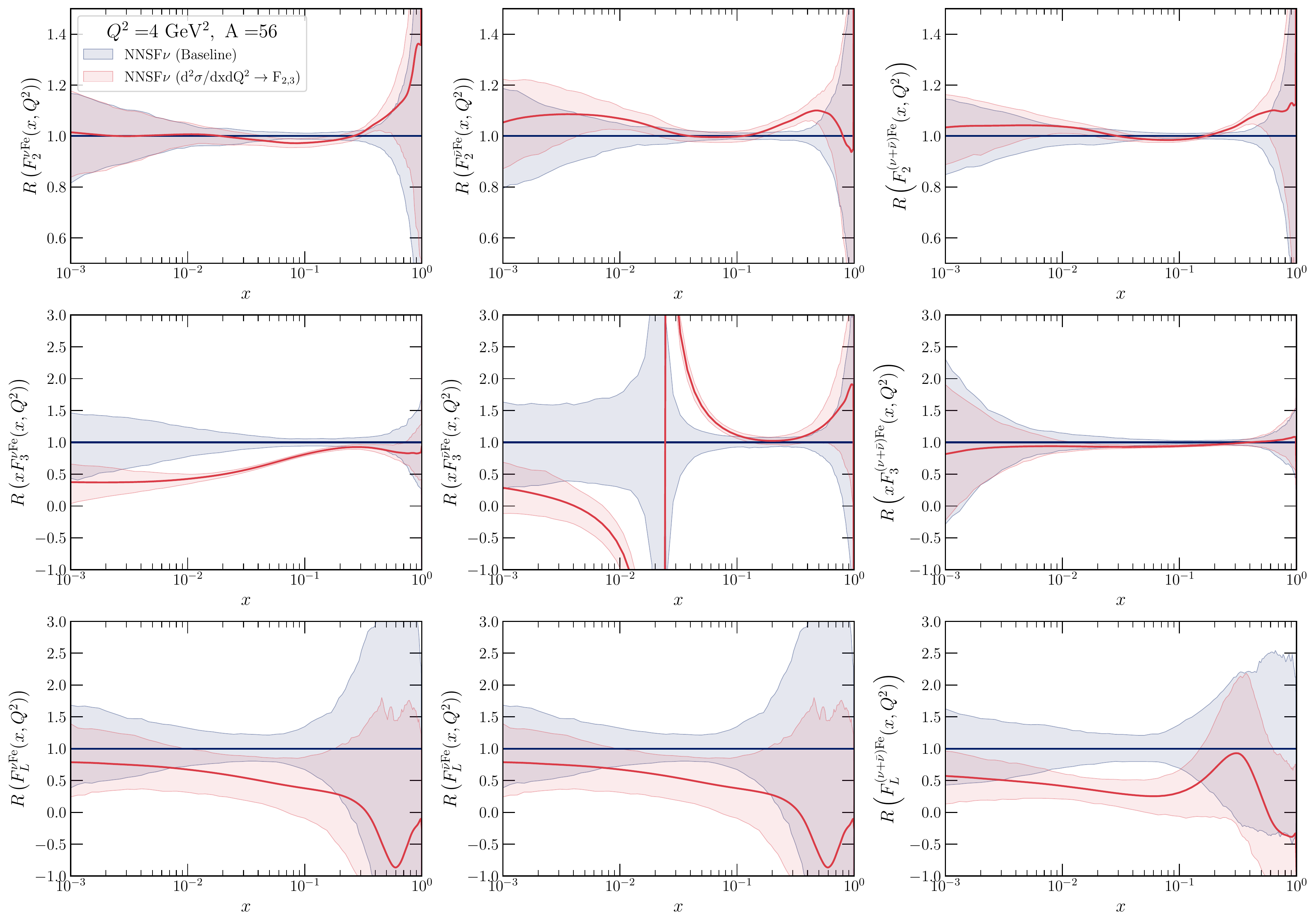}
	\includegraphics[width=0.96\linewidth]{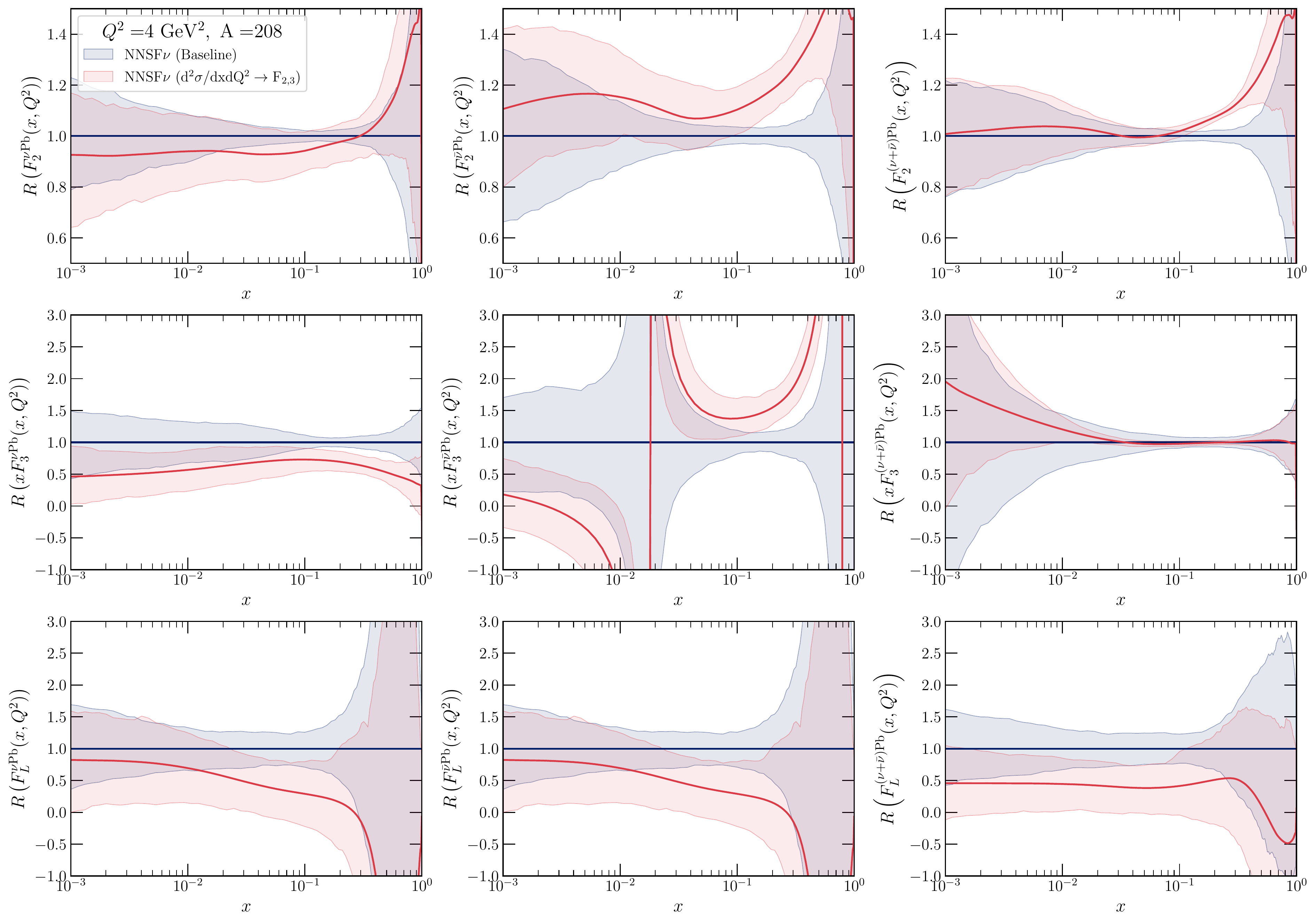}
	\caption{Same as Fig.~\ref{fig:qdat_dependence} for the NNSF$\nu$ fit variant in which
          the data on double-differential cross-sections for the CDHSW, CHORUS, and NuTeV experiments
          has been replaced by the corresponding measurements at the $\lp F_2, xF_3\rp$
        structure function level.}
	\label{fig:xsec_to_sfs}
\end{figure}

\paragraph{Reduced cross-sections vs structure functions.}
Fig.~\ref{fig:xsec_to_sfs} presents the same comparison as
that in Fig.~\ref{fig:qdat_dependence} now for a
 NNSF$\nu$ fit variant in which
 the data on the double-differential cross-sections for the CDHSW, CHORUS,
 and NuTeV experiments
 has been replaced by the corresponding measurements at the level
 of the individual structure function $F_2$ and  $xF_3$,
see also the discussion in Sect.~\ref{subsec:expdata}.
We note that since in these experiments the separate
structure functions
are extracted from the differential cross-sections by means
of a theory-assisted averaging procedure, this replacement
entails removing 3805 cross-section data points and replace them by
a much smaller number, 490, of structure functions data points.

In terms of the fit quality, from the outcome of this fit variant
we find a marked deterioration of the $\chi^2_{\rm exp}$,
which increases to $1.450$ as compared to the value of 1.287 for the baseline
(see Table~\ref{tab:chi2-baseline}).
This fit quality worsening can be traced back to the effect of the CHORUS and NuTeV
experiments in particular, where for instance the
$F_2$ structure function data has $\chi^2_{\rm exp}=2.073$
and 2.445 per data point respectively.
We note that the poor fit quality to the  NuTeV $F_2$ and $xF_3$ structure function
data has already been reported and studied in the
literature (see~\cite{Muzakka:2022wey,Kovarik:2010uv,Paukkunen:2010hb}
and references therein), and concerns
about the internal consistency of this dataset have been raised.
While our choice of reduced cross-sections $d^2\sigma/dxdQ^2$
for the baseline dataset is motivated by a priori considerations
based on it being a more robust, less theory-dependent observable,
the poor description of the $F_2$ data from  CHORUS and NuTeV provides
a further argument in favour of the choice adopted here.

Concerning the impact that this dataset variation has at the level of
the NNSF$\nu$ output, from Fig.~\ref{fig:xsec_to_sfs}
we find that in general results are  compatible at the
one-sigma level,
specially when considering structure functions averaged over neutrinos
and antineutrinos.
Nevertheless, non-negligible differences, not covered by the respective
uncertainties, are observed for instance for $F_2^{\bar{\nu}}$ at large-$x$ both
for Fe and Pb,
for $xF_3$ for $A=56$ (the NuTeV target) when separated into
neutrino and antineutrino predictions, and for $F_L$ for
intermediate $x$ and also for $A=56$.
We note that in the latter case, this fit variant does not include direct
experimental constraints on $F_L$ and hence the only information provided
by the fit comes from the QCD boundary conditions in Region II.

Taking into account both the deterioration at the fitted  $\chi^2_{\rm exp}$ level,
the lack of agreement in the fitted structure functions for specific regions
of $(x,Q^2,A)$,
and their poor description when the baseline NNSF$\nu$ predictions are used,
one concludes the separate $F_2$ and $xF_3$ structure functions
are not equivalent, and may be inconsistent, as compared
with their differential cross-sections counterparts which is our default choice.
  Even so,  we note that at the level of $F_2$ and $xF_3$ averaged
  over neutrinos and antineutrinos, except for $F_2$ at $x\gsim 0.2$, the baseline
  and the variant fits are in agreement at the 68\% CL and hence predictions
  obtained from them, for instance for inclusive cross-sections, are also likely
  to be in agreement, the only possible exception being low $E_\nu$ values where the
  large-$x$ region dominates.

\paragraph{Interpolation in atomic number $A$.}
One of the key advantages of the NNSF$\nu$ strategy is the ability to interpolate the predictions
for neutrino structure functions to other targets, with different atomic mass numbers,
beyond those directly considered in the fit and that may be relevant
for neutrino phenomenology, such as oxygen ($A=16$), argon ($A=39$), calcium ($A=40$), and
tungsten ($A=184$) among others.

Here we validate the NNSF$\nu$ interpolation in $A$ by comparing the baseline fit
with two variants.
First, we compare the baseline fit results with those of a
variant in which the datasets with
atomic mass number $A\simeq 20$, namely BEBCWA59 (Ne) and CHARM (CaCO$_3$),
are excluded.
In such case, the predictions
for $A=20$ of the fit variant in Region I will be obtained
from a extrapolation of the constraints provided
by the fitted data with $A \ge 56$ for iron and lead targets.
Second, we compare the NNSF$\nu$ predictions from the baseline fit in Region II
with the YADISM+nNNPDF3.0  calculations for calcium $(A=40)$, a nuclear
species which is not contained in the baseline analysis.
These two tests make possible validating the interpolation
(and extrapolation) of the  NNSF$\nu$ predictions
to new values of $A$ not used in the fit neither in Region I nor in Region II.

\begin{figure}[t!]
	\centering
	\includegraphics[width=1\linewidth]{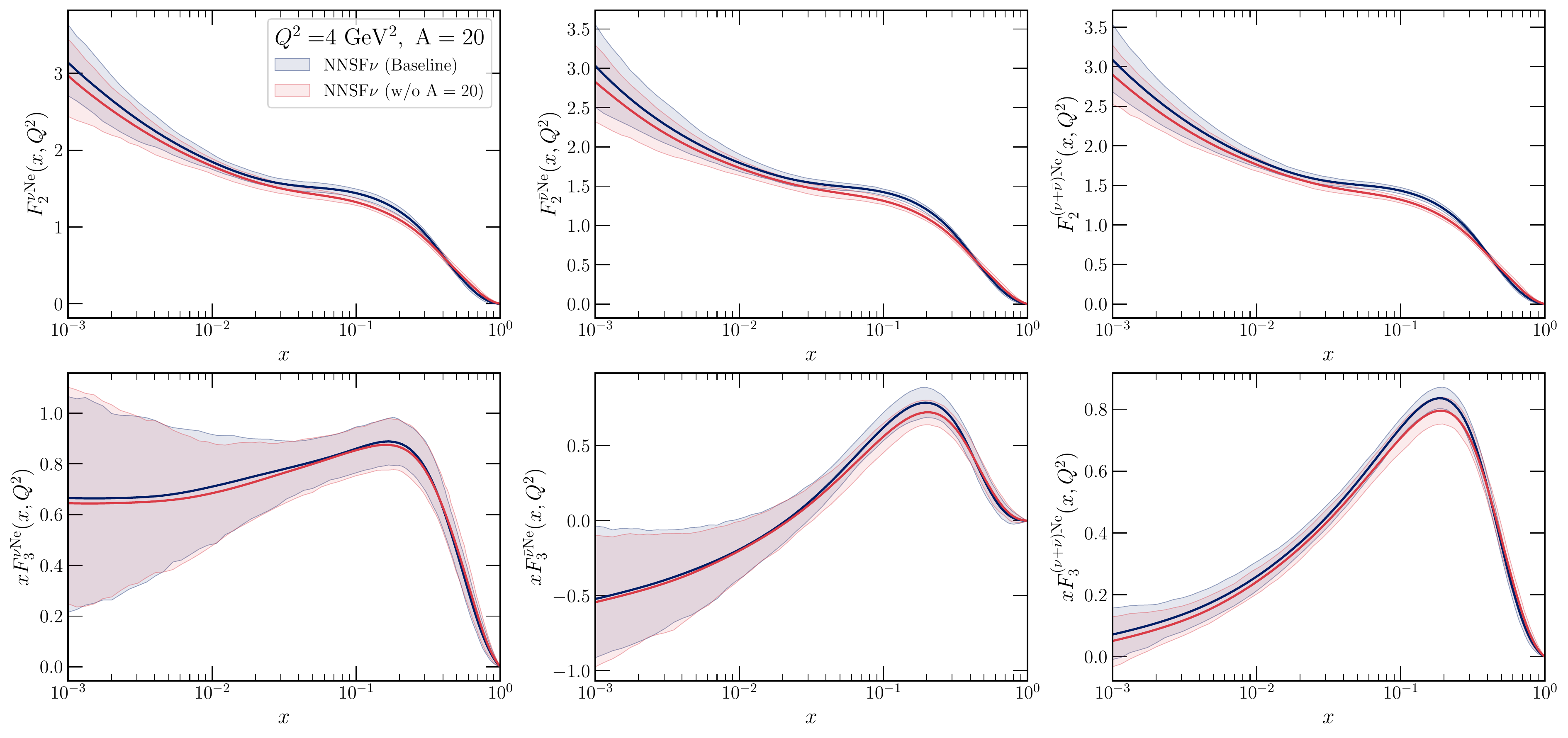}
	\caption{Comparison between the results of the NNSF$\nu$ baseline fit
          and those of a fit variant in which
          the data from the $A \simeq 20$ datasets, BEBCWA59 (Ne) and CHARM (CaCO$_3$),
are excluded.
          We display the $F_2$ and $xF_3$ structure functions for $A=20$
          for $Q^2=4$ GeV$^2$ as a function of $x$, separately for the neutrino,
          antineutrino, and their sum initial states.
          In this kinematic region, the predictions
          for $A=20$ of the fit variant are obtained
          from extrapolating the constraints provided
          by the fitted data with $A \ge 56$ (for iron and lead targets).
     }
	\label{fig:interpolation_A100}
\end{figure}

First, Fig.~\ref{fig:interpolation_A100} displays
the results comparing the baseline fit with the
variant in which the $A \simeq 20$ datasets, namely BEBCWA59 (Ne) and CHARM (CaCO$_3$),
are excluded.
 We display the $F_2$ and $xF_3$ structure functions for $A=20$
 for $Q^2=4$ GeV$^2$ as a function of $x$, separately for the neutrino,
 antineutrino, and their sum initial states.
 In this kinematic region, the predictions
 for $A=20$ of the NNSF$\nu$ variant arise entirely
 from extrapolating the constraints provided
 by the fitted data with $A \ge 56$ for iron and lead targets.
 As one can observe from the comparison,
 the two fits are compatible within the respective uncertainties,
 with the possible exception
 of $F_2^{\nu A}$ for $x\simeq 0.2$ where
 the baseline and its variant overlap at the $2\sigma$ level.
 The good agreement between the baseline fit and its variant,
 specially for $F_2$ for $x\lsim 0.1$ and for $xF_3$,
 confirms that the methodology extrapolates in the atomic mass number $A$
 in a way that is consistent with the available
 experimental constraints in Region I,
 and that the predicted
 structure functions are stable upon removal
 of a subset of the fitted datasets.
 
\begin{figure}[t!]
	\centering
	\includegraphics[width=1\linewidth]{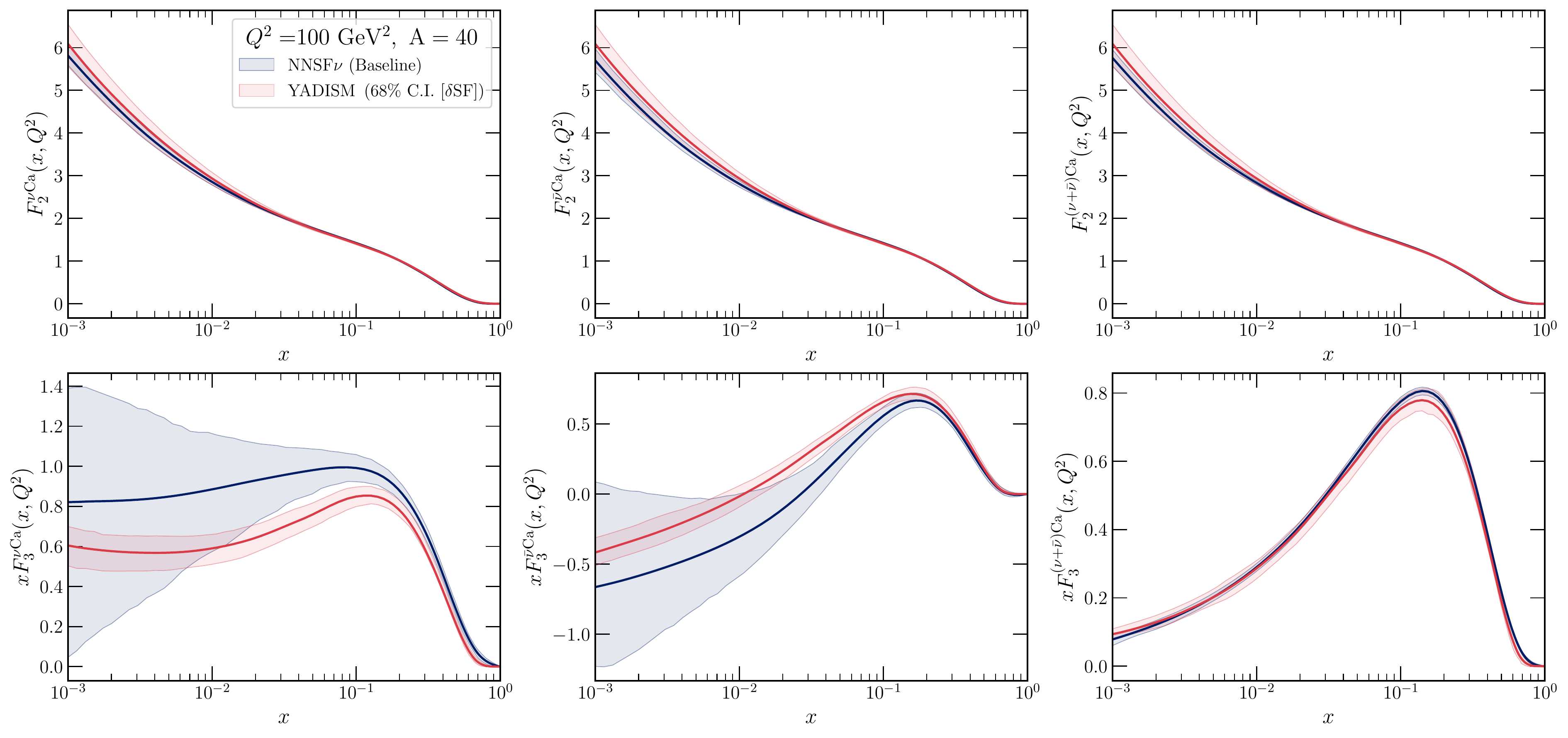}
	\caption{Same as Fig.~\ref{fig:interpolation_A100},
          now for the structure functions
          of an $A=40$ target
          comparing the outcome of
           NNSF$\nu$ with that of {\sc\small YADISM} (with nNNPDF3.0 as input).
          We display results at $Q^2 = 100~\rm{GeV^2}$ as a function of $x$.
          Since neither the fitted data (in Region I) nor the QCD boundary conditions
          (in Region II) include information on $A=40$,
          this comparison tests the interpolation of NNSF$\nu$ to $A$ values not
          considered in the fit.
          The error band in the {\sc\small YADISM} predictions contains the
          PDF uncertainties, but not the MHOU ones.
          }
	\label{fig:interpolation_A40}
\end{figure}

Fig.~\ref{fig:interpolation_A40} then compares the NNSF$\nu$
predictions for $A=40$ with the corresponding YADISM
calculations, with nNNPDF3.0 as input, for
a calcium $\left( { }^{40} \mathrm{Ca} \right)$ target.
As in Fig.~\ref{fig:interpolation_A100}, we compare the $F_2$
and $xF_3$ neutrino structure functions
but this time at $Q^2 = 100~\rm{GeV^2}$,
which corresponds to Region II.
 Since neither the fitted data (in Region I) nor the QCD boundary conditions
 (in Region II) include information on $A=40$,
 this comparison assesses the interpolation
 capabilities of NNSF$\nu$ to atomic mass numbers not
 considered in the fit.
 The error band in the {\sc\small YADISM} predictions of
 Fig.~\ref{fig:interpolation_A40}
 contains the PDF uncertainties but not the MHOU ones.

 One finds that the  interpolated predictions
 of  NNSF$\nu$ for $A=40$ are in very good
 agreement with the YADISM calculation, which do not enter the fit.
 While excellent agreement can be seen at the averaged level, some
 differences are noticed when looking at large-$x$ of the separate $\nu$- and 
 $\bar{\nu}$-results for $xF_3$. These discrepancies could however be
 attributed to the small constraints on the sea-quark distributions that
 cancel out when taking the average.
 This result implies that  the NNSF$\nu$ parametrisation
 interpolates in $A$ in a manner consistent with the
 underlying behaviour of the QCD prediction in Region II,
 and furthermore that the fit results would not be affected in the case that
 QCD data with $A=40$ is added to the NNSF$\nu$ fit.
 These results demonstrate that NNSF$\nu$ is
 able to reliably interpolate between
 nuclear targets and therefore capable of providing predictions for
 $A$ values for which 
 no direct experimental measurements are available.
 We use this feature to provide
 {\sc\small LHAPDF} sets for all values
 of $A$ relevant for neutrino phenomenology as listed in  App.~\ref{app:delivery}.
 
\subsection{The Gross-Llewellyn Smith sum rule}

Finally, we use the NNSF$\nu$ determination to evaluate the  Gross-Llewellyn-Smith sum rule
of neutrino structure functions, Eq.~(\ref{eq:GLS_sumrule}), and assess its agreement with respect to the
perturbative QCD calculation.
The latter prediction indicates that for an isoscalar target $A$ one expects to find
\be
\label{eq:GLS_sumrule_truncated}
\int_0^1 \frac{dx}{x} xF_3^{\nu A}(x,Q^2) = 3\lp 1+ \sum_{k=1}^3 \lp \frac{\alpha_s(Q^2)}{\pi}\rp^k c_k(n_f)\rp \, ,
\ee
with $n_f$ being the number of active flavours at the scale $Q^2$ and the coefficients $c_k$
known up to third order in perturbation theory.
Evaluating Eq.~(\ref{eq:GLS_sumrule_truncated}) requires extrapolating down to the $x\to 0$ region
which is not accessible experimentally, and hence instead we compute a truncated variant
of the GLS sum rule,
\be
\label{eq:GLS_sumrule_truncated_v2}
{\rm GLS }\lp Q^2,A,x_{\rm min}\rp\equiv \int_{x_{\rm min}}^1 \frac{dx}{x} xF_3^{\nu A}(x,Q^2) \, ,
\ee
for different values of the lower integration limit $x_{\rm min}$ and the atomic mass number
$A$.
The truncated sum rule Eq.~(\ref{eq:GLS_sumrule_truncated_v2}) should converge
to the QCD prediction in Eq.~(\ref{eq:GLS_sumrule_truncated}) in the
$x_{\rm min}\to 0$ limit.
Furthermore, the GLS sum rule should hold true irrespective of the value
of the mass number $A$ entering the calculation.

\begin{figure}[!t]
\centering
\includegraphics[width=0.325\linewidth]{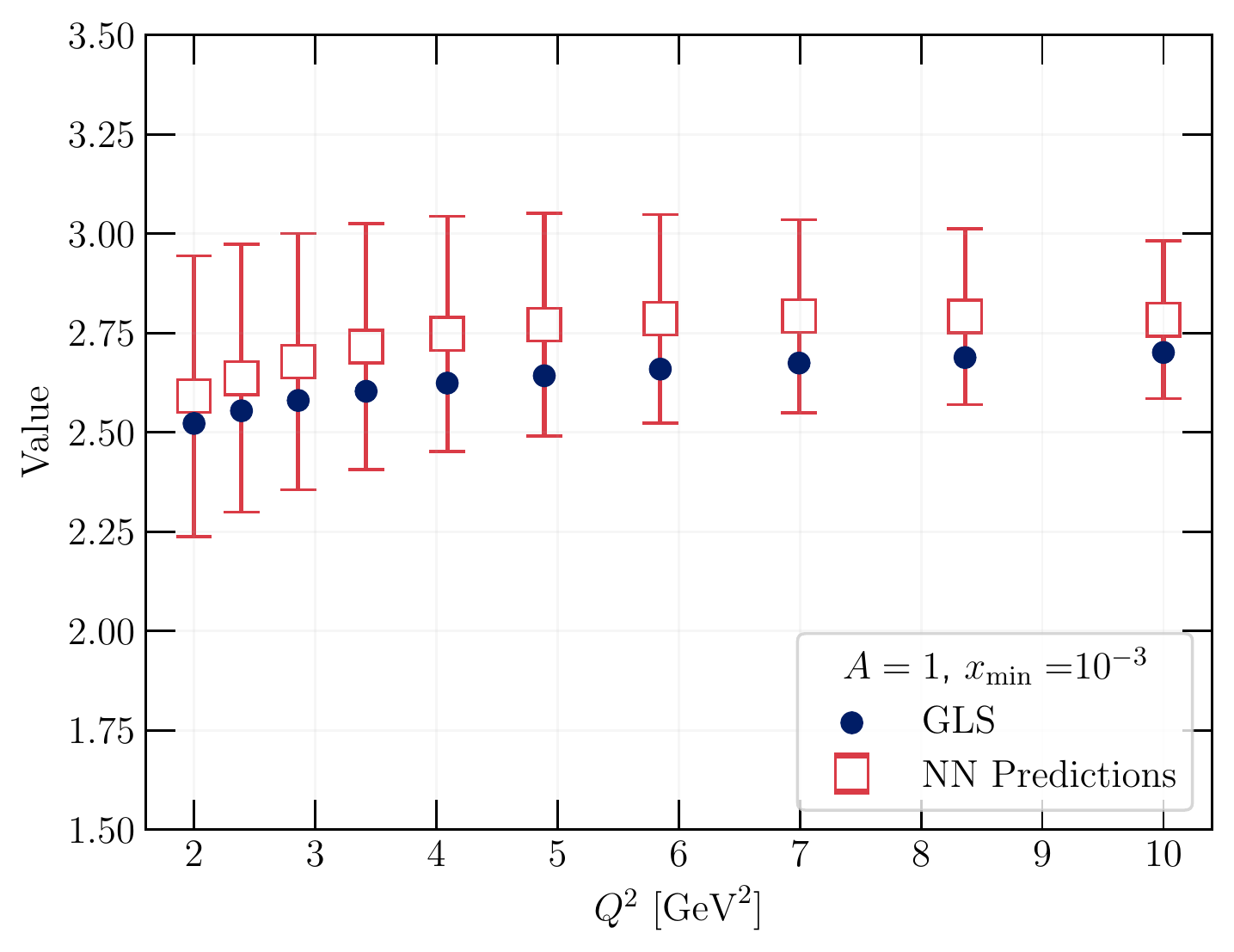}
\includegraphics[width=0.325\linewidth]{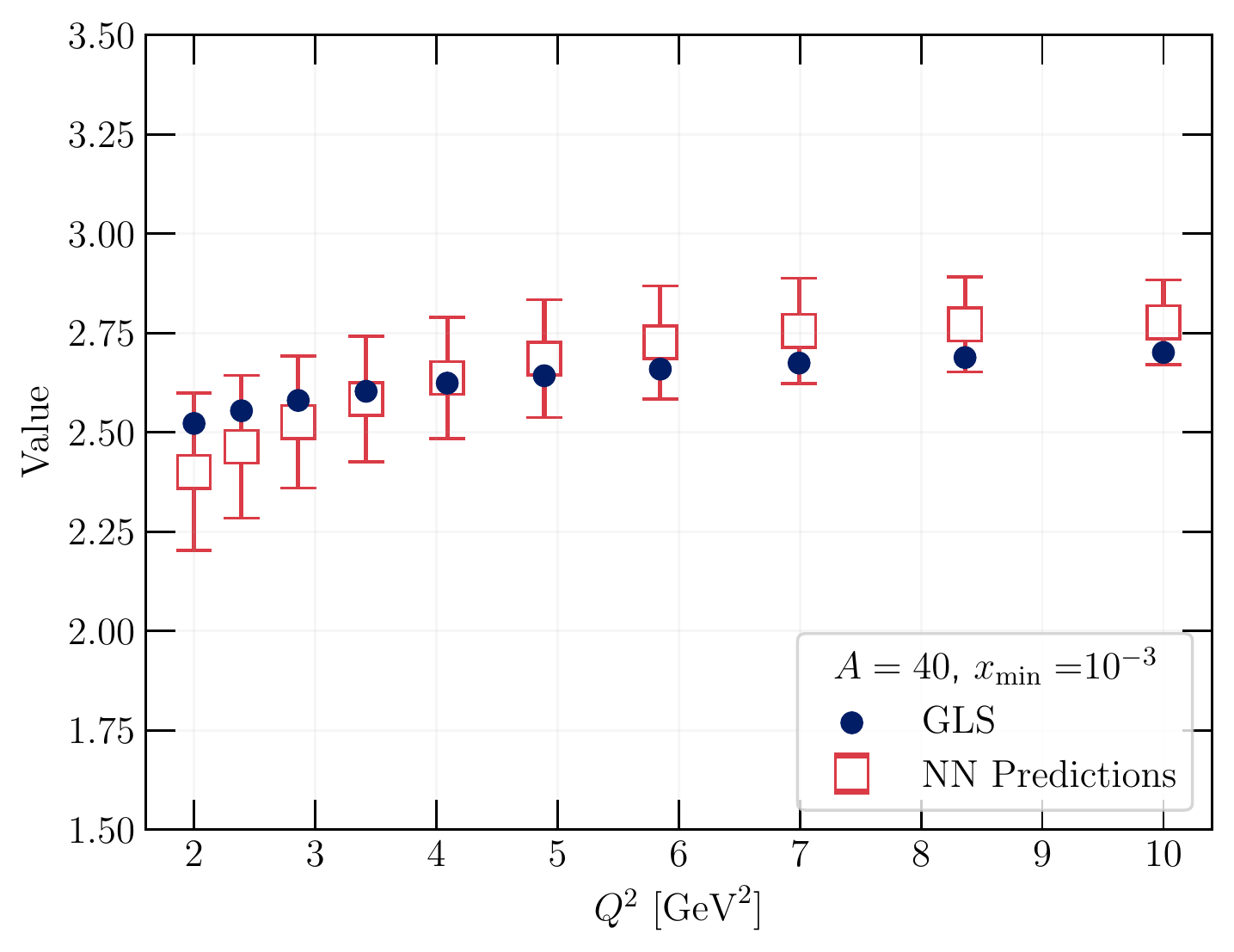}
\includegraphics[width=0.325\linewidth]{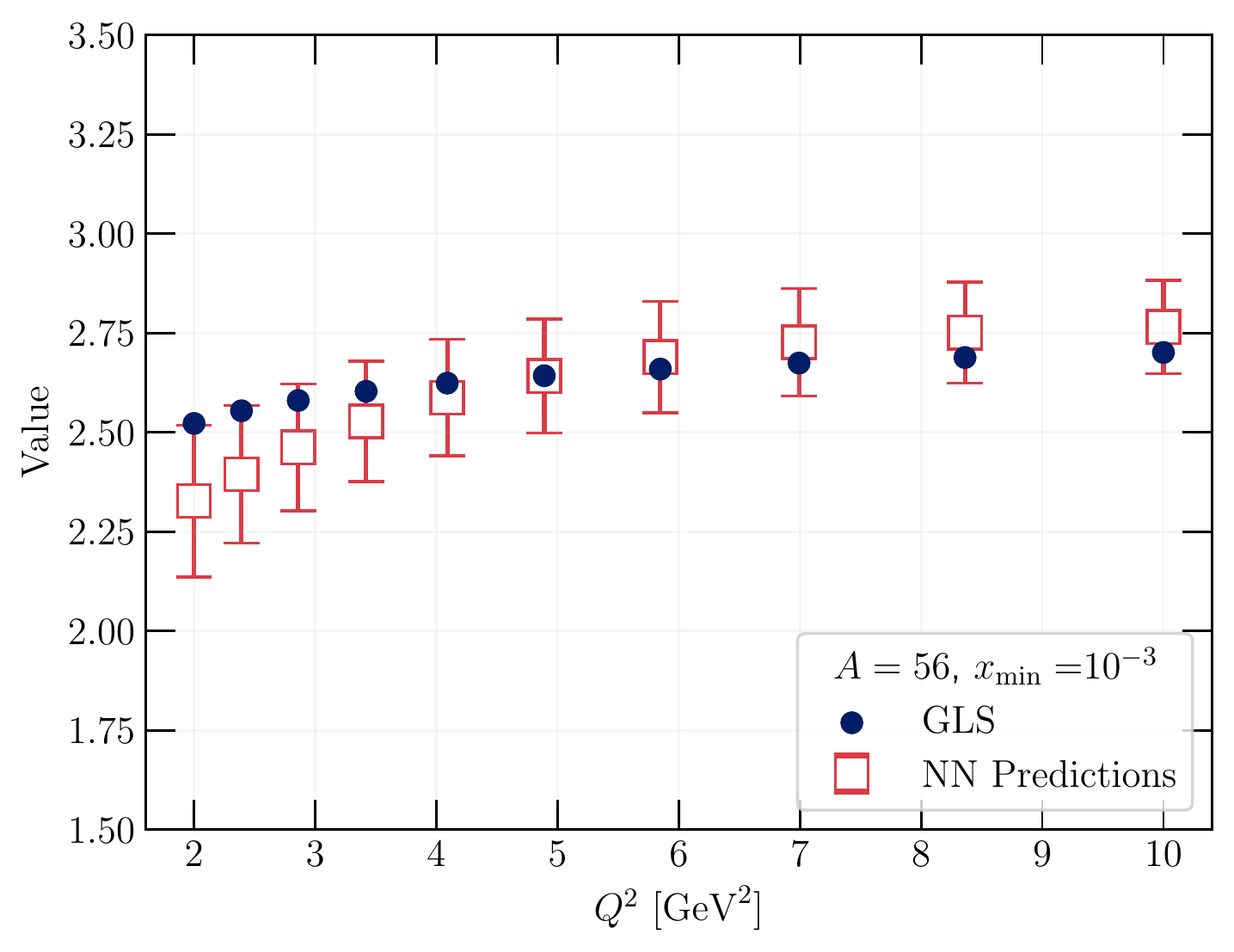}
\includegraphics[width=0.325\linewidth]{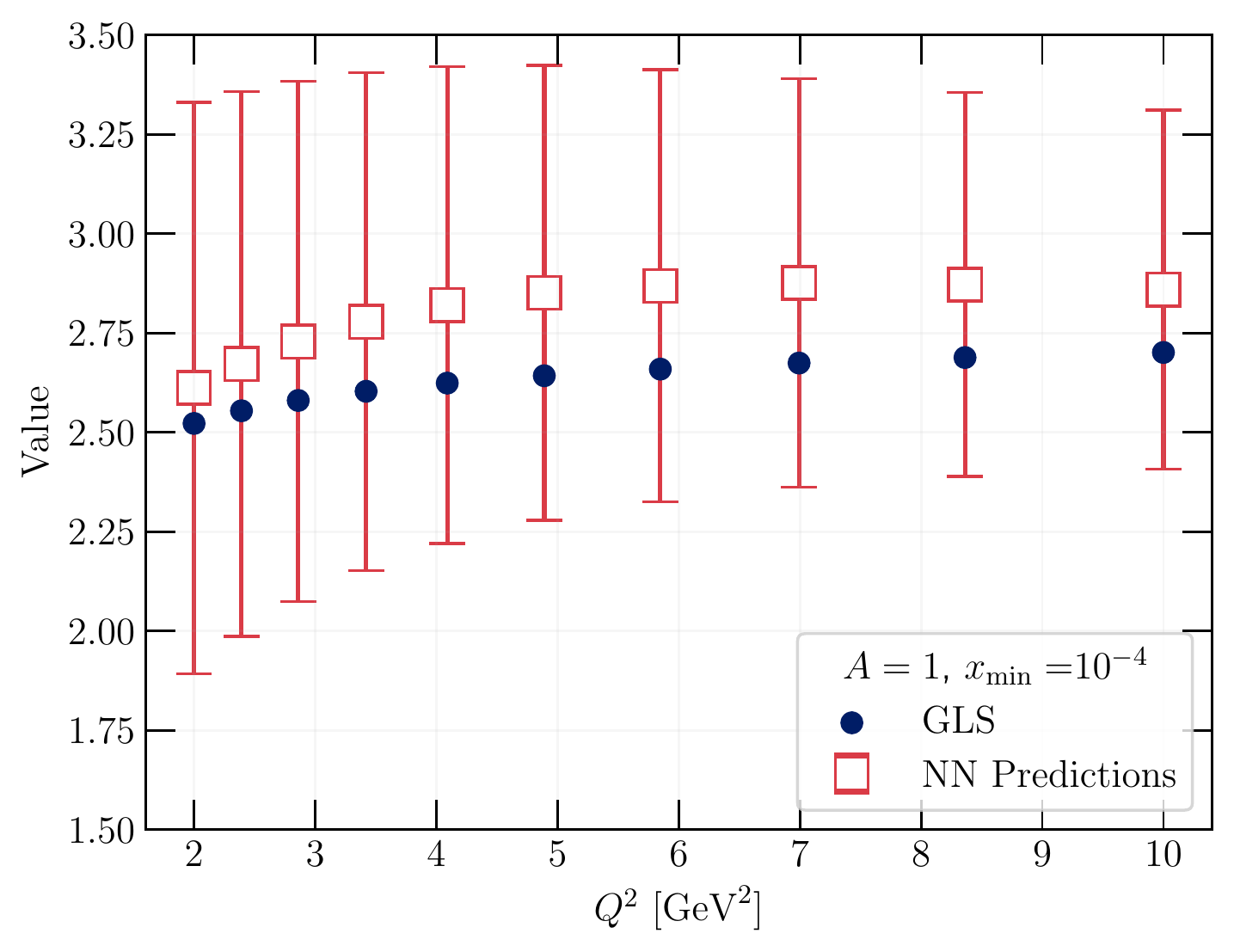}
\includegraphics[width=0.325\linewidth]{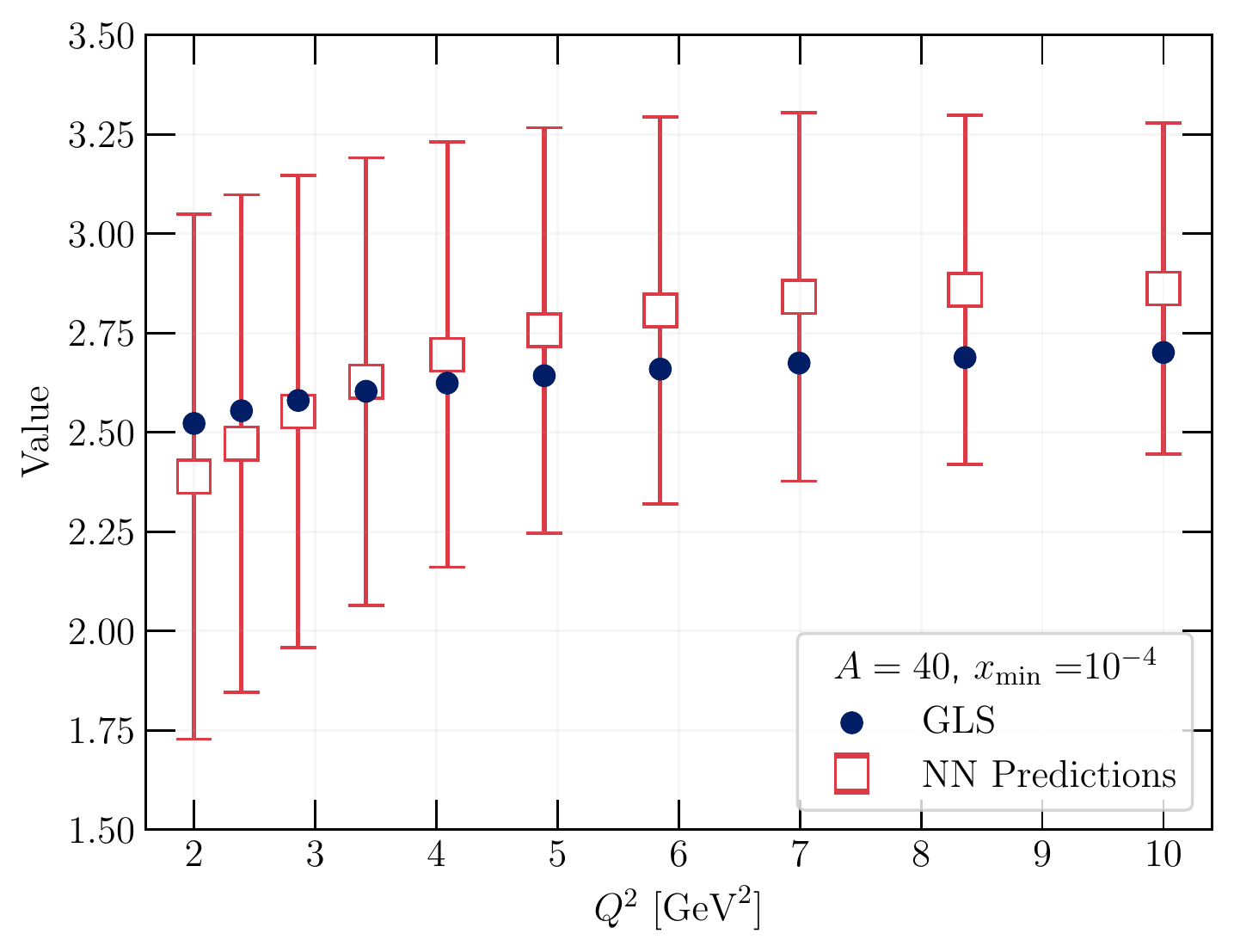}
\includegraphics[width=0.325\linewidth]{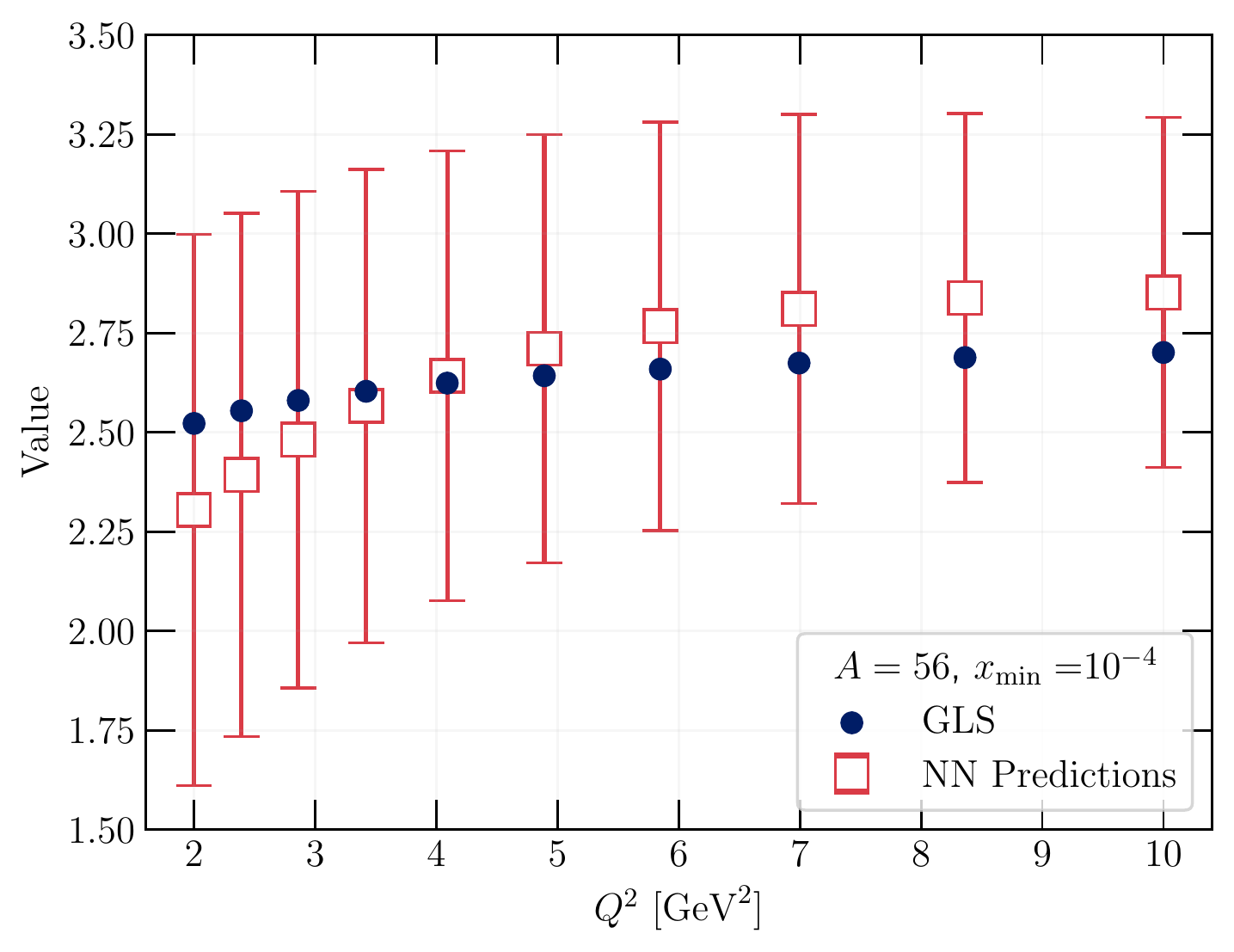}
\caption{The Gross-Llewellyn-Smith  sum rule evaluated as a function 
  of $Q^2$.
  The predictions of  NNSF$\nu$ are compared with
  the perturbative QCD calculation evaluated at N$^3$LO.
  We display results corresponding to a lower 
  integration limit of $x_{\rm min}=10^{-3}$ (top)
  and of $x_{\rm min}=10^{-4}$ (bottom panels) for $A=1$ (left), $A=40$  (middle),
  and for $A=56$ (right panel).
  Note that the QCD prediction  for the GLS sum rule is the same in all panels, since it
  is independent of both $x_{\rm min}$ and $A$.
}    
\label{fig:GLS-sumrule}
\end{figure}

Fig.~\ref{fig:GLS-sumrule} displays the outcome of the calculation of the (truncated)
Gross-Llewellyn-Smith sum rule, Eq.~(\ref{eq:GLS_sumrule_truncated_v2}),
evaluated with the NNSF$\nu$ baseline fit as a function of $Q^2$.
We display results corresponding to lower 
integration limits of $x_{\rm min}=10^{-3}$ and of $10^{-4}$  for $A=1$, $A=40$, and $A=56$ nuclei.
We compare the NNSF$\nu$ baseline truncated results with the corresponding QCD predictions
for the exact GLS sum rule.
Note that the latter is the same in all panels, since it
  is independent of both $x_{\rm min}$ and $A$, and that its $Q^2$ dependence is
  entirely dictated by that of the running of the strong coupling $\alpha_s(Q^2)$.
  As in the rest of this section, the uncertainty in the  NNSF$\nu$ results is computed
  as the 68\% CL intervals from the $N_{\rm rep}=200$ replicas
  that constitute the representation of its probability density.

From this comparison, one can conclude that there is agreement within uncertainties
between the (truncated) calculation of the GLS sum rule
from the phenomenological NNSF$\nu$ determination  of the $xF_3$
neutrino structure function and the corresponding perturbative QCD
prediction.
The agreement in central values, especially in the large-$Q^2$ region, slightly deteriorates
once $x_{\rm min}$ is reduced from $10^{-3}$ to $10^{-4}$.
But at the same time, the
uncertainties in the truncated sum rule calculation increase as $x_{\rm min}$
decreases, as expected since NNSF$\nu$ does not have direct experimental constraints
below $x\simeq 0.01$, see also Table~\ref{tab:neutrino-DIS}.
Even more remarkably, not only the value but also the slope with $Q^2$,
which in the QCD calculation is dictated by the $\alpha_s(Q)$ running,
is correctly reproduced by the data-driven NNSF$\nu$ analysis.
The analysis of Fig.~\ref{fig:GLS-sumrule} hence indicates that the
NNSF$\nu$
data-driven determination of $xF_3$ is consistent with the QCD expectations,
and in particular that extrapolates to the small-$x$, low-$Q^2$ region
in a manner which improves the agreement with the N$^3$LO result
given by Eq.~(\ref{eq:GLS_sumrule_truncated}).

The agreement between the truncated GLS sum rule
computed with NNSF$\nu$ and the QCD calculation holds true for the three values of
$A$ considered, namely $A=1,40$, and $56$.
The main effect of the $A$ dependence appears to be the increase in the uncertainties
in the $A=1$ case as compared to the heavier nuclei, due to the lack
of direct experimental constraints on light nuclear targets.
This insensitivity of the results of the truncated GLS sum rule
with respect to the value of $A$ is consistent with the expectation that this sum rule
is related to the nucleon valence sum rules, which are satisfied and take
the same values irrespective of the value of $A$ entering the calculation
of the neutrino structure functions.

\section{Inclusive neutrino cross sections}
\label{sec:inclusive_xsec}

Here we deploy the NNSF$\nu$ determination of neutrino
structure functions 
to evaluate inclusive
neutrino scattering cross sections
as a function of $E_\nu$ for different projectiles and targets.
Specifically, we provide predictions for the complete range of energies
relevant for neutrino phenomenology, from the GeV region
entering accelerator experiments up to the multi-EeV energies of cosmic neutrinos.
Our predictions come accompanied with an estimate
of the dominant uncertainties relevant in each energy region.

First, we compare the  NNSF$\nu$ determination with previous results in the literature,
in particular with the Bodek-Yang, BGR18, and CSMS11 calculations.
We evaluate  the mean inelasticity $\la y \ra$ for different neutrino energies,
compare with experimental measurements of the inclusive
cross-section from NuTeV,
assess the sensitivity of our calculation to the low-$Q$ region, and  quantify the
impact of nuclear corrections for the most relevant target materials.
Furthermore,  we provide dedicated predictions for the energy range and target materials
required for the interpretation of far-forward neutrino scattering  experiments at the LHC.

\subsection{The NNSF$\nu$ cross sections}
\label{sec:nnsfnu_pheno_xsec}

Starting from the double-differential neutrino-nucleus
scattering cross sections of Eqns.~(\ref{eq:neutrino_DIS_xsec_FL})
and (\ref{eq:antineutrino_DIS_xsec_FL}), the inclusive cross section
is obtained~\cite{Bertone:2018dse} by integrating over
the kinematically allowed range in $x$ and $Q^2$,
\be
\label{eq:sigma_inclusive}
\sigma^{\nu N}(E_\nu) = \int_{Q^2_{\rm min}}^{Q^2_{\rm max}} {\rm d}Q^2
\lc  \int_{x_0(Q^2)}^1{\rm d}x \frac{{\rm d}^2\sigma^{\nu N}}{{\rm d}x{\rm d}Q^2}(x,Q^2,y)\rc \, ,
\ee
where the integration limits are given by
\be
Q^2_{\rm max}=2m_NE_\nu\,, \qquad
x_0(Q^2)=\frac{Q^2}{2m_NE_\nu} \, ,
\ee
and with
the inelasticity related to the neutrino energy by
$y=Q^2/(2m_NxE_{\nu})$.
In an inclusive calculation,
the lower integration limit $Q^2_{\rm min}$ should go all the way down to  $Q^2_{\rm min}=0$.
While most previous studies impose a cut in $Q^2_{\rm min}$ to restrict
the integration in Eq.~(\ref{eq:sigma_inclusive}) to the perturbative
region, this is not required within the NNSF$\nu$ approach since our structure
function predictions are valid for all $Q^2$ values.

For the calculations presented in this section we take $Q_{\rm min}=0.03$ GeV
and study the dependence of the predictions with respect to variations
of this choice.
A kinematic cut in the final-state invariant mass of $W\ge 2$ GeV
is applied to restrict
the calculation to the inelastic scattering region.
See App.~\ref{app:delivery} for technical details about the implementation
of Eq.~(\ref{eq:sigma_inclusive}) and the estimate of the associated
uncertainties.

\begin{figure}[!t]
 \centering
 \includegraphics[width=1\linewidth]{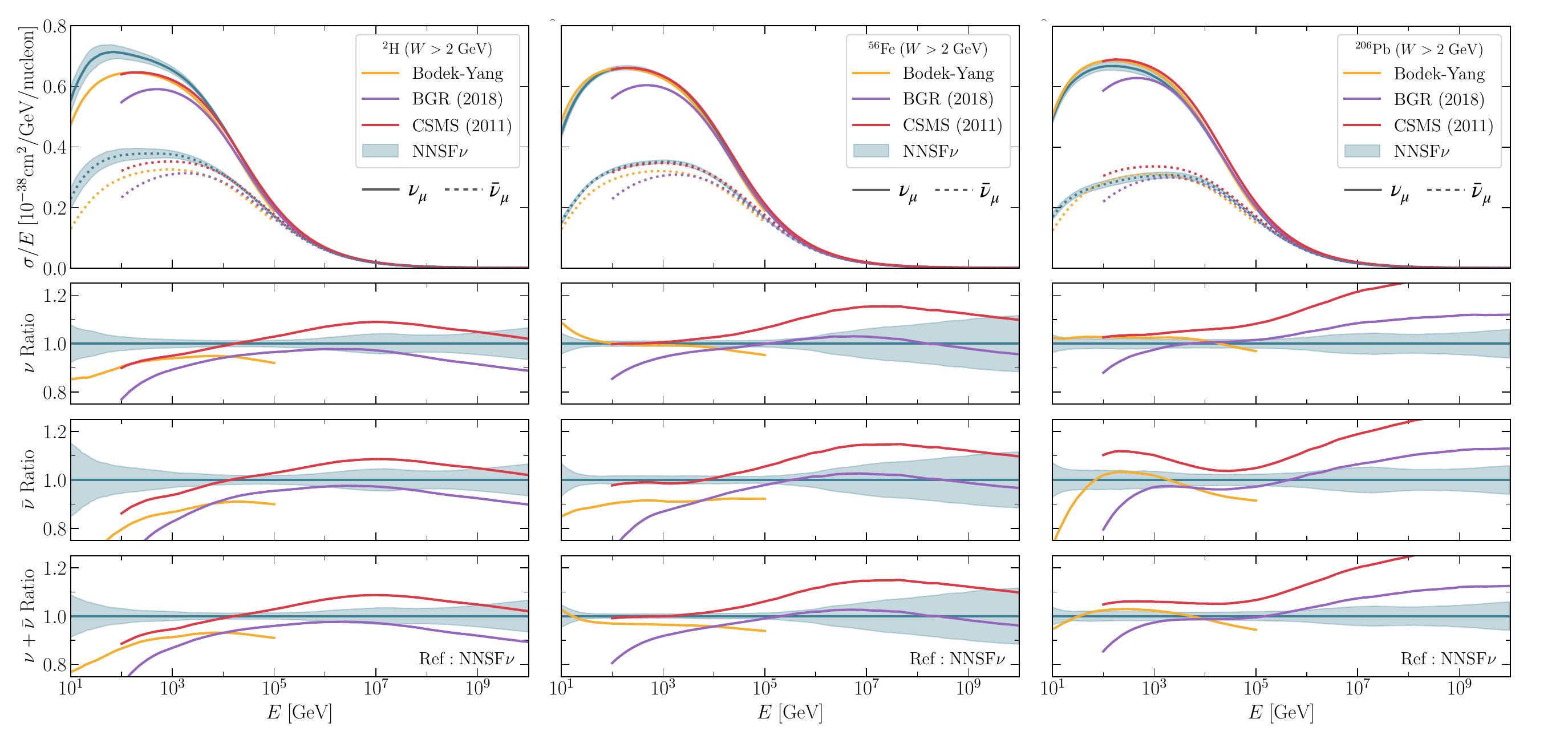}
 \caption{The inclusive neutrino-nucleus inelastic cross section, Eq.~(\ref{eq:sigma_inclusive}),
 divided by the neutrino
   energy $E_\nu$ as a function of the latter.
   The NNSF$\nu$ prediction, together with the associated 68\% CL
   uncertainty band, is compared with the central values of the
   Bodek-Yang, BGR18, and CSMS11 calculations.
   The top panels displays the results for neutrinos
   and antineutrinos scattering on deuterium (left), iron (middle), and lead (right panel) targets.
   The bottom panels display the corresponding ratios for neutrino, antineutrino,
   and their sum  with respect to the central value of the NNSF$\nu$ baseline calculation.
   We display the energy region covering from $E_\nu=10$ GeV up to $E_\nu=10^{11}~{\rm GeV}$.
 }    
 \label{fig:nu_xsec_tot_models}
\end{figure}

\paragraph{Comparison with Bodek-Yang, BGR18, and CSMS11.}
Fig.~\ref{fig:nu_xsec_tot_models} displays
the ratio of the inclusive neutrino-nucleus inelastic cross section, Eq.~(\ref{eq:sigma_inclusive}),
over  the neutrino energy $E_\nu$ as a function of the latter.
   The NNSF$\nu$ prediction for the ratio $\sigma^{\nu N}/E_\nu$, together with the associated 68\% CL
   uncertainty band
   is compared with the central values of the
   Bodek-Yang, BGR18, and CSMS11 calculations.
   The top panels display the absolute predictions for neutrinos
   and antineutrinos scattering on deuterium (understood here as
   isoscalar $^2$H without nuclear effects), iron, and lead  targets.
   The bottom panels show
  the corresponding ratios for neutrinos, antineutrinos,
   and their sum  with respect to the central  NNSF$\nu$ baseline.
The ratio $\sigma^{\nu N}/E_\nu$ is provided in units of  $10^{-38}$ cm$^2$/GeV
per nucleon, hence  assuming the proton and neutron content of the average nucleon
in the target nuclei.
   We display  neutrino energies from $E_\nu \simeq 10$ GeV
   up to $E_\nu \simeq 10^{11}~{\rm GeV} =100~{\rm EeV}$.

The NNSF$\nu$ determination is the only available theory prediction applicable in the complete
range of $E_\nu$ relevant for neutrino phenomenology.
 The region of formal applicability of the BY calculation is $E_\nu \le 10^5$ GeV,
 while CMS11 and BGR18 are restricted to $E_\nu > 100$ GeV.
 Furthermore, BGR18 was optimised to high-energy scattering and does not provide
 a good description of the energy region sensitive to low-$Q$ values.
In the comparisons of Fig.~\ref{fig:nu_xsec_tot_models},
the CMS11 and BGR18 predictions are provided for a free isoscalar
 target and hence do not account for nuclear modification effects.
The latter could be included by means of an external nuclear PDF analysis
within a  factorised  approach,
but only in the perturbative QCD region. 
Nuclear corrections are data-driven and model-independent in NNSF$\nu$,
while they are model-dependent and neglect the constraints from proton-ion
collisions at the LHC in the  Bodek-Yang case.
These nuclear effects are specially significant for heavy nuclei such as lead,
as we discuss below.

The uncertainty band of the  NNSF$\nu$ prediction
varies from a few percent up to a maximum of 15\%, depending on
$E_\nu$ and the nuclear target.
At low energies, it is the largest for $^2$H, given the lack of direct
experimental constraints on low-$Q$ neutrino-hydrogen structure functions.
At very high energies, it is the largest in iron since both for a free nucleon
and for a lead target  the nNNPDF3.0-based calculation accounts for
the constraints on small-$x$ PDFs provided by charm production at LHCb.
On the intermediate energy region with 100 GeV $\lsim E_\nu \lsim 100$ TeV, uncertainties in
the NNSF$\nu$ calculation of the inclusive cross-sections are at the few percent level at most.

Concerning the comparison between NNSF$\nu$ and the theory predictions from other groups,
starting with the case of lead nuclei one observes agreement within uncertainties with the Bodek-Yang
calculation for $E_\nu \lsim 10$ TeV, except for antineutrinos with very low energies.
For higher energies,  NNSF$\nu$ agrees with BGR18 for energies above 1 TeV up to $10^6$ TeV,
with the latter overshooting the former for $E_\nu \ge 10^6$ TeV due to the missing nuclear
corrections related to the strong small-$x$ quark and gluon shadowing in lead nuclei
found in nNNPDF3.0.
The differences with CSMS11 at high energies are explained from the absence of nuclear effects,
the choice of proton PDFs and the corresponding small-$x$ behaviour, and the treatment of
top quark mass effects, see the discussion of~\cite{Bertone:2018dse,Garcia:2020jwr}.

In the case of an iron target, there is also good agreement between NNSF$\nu$ and the Bodek-Yang
predictions for neutrinos, while for antineutrinos the latter is suppressed by a factor around 10\%.
As demonstrated below in Fig.~\ref{fig:nu_xsec_nutev}, the NNSF$\nu$ antineutrino predictions on Fe are preferred
by the NuTev measurements of the inclusive cross-sections.
In the case of the $\nu+\bar{\nu}$ sum, the differences between NNSF$\nu$ and BY are relatively
moderate.
The NNSF$\nu$ predictions are consistent with BGR18 for $E_\nu \gsim 100$ TeV, as expected
from the use of common settings for the QCD structure functions and the similarities
in the input (n)PDFs, given that no strong nuclear small-$x$ shadowing effects have been identified in the case of
lead.
The CSMS11 calculation agrees with  NNSF$\nu$ until around 10 TeV, and then overshoots
it for the reasons discussed in the case of lead~\cite{Bertone:2018dse,Garcia:2020jwr}.

Finally, concerning the isoscalar free-nucleon target $^2$H,
NNSF$\nu$ predicts a larger cross-section as compared to Bodek-Yang by a factor
between 10\% and 20\%, depending on the $E_\nu$ value.
For this target, there exist no direct
experimental constraints on neutrino-nucleon scattering,
and hence in NNSF$\nu$ the  $A=1$, low-$Q$ behaviour arises from
the extrapolation from the data  with $A\gsim 20$ included in the fit and
from the  QCD constraints in Region II for $Q \ge Q_{\rm thr}$ with $A=1$.
Despite the absence of direct constraints,
thanks to  the  NNSF$\nu$ matching procedure
for energies within $E_\nu \gsim 1$ TeV a reliable prediction
can be obtained by means of the perturbative QCD calculation using
the reasonably accurate proton PDFs.
In this high-energy region, the NNSF$\nu$ prediction is bracketed by the CSMS11 from above
and BGR18 from below.
The differences with BGR18 are explained by a combination of the input
PDFs and the treatment of top quark mass effects~\cite{Garcia:2020jwr}, since  NNSF$\nu$ uses
FFNS5 and BGR18 is based on FONLL instead, see also Sect.~\ref{sec:theory}. 

Nevertheless, neutrino cross-sections on a free-nucleon target
are of limited phenomenological interest
due to the absence of data for this target material.
This said, the large spread in the theory predictions for a free-nucleon target motivates
future experimental analyses of neutrino scattering in the region $E_\nu \lsim 1$ TeV with data taken
on hydrogen, deuterium, or other light nuclei as targets.

\paragraph{Comparison with experimental data.}
Several experiments have measured the inclusive neutrino-nucleus interaction cross section
for different nuclear targets and energy ranges.
To further validate the NNSF$\nu$ predictions on neutrino cross-section measurements,
Fig.~\ref{fig:nu_xsec_nutev} compares NNSF$\nu$  with the NuTeV~\cite{NuTeV:2005wsg} experimental data on inclusive
neutrino cross sections an iron target, separately for neutrinos and antineutrinos.
We focus on the energy region (between 20 GeV and 380 GeV) relevant for the interpretation
of the NuTeV data.
 In addition to NNSF$\nu$, we also include the predictions
 from the Bodek-Yang, CSMS11 and BGR18 calculations.

\begin{figure}[!t]
 \centering
 \includegraphics[width=0.68\linewidth]{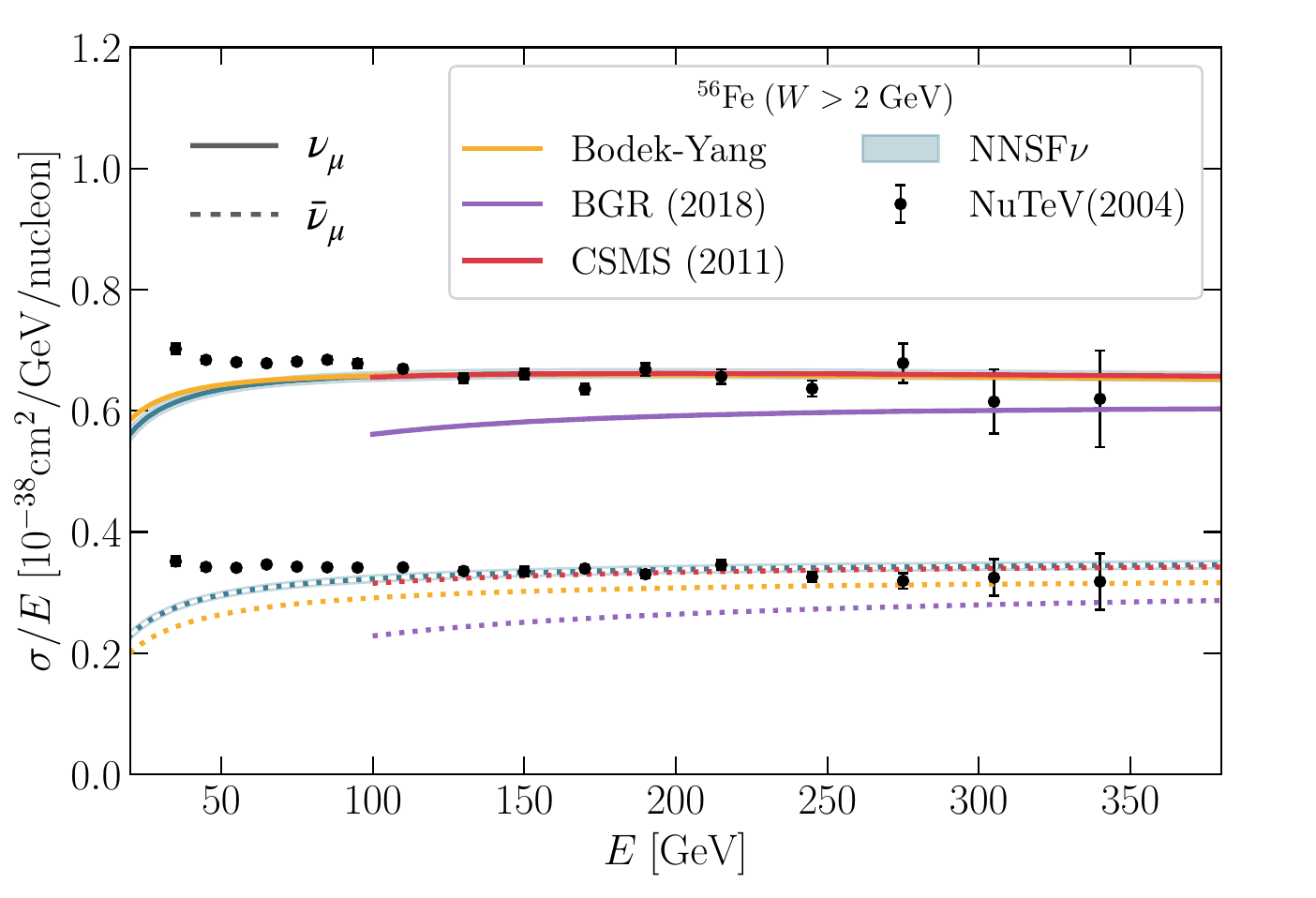}
 \caption{Same as the upper panel of Fig.~\ref{fig:nu_xsec_tot_models}, restricted to an iron
 target and focusing on the region between 20 GeV and 380 GeV.
 The predictions of NNSF$\nu$, Bodek-Yang, CSMS11 and BGR18 are compared with the experimental
 data from NuTeV~\cite{NuTeV:2005wsg} separately for neutrinos and antineutrinos.
  The invariant mass cut $W\ge 2$ GeV
   is applied to the theoretical calculations but not to the NuTeV measurement,
   explaining the disagreement for $E_\nu \lsim 100$ GeV where other contributions
   in addition to inelastic scattering become significant.
 }    
 \label{fig:nu_xsec_nutev}
\end{figure}

For the energy region in which inelastic scattering dominates, $E_\nu \gsim 100$ GeV,
we find excellent agreement between the NuTeV data and the NNSF$\nu$
predictions, separately for neutrinos and antineutrinos.
The Bodek-Yang calculation undershoots the NuTeV measurements on antineutrinos by up to 10\%.
CSMS11 is in good agreement with  NNSF$\nu$ and the NuTeV data,
while the BGR18 calculation, optimized for
high-energy scattering, undershoots it for both neutrinos and antineutrinos.
We note that the invariant mass cut $W\ge 2$ GeV
is applied only to the theoretical calculations but not to the NuTeV measurement.
This difference explains the disagreement for $E_\nu \lsim 100$ GeV where other contributions
   in addition to inelastic scattering become significant, in particular resonant scattering.
In this low-energy region, NNSF$\nu$ and Bodek-Yang exhibit the same qualitative behaviour.

\paragraph{Inelasticity.}
In neutrino-hadron scattering, the inelasticity variable  $y=Q^2/(2m_NxE_{\nu})$ represents
the fraction of the incoming neutrino energy which is transferred to the hadronic final state
and hence to the associated hadronic shower.
The higher the value of $y$, the more energy that is  transferred to the hadronic shower,
facilitating the experimental measurement and characterisation of the latter.
At intermediate energies, with $E_\nu \lsim 100$ TeV, neutrino interactions are expected to produce on average 
more energetic hadronic showers than their antineutrino counterparts.
Therefore, a measurement of the event-by-event inelasticity provides
a useful handle in order to statistically separate $\nu$- from $\bar{\nu}$-bar
initiated interactions, a feature exploited in the by IceCube analysis
of~\cite{IceCube:2018pgc}.
The expected mean value of the inelasticity as a function of the neutrino energy
can be computed as
\be
\label{eq:sigma_inclusive_y}
\la y \ra (E_\nu) = \lp \int_{Q^2_{\rm min}}^{Q^2_{\rm max}} {\rm d}Q^2
\lc  \int_{x_0(Q^2)}^1{\rm d}x\, y\,\frac{{\rm d}^2\sigma^{\nu N}}{{\rm d}x{\rm d}Q^2}(x,Q^2,y)\rc\rp  \Bigg/ \sigma^{\nu N}(E_\nu)   \, ,
\ee
using as input a given prediction for the double-differential scattering cross-section.

\begin{figure}[!t]
 \centering
 \includegraphics[width=1.0\linewidth]{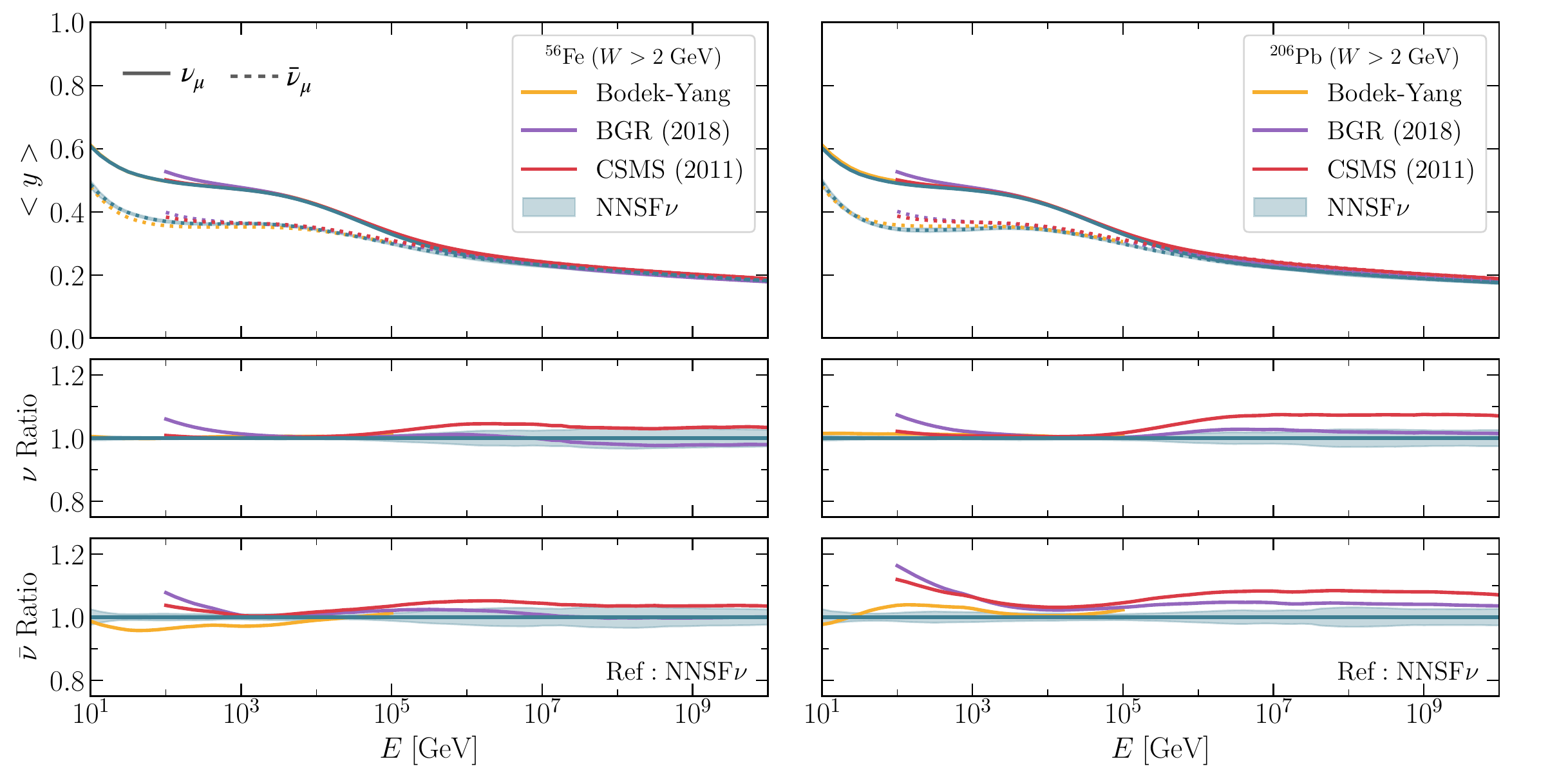}
 \caption{The mean value of the inelasticity, $\la y \ra$,
relevant for inelastic neutrino scattering as a function 
 of the neutrino energy $E_\nu$.
   The NNSF$\nu$ prediction, together with the associated 68\% CL
   uncertainty band, is compared with the central values of the
   Bodek-Yang, BGR18, and CSMS11 calculations.
     The top panels displays the results for neutrinos
   and antineutrinos for both iron (left) and lead (right panel) targets,
   while the bottom ones display the ratios to the NNSF$\nu$ baseline
   separately for neutrino and antineutrino.
 }    
 \label{fig:mean_y}
\end{figure}

Fig.~\ref{fig:mean_y} displays the mean value of the inelasticity,
Eq.~(\ref{eq:sigma_inclusive_y}),
as a function of the neutrino energy for both iron and lead targets.
   The NNSF$\nu$ prediction for $\la y \ra$, together with the associated 68\% CL
   uncertainty band, is compared with the central values of the
   Bodek-Yang, BGR18, and CSMS11 calculations.
   The bottom panels display the ratio of the different calculations
   with respect to the NNSF$\nu$ central value.
  As mentioned above, for $E_\nu \lsim 100$ TeV the average inelasticity is markedly larger
   for neutrino-initiated scattering, while at higher energies the prediction
   for $\la y\ra$ becomes projectile-independent. 
An overall good agreement is observed between the different predictions,
with differences up to the 10\% level depending on the target and $E_\nu$ ranges.
Specifically, at high  neutrino energies, the NNSF$\nu$ prediction for $\la y\ra$
in a lead target is around 10\% smaller as compared to CSMS11, partially explained by the nuclear corrections accounted for
in the former but not in the latter.
Good agreement between the Bodek-Yang and the NNSF$\nu$ calculations of $\la y \ra$
is obtained in the region of applicability of the former,
with residual differences at the few
percent level.

\begin{figure}[!t]
 \centering
 \includegraphics[width=0.65\linewidth]{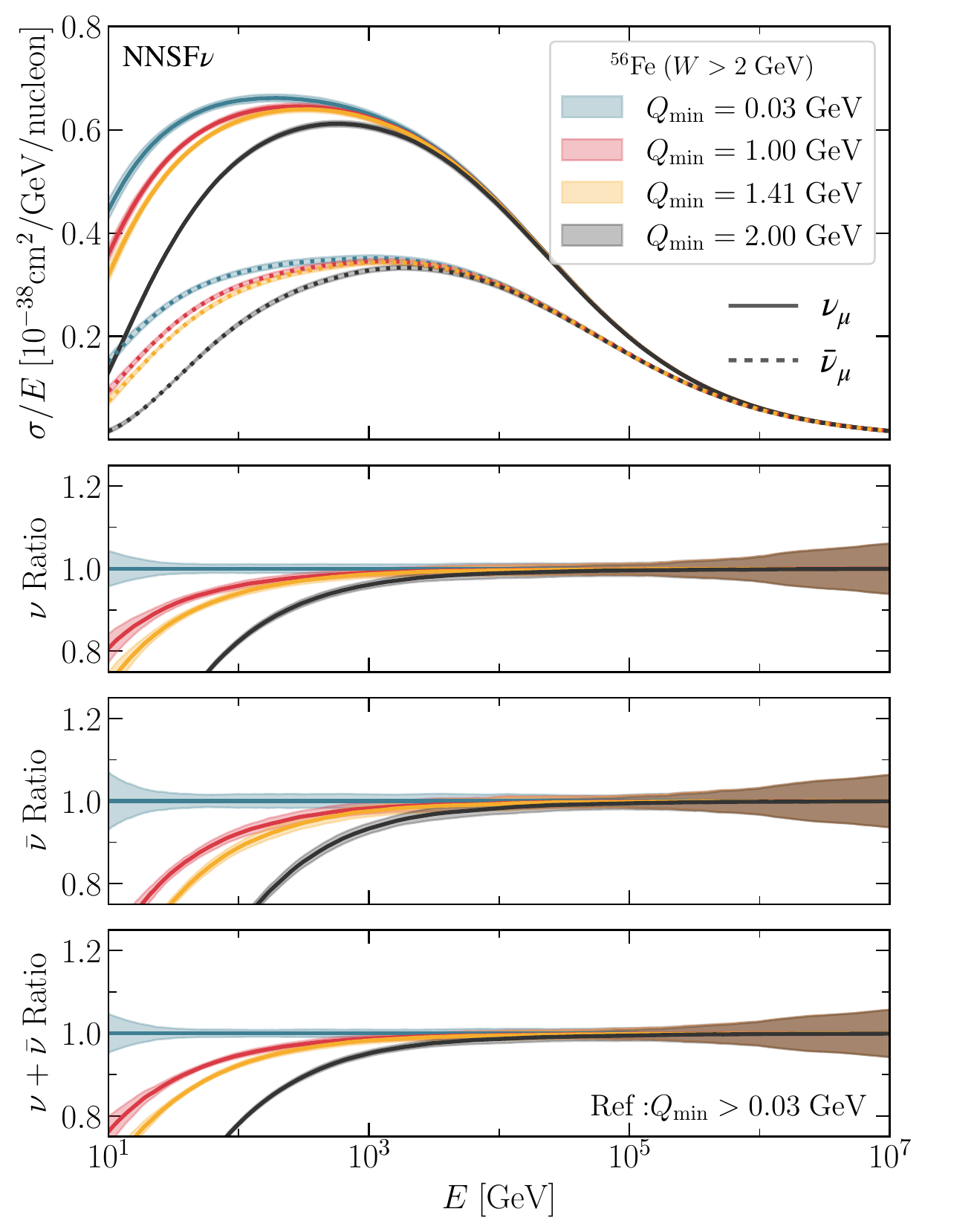}
 \caption{Same as Fig.~\ref{fig:incl_xsec_nuclearratios} for an iron target
 and restricted to the energy region with $E_\nu \le 10^7$ GeV.
 We assess the dependence of the inclusive cross-section with respect to variations
 in the lower integration limit $Q_{\rm min}$ in
   Eq.~(\ref{eq:sigma_inclusive}).
The default value of  $Q_{\rm min}=0.03$ GeV
is compared to calculations based on $Q_{\rm min}=$
1, 1.41, and 2 GeV.
 }    
 \label{fig:incl_xsec_kindep1}
\end{figure}

\paragraph{Sensitivity to the low-$Q$ region.}
Fig.~\ref{fig:incl_xsec_kindep1} displays the  NNSF$\nu$ inclusive neutrino scattering
cross-sections on  an iron target for energies between 10 GeV and  $10^7$ GeV,
for different values of  he lower integration limit $Q_{\rm min}$ in
   Eq.~(\ref{eq:sigma_inclusive}).
Specifically, the default value of  $Q_{\rm min}=0.03$ GeV
is compared to calculations based on $Q_{\rm min}=$
1, 1.41, and 2 GeV respectively.
This comparison allows one to determine the relative contribution
of the low-$Q$ region to the inclusive cross-sections as a function of $E_\nu$,
an effect which is similar for neutrinos and antineutrinos.

From the analysis of Fig.~\ref{fig:incl_xsec_kindep1} one
determines that neutrinos with energies
around $E_\nu \simeq 100$ GeV receive a contribution of up to 10\%
from the region with $Q \le 1.41$ GeV and up to 25\% from the region
with $Q \le 2$ GeV.
As the neutrino energy is increased, the contribution from the low-$Q$ region decreases,
but even for $E_\nu \approx 1$ TeV neutrinos, up to 10\% of the inclusive
cross-section can arise from momentum transfers of $Q \le 2$ GeV.
As discussed in Sect.~\ref{sec:theory}, this low-$Q$ region
is affected by sizable theory uncertainties
(MHOUs, higher twists, mass effects, factorisation breakdown, ....),
and hence Fig.~\ref{fig:incl_xsec_kindep1} illustrates how
the NNSF$\nu$ data-driven method makes it possible bypassing the limitations
of the perturbative QCD calculation of neutrino inclusive cross-sections
in the region 
$E_\nu \le$ few TeV.
It is interesting to highlight that for neutrino energies between  300 GeV and a few TeV,
relevant for the interpretation of 
 LHC far-forward neutrino scattering experiments, neither a data-driven, theory-agnostic
 calculation nor a purely QCD calculation can accurately predict
 the inclusive cross-sections.
For higher neutrino energies, $E_\nu \gsim 10$ TeV, the NNSF$\nu$ prediction
becomes independent of structure functions in the $Q \le 2$ GeV region and hence a
perturbative
QCD calculation of the cross-section such as that provided by {\sc\small YADISM}
can be reliably deployed.

\subsection{Impact of nuclear effects}
\label{sec:impact_nuclear}

\begin{figure}[!t]
 \centering
 \includegraphics[width=0.65\linewidth]{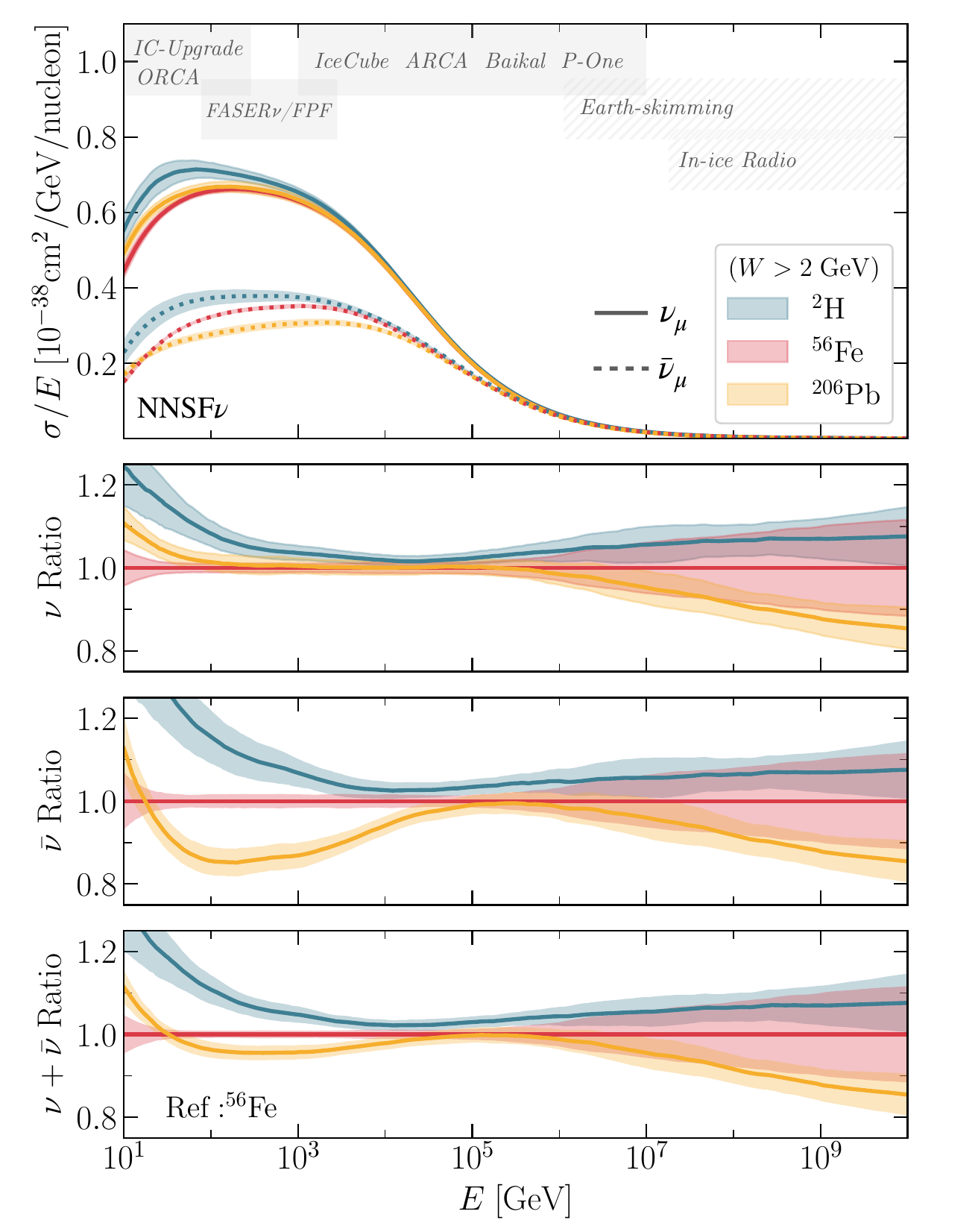}
 \caption{The NNSF$\nu$ predictions for the inclusive inelastic scattering
 cross-sections comparing the results of  neutrino and antineutrino
   projectiles on deuterium, iron, and lead targets.
    The grey bands indicate the $E_\nu$ regions relevant for different experiments and detection techniques.
   The top panel displays the absolute cross-sections while the bottom ones
   the ratio to the iron target baseline sequentially for neutrinos,
   antineutrinos, and their sum.
 }    
 \label{fig:incl_xsec_nuclearratios}
\end{figure}

As illustrated by Fig.~\ref{fig:nu_xsec_tot_models}, the NNSF$\nu$ scattering cross section (per nucleon)
on targets such as iron or lead is in general different to that evaluated
on a isoscalar free-nucleon target.
To further scrutinise the role that nuclear effects have on the inclusive neutrino cross-sections,
Fig.~\ref{fig:incl_xsec_nuclearratios} displays the NNSF$\nu$ predictions for the
 inelastic 
 cross-section (divided by $E_\nu$) comparing the results of  neutrino and antineutrino
projectiles on deuterium (isoscalar free-nucleon), iron, and lead targets.
The top panel displays the absolute cross-sections while the bottom ones show
the ratio to the iron target baseline, sequentially for neutrinos,
antineutrinos, and their sum.
As in previous comparisons, the band in the  NNSF$\nu$ predictions
indicates the associated 68\% CL uncertainties.

To facilitate the identification of the different energy regions of interest
for neutrino phenomenology, Fig.~\ref{fig:incl_xsec_nuclearratios}
also displays grey bands indicating the $E_\nu$ regions
relevant for different experiments and detection techniques.
Specifically, from low to high energies, we indicate the coverage of:
\begin{itemize}

\item $10~{\rm GeV}\lsim E_\nu \lsim 300$ GeV: KM3NET-ORCA and DeepCore/IceCube-upgrade.

\item $100~{\rm GeV}\lsim E_\nu \lsim 4$ TeV: FASER$\nu$, SND@LHC, Forward Physics Facility (LHC far-forward neutrinos).

\item $1~{\rm TeV}\lsim E_\nu \lsim 10$ PeV:  IceCube, KM3NET-ARCA, Baikal, P-One.

\item $1~{\rm PeV}\lsim E_\nu \lsim 10$ EeV: Earth-skimming neutrinos detection.

\item $10~{\rm PeV}\lsim E_\nu \lsim 10$ EeV: In-ice radio neutrino detection.

\end{itemize}

This list is not exhaustive, and has been added to the plot for illustrative purposes only.

As further discussed in App.~\ref{app:nuclear}, two different types
of nuclear effects are responsible for  differences between scattering rates on a heavy nucleus
and on a free proton (hydrogen target).
The first is related to the different content in protons and neutrons,
with  iron (lead) containing $Z=26$ ($Z=82$) protons
and $A-Z=30$~(126) neutrons.
Therefore, as compared to a hydrogen target, the heavy nuclear targets display an enhanced (suppressed)
content of down (up) valence quarks which leads
in turn to an enhancement (suppression) of the associated
structure functions for neutrino (antineutrino) scattering.
This effect is most relevant in the valence structure function region,
while as $E_\nu$ increases the cross section becomes dominated by small-$x$ scattering
involving isospin-symmetric  sea quarks and gluons.
In the comparisons in Fig.~\ref{fig:incl_xsec_nuclearratios} this effect is
partially factored out since we normalise to an isoscalar $^2$H target, though 
the non-isoscalarity of heavy nuclei, specially in lead, still plays a role.

The second type of nuclear effect is the modifications of the structure of bound
nucleons as compared to their free-nucleon counterparts.
In the perturbative QCD region, these  modifications are encoded
by the nuclear PDFs, here taken from the nNNPDF3.0 determination.
The corrections, quantified in App.~\ref{app:nuclear},
are similar for neutrinos and antineutrinos and become most important
in the shadowing region at medium and small-$x$, where a strong suppression
in heavy nuclei is preferred.
Nuclear structure modification effects are also present for $E_\nu \lsim 1$ TeV,
a region where they are partially described by the DIS calculations but which also
 receives contributions from the SIS region which cannot be expressed
in terms of factorised nuclear PDFs.
Within the NNSF$\nu$ framework, one estimates the nuclear modifications in Region I by allowing
a free dependence of $F_i(x,Q^2,A)$ to be directly constrained from the data, and then matched
to the QCD calculation in Region II.

From the comparisons in Fig.~\ref{fig:incl_xsec_nuclearratios}, one observes that
differences between inclusive cross-sections on Fe and Pb targets and those on $^2$H targets
are moderate (a few percent at most) for energies in the intermediate region between 10 TeV and a few PeV.
In the high energy region, for $E_\nu \gsim 10$ TeV, the most marked effect is the strong suppression
of the cross-sections in lead as compared to those in a free-nucleon target, reaching
up to 20\% at 100 EeV.
This suppression is a consequence of the small-$x$ shadowing of quark and gluons in nNNPDF3.0.
Given the large nPDF uncertainties associated to an iron target, high-energy cross-sections for Fe
are compatible within errors with those for both $^2$H and Pb.

In the region where the non-DIS contribution is sizable, $E_\nu \lsim 1$ TeV,
differences between Fe and Pb targets
for the $\nu+\bar{\nu}$ cross-section
remain at the few percent level, with again the latter being suppressed in
comparison to the former.
The cross-sections on a $^2$H target are enhanced at low energies, with the difference
being up to 20\% in the $E_\nu=10$ GeV where the inelastic component is very small.
In the same energy region, differences between the three nuclear targets
appear to be less (more) significant once we consider separately the predictions for neutrinos
(antineutrinos).

While in this section we discuss explicitely nuclear effects only for Pb and Fe targets,
NNSF$\nu$ structure functions and inclusive cross-section predictions are provided
for all targets relevant for neutrino phenomenology and summarised in Table~\ref{tab:sf_lhapdf}.
In addition, below we provide dedicated predictions for the energy ranges and
target materials relevant for far-forward neutrino experiments at the LHC.

\subsection{Far-forward neutrino scattering at the LHC}
\label{sec:fpf}

Precise predictions for neutrino scattering rates are a key ingredient
for the interpretation of data from experiments aiming to detect
and study the far-forward neutrinos produced in LHC collisions.
These include the current
SND@LHC and FASER$\nu$ experiments as well as the proposed AdvSND@LHC, FLArE, and FASER$\nu$2,
which would be installed in the Forward Physics Facility operating concurrently
with the HL-LHC.
The dominant component of these far-forward LHC neutrino fluxes lies in the region
between a few hundreds of GeV and a few TeV~\cite{Kling:2021gos,Bai:2020ukz}, with the high-energy
component most
sensitive to neutrinos produced from charm meson decays.
Here we provide dedicated predictions for the inclusive neutrino
scattering cross-sections in the FPF energy range, specifically
between $E_\nu=100$ GeV and 10 TeV, assuming a tungsten (W) nuclear target
with $A=184$, which is the current target of  FASER$\nu$
and the intended target for the FASER$\nu$2 experiment.
For completeness, we also provide predictions for neutrino scattering
on an oxygen (O) target, which are relevant for the intepretation
of ongoing and future neutrino
oscillation measurements taking place at
the DeepCore/IceCube and KM3NET-ORCA experiments in a partially overlapping energy range
as the FPF.

\begin{figure}[!t]
 \centering
 \includegraphics[width=1.00\linewidth]{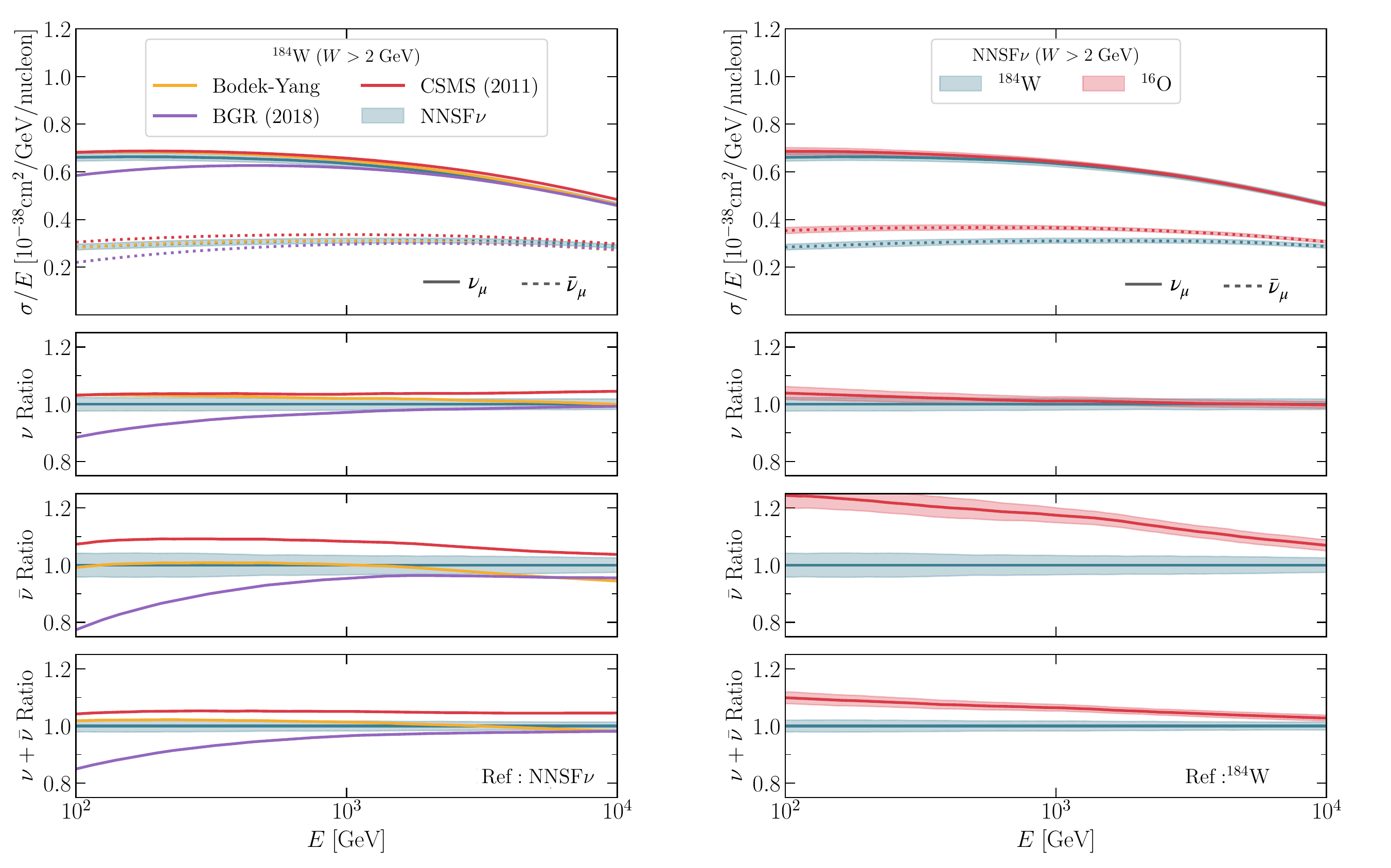}
 \caption{Left: same as Fig.~\ref{fig:nu_xsec_tot_models} on a tungsten (W)
   target and restricted to the energy region relevant for
   the FPF experiments.
   Right: same as Fig.~\ref{fig:incl_xsec_nuclearratios}
   comparing the NNSF$\nu$ predictions for W and O targets,
   again in the range covered by neutrino scattering at the FPF.
 }    
 \label{fig:nu_xsec_fpf}
\end{figure}

With this motivation, Fig.~\ref{fig:nu_xsec_fpf} (left)
presents the same comparison
as in Fig.~\ref{fig:nu_xsec_tot_models} now on a tungsten
target and  restricted to the energy region relevant for
the FPF experiments.
In this energy region and for this specific nuclear target, one observes excellent agreement
between the NNSF$\nu$ determination and the Bodek-Yang model
predictions, except perhaps for antineutrinos with $E_\nu$ of several TeV.
Given that  the two calculations are completely different
in terms of both data and QCD input as well as in methodological
assumptions, such agreement at the $1\%$ level is remarkable.
The CSMS11 (BGR18) predictions tend to overestimate (underestimate) the
FPF neutrino scattering rates in comparison to the NNSF$\nu$ baseline,
specially for BGR18 at low energies where the calculation falls outside
its regime of applicability.

All in all, Fig.~\ref{fig:nu_xsec_fpf} demonstrates that the
uncertainties on the inclusive neutrino cross-section predictions
for  a W target
at FPF energies are estimated to be at most at the 5\% level,
and even less for the combined $\nu+\bar{\nu}$ cross-section.
We conclude that state-of-the-art calculations of neutrino scattering
in the SM at the FPF energies can achieve a precision of
a few percent, thus providing
an excellent starting point for further investigations of neutrino
interactions in this unexplored energy range, for instance in terms
of anomalous couplings or of EFT effects.

The right panel of Fig.~\ref{fig:nu_xsec_fpf} displays
a similar comparison as that of Fig.~\ref{fig:incl_xsec_nuclearratios},
now for the  NNSF$\nu$ predictions for tungsten and oxygen targets.
While neutrino cross-sections display almost no differences between the two
targets, for the antineutrino case we observe sizable nuclear effects, of up to 20\% and
specially for energies below 1 TeV.
As the energy is increased, the  differences related to nuclear
corrections wash away but even at 10 TeV they can represent a 10\% effect.
The predictions for the combined $\nu+\bar{\nu}$ cross-section, averaged
over the two projectiles, lead to more moderate differences but still
up to the 10\% level at energies of few hundreds of GeV.
Fig.~\ref{fig:nu_xsec_fpf} indicates that properly accounting for nuclear effects
associated to the target material
is required in order to translate constraints on neutrino interactions
obtained from the FPF experiments to those from KM3NET and IceCube analyses and vice-versa,
specially
in those cases where one is sensitive to the separation between
neutrinos and antineutrinos in the initial state.

\section{Summary and outlook}
\label{sec:summary}

We have presented a novel approach to the determination
of neutrino inelastic structure functions based on the combination of
a data-driven parametrisation
at low and moderates values of $Q^2$
matched to perturbative QCD calculations at high $Q^2$.
The resulting structure functions, dubbed NNSF$\nu$, enable the evaluation of inclusive
neutrino scattering cross sections over 12 orders of magnitude in energy $E_{\nu}$ from
a few GeV up to the multi-EeV region.
In particular, it makes possible the accurate evaluation of scattering event rates
for neutrinos with energies between 100 GeV and a few TeV, relevant for
both far-forward neutrino detection at the LHC and for atmospheric neutrino oscillation
experiments, without the need to impose acceptance cuts in $Q^2$.
The NNSF$\nu$ determination comes accompanied 
with a faithful estimate of the associated uncertainties and  accounts for the
constraints provided by state-of-the-art determinations
of nucleon and nuclear structure.

We have compared the  NNSF$\nu$ determination with existing calculations
in the literature, such as the popular Bodek-Yang model,
both at the level
of structure functions and of inclusive cross sections.
We have demonstrated the agreement between our parametrisation
and the experimental data, highlighted the
smooth matching between the data-driven extraction and the QCD boundary conditions,
and assessed the behaviour of  NNSF$\nu$ in the extrapolation regions.
We have  quantified the impact
of nuclear PDF modifications, finding that these
are most significant for heavy nuclei such as tungsten and lead.
We have also studied the implications of the NNSF$\nu$ analysis for key observables
in neutrino DIS, such as the Gross-Llewellyn-Smith sum rule.

The NNSF$\nu$ determination is made available in terms of {\sc\small LHAPDF} interpolation
tables~\cite{Buckley:2014ana} for neutrino structure functions corresponding
to the relevant nuclear targets, as described in App.~\ref{app:delivery}.
Two different sets of grids are provided, one associated to the neural network determination
and the other to the {\sc\small YADISM} calculation, and a {\sc\small Python} script
showcasing how to use and combine these two grids is provided.
We also provide a script that takes as input these  {\sc\small LHAPDF}  structure
function grids and integrates them to evaluate the inclusive cross section and
the corresponding uncertainties.
For completeness, we also make available look-up tables with the NNSF$\nu$ predictions
for inclusive cross sections $\sigma^{\nu A}(E_{\nu})$ in the relevant range of $E_{\nu}$ and
target nuclei.
The {\sc\small LHAPDF} grids, {\sc\small Python} driver scripts, and
look-up tables are available from the project {\sc\small GitHub} repository
\begin{center}
\url{https://nnpdf.github.io/nnusf}
\end{center}
together with the data, code, and theory tables that enter the structure
function determination.
The NNSF$\nu$ structure functions have
also been implemented in the  {\sc\small GENIE}~\cite{Andreopoulos:2009rq} 
event generator, where it can be accessed in a similar manner as the BGR18 and
Bodek-Yang calculations.
It is hence readily available in a software
framework already integrated in the analysis pipeline of many neutrino experiments.

The results presented in this work could be expanded in different directions.
First, far-forward neutrino measurements taking place at LHC experiments
such as FASER$\nu$ and SND@LHC, and in the longer term the Forward Physics Facility,
will provide further constraints on the modelling of neutrino scattering at low and moderate $Q^2$.
Second, the NNSF$\nu$  calculation could be integrated into other neutrino event generators
to streamline its usage, also
for neutrino experiments where  inelastic scattering,
without dominating, still represents a sizable contribution to the total event rates.
Third, by incorporating updated determinations of proton and nuclear PDFs
benefitting from new experimental constraints and higher-order
theory calculations, such as the information provided by charm production
at LHCb combined with small-$x$ BFKL resummation to improve
predictions of neutrino structure functions at UHE energies~\cite{Gauld:2016kpd,Ball:2017otu,Bertone:2018dse}.
Fourth, by considering the information provided by lattice QCD calculations~\cite{Constantinou:2020hdm}
to constrain structure functions in the non-perturbative low $Q^2$ region.

All in all, the NNSF$\nu$ analysis bridges a significant gap 
in current neutrino phenomenology by providing a consistent
determination of structure functions enabling the calculation of cross sections
in the full range of $E_{\nu}$ relevant for inelastic scattering.
For the first time, a determination of structure functions can be used
for predictions involving
accelerator neutrinos with $E_{\nu}\sim {\rm few~GeV}$, atmospheric and collider neutrinos with
energies between
$\sim{\rm 100~GeV}$ and ${\rm few~TeV}$,
and ultra-high energy neutrino scattering at EeV energies.
The availability of NNSF$\nu$ allows bypassing a major limitation
of current analyses where different
cross-section calculations, inconsistent among them, have
to be adopted depending on the range of energies involved.

\subsection*{Acknowledgments}

We thank Jonathan Feng, Felix Klein, Hallsie Reno, and Dennis Soldin for
productive discussions concerning neutrino scattering in the context
of the Forward Physics Facility working group.
We are grateful to Nusch Mortazavi for collaborations in an early stage of the project.
A.~C. and partially R.~S. are supported by
the European Research Council under 
the European Union's Horizon 2020 research and innovation Programme
(grant agreement n.740006).
R.~S. is partially supported by the U.K. Science and Technology Facility Council (STFC) grant ST/P000630/1.
J.~R. and G.~M. are partially supported by NWO (Dutch Research Council).
J.~R. and T.~R. are supported by an ASDI (Accelerating Scientific Discoveries) grant
from the Netherlands eScience Center.
A.~G. acknowledges support from the European
Union’s H2020-MSCA Grant Agreement No.101025085
and the Faculty of Arts and Sciences of Harvard University.

\appendix
\section{Usage and delivery of the NNSF$\nu$ determination}
\label{app:delivery}

The software framework used to produce the  NNSF$\nu$ determination
of neutrino inelastic structure functions, together with the results
obtained from it, can be obtained from the project website:

\begin{center}
  \url{https://nnpdf.github.io/nnusf}
\end{center}
Specifically, there one can find the following:
\begin{itemize}

\item Installation instructions, code documentation, and user-friendly examples
  of the  NNSF$\nu$ software framework, including the data and theory runcards
  that allow reproducing the results of this paper.

\item Fast interpolation grids in $x$ and $Q^2$ for the inelastic structure functions $F_2$, $xF_3$,
  and $F_L$ for neutrinos, antineutrinos, and their sum.
  These structure function grids can be accessed by means of the
  {\sc\small LHAPDF} interface and are provided for all nuclear targets of phenomenological relevance
  for neutrino scattering experiments.

\item A driver code that evaluates, taking these structure functions grids as inputs,
  the corresponding central values, uncertainties, and correlations.
  As explained below,
  two structure function parametrisations with a different coverage in the $(x,Q^2)$ plane
  are provided and the driver code takes care of their combination.
  
\item A second {\tt Python} driver code that evaluates the inclusive neutrino cross section
  Eq.~(\ref{eq:sigma_inclusive}) as a function of $E_{\nu}$ from the {\sc\small LHAPDF} grids and
  provides the corresponding uncertainty estimate.

\item Look-up tables compiling the NNSF$\nu$ predictions for the inclusive cross sections $\sigma^{\nu A}$
  and $\sigma^{\bar{\nu} A}$
  for  phenomenologically relevant nuclei as a function of $E_{\nu}$, together with
  their uncertainty estimates.

\end{itemize}

In this appendix we provide details on these deliveries
associated to the NNSF$\nu$ determination, from the code and documentation to the
inclusive cross-section look-up tables.
In particular, we describe the prescriptions required to evaluate
central values and uncertainties for the 
structure functions and inclusive cross sections from the {\sc\small LHAPDF} grids provided.

\paragraph{The  NNSF$\nu$ framework.}
The framework developed in this work provides a stand-alone code to parametrise structure functions
from experimental data in the presence of general theory constraints.
It provides an independent implementation of the NNPDF fitting methodology, and shares
the main methodological aspects with the NNPDF4.0 and nNNPDF3.0 proton and nuclear PDF determination
such as the use of Stochastic Gradient Descent for the minimisation.
For the first time in this context, neural networks with three independent inputs
$(x,Q^2,A)$ are used to parametrise unknown functions from experimental data.

\paragraph{Structure functions.}
Table~\ref{tab:sf_lhapdf} lists the NNSF$\nu$ structure function grids released in the {\sc\small LHAPDF}
format.
We indicate the values of $(Z,A)$ for each nuclear target and the names
of the corresponding low-$Q$ and high-$Q$ grids.
Each grid provides, as a function of $x$ and $Q^2$,
as outputs the neutrino and antineutrino structure functions as well as their sum, namely
\bea
&&F_2^{\nu A}(x,Q^2)\, , F_L^{\nu A}(x,Q^2)\,  ,xF_3^{\nu A}(x,Q^2)\, ,\nonumber \\
&&F_2^{\bar{\nu} A}(x,Q^2)\, , F_L^{\bar{\nu} A}(x,Q^2)\,  ,xF_3^{\bar{\nu} A}(x,Q^2)\,, \\
&&F_2^{(\nu+\bar{\nu}) A}(x,Q^2)\, , F_L^{(\nu+\bar{\nu}) A}(x,Q^2)\, ,xF_3^{(\nu+\bar{\nu}) A}(x,Q^2)\, ,\nonumber
\eea
with the PDG ID codes being respectively {\tt [1001, 1002, 1003, 2001, 2002, 2003, 3001, 3002, 3003]}.
We emphasize that the number of replicas,  coverage on $x$ and $Q^2$, and 
statistical interpretation of the low- and high-$Q$ grids is different, as discussed below.
The nuclear targets listed in Table~\ref{tab:sf_lhapdf} are selected for their
relevance in the interpretation of past and present experimental data as well
for theoretical predictions for upcoming experiments.
For instance, tungsten (W) is relevant for the event rate predictions
at  FASER$\nu$ and its eventual successor  FASER$\nu$2, while oxygen (O) enters
scattering rates for neutrinos on air, water, or ice targets.
We also provide a grid for $(A,Z)=(15,31)$, which correspond to the average
nuclear numbers for Earth matter relevant for calculations of UHE neutrino attenuation
when crossing the Earth~\cite{Garcia:2020jwr}.

\begin{table}[t]
\begin{center}
  \renewcommand{\arraystretch}{1.25}
  \begin{tabular}{lll}
    \toprule
   $(Z,A)$ [target]$\qquad$ $\qquad$  & low-$Q$ grid $\qquad$ $\qquad$$\qquad$ &  high-$Q$ grid$\qquad$ $\qquad$ \\
        \midrule
        $(1,2)$ [D]  & {\tt NNSFnu\_D\_lowQ} & {\tt NNSFnu\_D\_highQ}   \\
        $(2,4)$ [He]    & {\tt NNSFnu\_He\_lowQ} & {\tt NNSFnu\_He\_highQ}  \\
        $(3,6)$ [Li]    &  {\tt NNSFnu\_Li\_lowQ} & {\tt NNSFnu\_Li\_highQ} \\
        $(4,9)$ [Be]   &   {\tt NNSFnu\_Be\_lowQ} & {\tt NNSFnu\_Be\_highQ}\\
        $(6,12)$ [C]   &   {\tt NNSFnu\_C\_lowQ} & {\tt NNSFnu\_C\_highQ} \\
$(7,14)$ [N]    & {\tt NNSFnu\_N\_lowQ} & {\tt NNSFnu\_N\_highQ}  \\
$(8,16)$ [O]   &  {\tt NNSFnu\_O\_lowQ} & {\tt NNSFnu\_O\_highQ}\\
        $(13,27)$ [Al]   & {\tt NNSFnu\_Al\_lowQ} & {\tt NNSFnu\_Al\_highQ}\\
      $(15,31)$ [Ea]   & {\tt NNSFnu\_Ea\_lowQ} & {\tt NNSFnu\_Ea\_highQ}\\
$(18,39)$ [Ar]   &  {\tt NNSFnu\_Ar\_lowQ} & {\tt NNSFnu\_Ar\_highQ}\\
$(20,40)$ [Ca]   &  {\tt NNSFnu\_Ca\_lowQ} & {\tt NNSFnu\_Ca\_highQ}\\
$(26,56)$ [Fe]   &  {\tt NNSFnu\_Fe\_lowQ} & {\tt NNSFnu\_Fe\_highQ}\\
$(29,64)$ [Cu]  &  {\tt NNSFnu\_Cu\_lowQ} & {\tt NNSFnu\_Cu\_highQ}\\
$(47,108)$ [Ag]  & {\tt NNSFnu\_Ag\_lowQ} & {\tt NNSFnu\_Ag\_highQ}\\
$(50,119)$ [Sn] &  {\tt NNSFnu\_Sn\_lowQ} & {\tt NNSFnu\_Sn\_highQ}\\
  $(54,131)$ [Xe]  & {\tt NNSFnu\_Xe\_lowQ} & {\tt NNSFnu\_Xe\_highQ}\\
$(74,184)$ [W] & {\tt NNSFnu\_W\_lowQ} & {\tt NNSFnu\_W\_highQ}\\
  $(79,197)$ [Au]  & {\tt NNSFnu\_Au\_lowQ} & {\tt NNSFnu\_Au\_highQ}\\
  $(82,208)$ [Pb] & {\tt NNSFnu\_Pb\_lowQ} & {\tt NNSFnu\_Pb\_highQ}\\
\bottomrule
\end{tabular}
\end{center}
\caption{The NNSF$\nu$ structure function grids released in the {\sc\small LHAPDF}
  format.
  We indicate the values of $(Z,A)$ for each nuclear target and the names
  of the corresponding low-$Q$ and high-$Q$ grids.
  Each grid provides, as a function of $x$ and $Q^2$,
  as outputs the neutrino and antineutrino structure functions $F_2$, $xF_3$, and $F_L$
  as well as their sum.
  We note that both the number of replicas and the statistical interpretation thereof
  is different in the low- and high-$Q$ grids, see text for more details.
  \label{tab:sf_lhapdf}
}
\end{table}

The low-$Q$ and high-$Q$ NNSF$\nu$ structure function grids are defined as follows:

\begin{itemize}

\item {\bf low-$Q$ grid. }
  This grid encapsulates the direct output of the
  NNSF$\nu$ neural networks trained on the neutrino structure function
  data and supplemented by the theoretical constraints from the QCD calculation
  based on nNNPDF3.0.

  The {\sc\small LHAPDF} grid is constituted by a central set
  and $N_{\rm rep}=200$ Monte Carlo replicas,
  which represent the probability density in the space of fitted structure functions
  and from which statistical estimators such CL
  intervals can are evaluated with the usual NNPDF prescription.
  The central set is defined as the average over the $N_{\rm rep}$ replicas.
  The uncertainty band associated with these replicas
  receives two contributions: the experimental uncertainties  associated
  to the input fitted neutrino data and those associated to the
  methodology such as the functional uncertainty, in both cases
  subject to the constraints
  provided by the QCD boundary conditions imposed during the fit.

  This  low-$Q$ grid can be used in the kinematic region
  \be
  \label{sec:x_q_app_lowQ}
  0.01~{\rm GeV} \le Q \le 22~{\rm GeV} \, \quad {\rm and}\quad  10^{-5}\le x \le 1 \, .
  \ee
  We note that extending this coverage could be achieved by enlarging the $(x,Q^2)$
  coverage of the QCD constraints added to the fit, but this may distort the fit quality
  by decreasing the  weight given to the experimental data.
  If values of $(x,Q^2)$ outside the region defined
  by  Eq.~(\ref{sec:x_q_app_lowQ}) are requested,
  the output will be determined by the 
 {\sc\small LHAPDF} extrapolation algorithm.

\item {\bf high-$Q$ grid.}
  This grid tabulates the direct output of the {\sc\small YADISM} calculation
  of neutrino structure functions at NLO in QCD with nNNPDF3.0 together with
  the corresponding uncertainties.
  This calculation is independent of the output of the neural network
  parametrisation used for the low-$Q$ grid, though the respective central values are matched due
  to their  use 
  of the same  QCD calculation.

  The {\sc\small LHAPDF} grid is constituted by a central set,
  $\widetilde{N}_{\rm rep}=200$ replicas that correspond to
   the nNNPDF3.0 replicas and structure functions
obtained with central factorisation
  and renormalisation scales, $\mu_F=\mu_R=1$, and 9 additional ``replicas'' computed with the
  central nNNPDF3.0 set and with nine $\mu_R$ and $\mu_F$ scale variations following the procedure
  outlined in Sect.~\ref{sec:fitting}.
  The central set in this grid is evaluated as the average of the $\widetilde{N}_{\rm rep}=200$ replicas.
  Therefore this grid is constituted by 210 members (including the central predictions).
  PDF uncertainties and MHOUs can be obtained and combined by means of the procedure
  from~\cite{NNPDF:2019ubu}, where the latter are obtained with the 9-point prescription
  and then added in quadrature with the PDF uncertainties.
  The scripts delivered with the grids illustrate how the uncertainty calculation
  is carried out.

  This  high-$Q$ grid can be used in the kinematic region
  \be
  \label{sec:x_q_app_highQ}
  2~{\rm GeV} \le Q \le 10~{\rm TeV} \, \quad {\rm and}\quad  10^{-9}\le x \le 1 \, ,
  \ee
  and again outside this region the grid output is obtained from {\sc\small LHAPDF} extrapolation.
  Restricting the {\sc\small YADISM} calculation to this region ensures the validity of
  the pQCD calculation.
  
  We note that Eqns.~(\ref{sec:x_q_app_lowQ}) and~(\ref{sec:x_q_app_highQ}) overlap
  in the region of $(x,Q^2)$ where both the data-driven and pQCD approaches can be reliably
  applied.
  In this overlap region, we recommend using the outcome of the low-$Q$ grid as baseline.

\item   {\bf Grid matching.}
  Within their respective regions of applicability, one can evaluate the neutrino structure functions
  and the associated uncertainties using the appropriate prescription.
  If we denote by $F^{\nu A}_{i,lQ} (x,Q^2)$ and $F^{\nu A}_{i,hQ} (x,Q^2)$ the outcome of the $F_i$
  structure function obtained from the
  low-$Q$ and high-$Q$ grids respectively, to evaluate the neutrino structure function
  one should use the prescription:
  \be
  \label{eq:grid_matching}
   F_i^{\nu A} (x,Q^2)=
    \begin{cases}
      F^{\nu A}_{i,lQ} (x,Q^2), & \text{if}\ Q \le Q_{\rm thr}\,\, \& \,\, x \ge 10^{-3} \, , \\[0.3cm]
      F^{\nu A}_{i,hQ} (x,Q^2), & \text{otherwise} \, ,
    \end{cases}
    \ee
 with $Q_{\rm thr}= 22$ GeV,
 and the same holds for the corresponding uncertainties.
 We emphasize that the replicas of the low-$Q$ grids can be treated as correlated among them,
 and the same holds for those of the  low-$Q$ grids, but that replicas of the low-$Q$
 grid are uncorrelated with those of the high-$Q$ grid.
 The user can vary the matching parameters in Eq.~(\ref{eq:grid_matching}), for instance by choosing
 a lower $Q^2$ threshold $Q_{\rm thr}$ to switch to the high-$Q$ grid calculation.
 
 Since the same pQCD calculation that is tabulated in  $ F^{\nu A}_{i,hQ} $ enters the data-driven fit
 of  $F^{\nu A}_{l,hQ}$ as theoretical constraint, central values obtained from the prescription of
 Eq.~(\ref{eq:grid_matching}) should match within uncertainties as one crosses
 the threshold value $Q_{\rm thr}$.
 This does not necessarily hold for the structure function uncertainties, since these are typically larger
 in the data-driven fit than in the pQCD calculation.
 If required, the user may implement the matching of
 the structure function uncertainty across the threshold $Q_{\rm thr}$  by means
 of a sliding window prescription.
  
\end{itemize}

As mentioned above, in the project website we provide
a {\tt Python} code that evaluates neutrino structure functions
and their uncertainties for any value of $x$ and $Q^2$ following
the prescription outlined above.

\paragraph{Neutrino inclusive cross sections.}
Inclusive neutrino structure functions are evaluated by integrating the output
of the structure function grids described above using Eq.~(\ref{eq:sigma_inclusive}).
As highlighted by the analysis of Sect.~\ref{sec:nnsfnu_pheno_xsec}, for neutrino energies
of $E_{\nu}\gsim 10$ TeV the contribution from the kinematic region
involving $Q \lsim 2$ GeV is negligible, and in such case one can evaluate
the cross section in terms of only the high-$Q$ structure function grid.
Conversely, for lower neutrino energies, say $E_{\nu}\lsim 100$ GeV,
the integral in Eq.~(\ref{eq:sigma_inclusive})
will be dominated by momentum transfers with $Q \lsim 20$ GeV and hence it can be evaluated
in terms of entirely the output of the low-$Q$ structure
function grid.
For intermediate values of the neutrino energy, one can combines the low- and
high-$Q$ structure function grid.

\begin{figure}[t]
 \centering
 \includegraphics[width=1.0\linewidth]{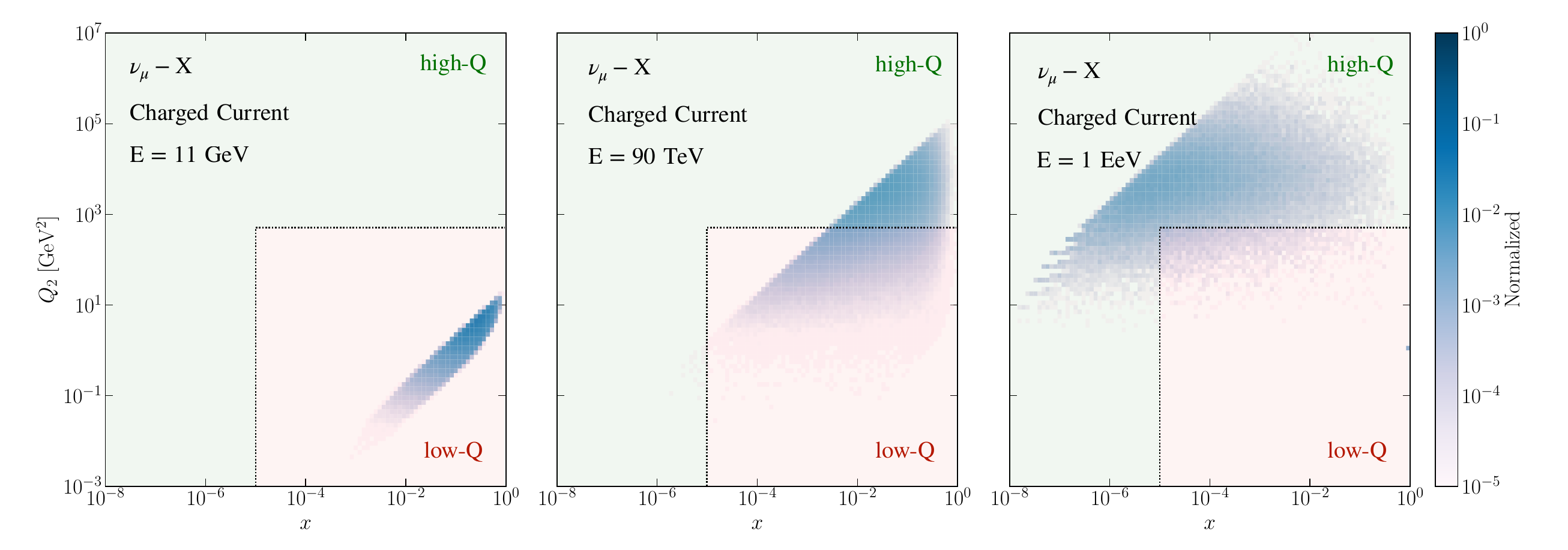}
 \caption{The coverage in the $(x,Q^2)$ plane of the inclusive neutrino cross section,
   Eq.~(\ref{eq:sigma_inclusive}),
   in muon-neutrino inelastic scattering.
   We consider 
   three values of the neutrino energy: from left to right,
   $E_{\nu}=11$ GeV, 90 TeV, and 1 EeV.
   The blue bins indicate the regions in $(x,Q^2)$ that contribute
   to the inclusive cross-section (normalised to the maximum value of the integrand), with darker bins
   dominating the integral.
   The region covered by red (green) indicates that the
   output of low-$Q$ (high-$Q$) structure function grid is being used
   to evaluate the inclusive cross section, see text for
   more details.
 }    
 \label{fig:x_Q_Enu_coverage}
\end{figure}

To further illustrate the kinematic coverage of the integrated neutrino cross-section
calculation,
Fig.~\ref{fig:x_Q_Enu_coverage}
displays
the coverage in the $(x,Q^2)$ plane of the inclusive cross section
   Eq.~(\ref{eq:sigma_inclusive})
   in muon-neutrino inelastic scattering.
   We consider 
   three values of the neutrino energy: from left to right,
   $E_{\nu}=11$ GeV, 90 TeV, and 1 EeV.
   The blue bins indicate the regions in $(x,Q^2)$ that contribute
   to the inclusive cross-section (normalised to the maximum value of the integrand), with darker bins
   dominating the integral.
   The region covered by red (green) indicates that the
   output of low-$Q$ (high-$Q$) structure function grid is being used
   to evaluate the inclusive cross section.
For $E_{\nu}=11$ GeV, the cross section is determined entirely by the output of the low-$Q$ grid.
For $E_{\nu}=90$ TeV, the output of the two grids are required, carrying
a similar weight in the calculation.
For $E_{\nu}=1$ EeV, the cross section depends only on perturbative structure functions
and the bulk of the contribution is associated to the region around $x\simeq 10^{-5}$ and $Q \simeq 100$ GeV.
From this last case, one can see how at high neutrino energies it is important to also account for the region
with $x\lsim 10^{-5}$ and $Q \lsim Q_{\rm thr}$, which is covered by the high-$Q$ grid
but not by the low-$Q$ one.

Taking into account the contribution from the relevant kinematics,
 the integration is Eq.~(\ref{eq:sigma_inclusive})
is separated into three disjoint regions
\begingroup
\allowdisplaybreaks
\begin{align}
	\label{eq:sigma_inclusive_separated} \nonumber
	\sigma^{\nu N}(E_\nu) = \int_{0}^{Q^2_{\rm max}} {\rm d}Q^2
	\lc  \int_{x_0(Q^2)}^1{\rm d}x \frac{{\rm d}^2\sigma^{\nu N}}{{\rm d}x{\rm d}Q^2}(x,Q^2,y)\rc
	=\sigma^{\nu N}_{lQ}(E_\nu) + \sigma^{\nu N}_{hQ}(E_\nu) \equiv  \qquad  \\
	\int_{0}^{Q^2_{\rm thr}} {\rm d}Q^2
	\lc  \int_{{\rm max}\lp x_0(Q^2),10^{-5}\rp}^1{\rm d}x \frac{{\rm d}^2\sigma_{lQ}^{\nu N}}{{\rm d}x{\rm d}Q^2}(x,Q^2,y)\rc
	+ \int_{Q^2_{\rm thr}}^{ Q^2_{\rm max} } {\rm d}Q^2
	\lc  \int_{x_0(Q^2)}^1{\rm d}x \frac{{\rm d}^2\sigma_{hQ}^{\nu N}}{{\rm d}x{\rm d}Q^2}(x,Q^2,y)\rc \\
	+ \int_{Q^2_{\rm min}}^{ Q^2_{\rm thr} } {\rm d}Q^2
	\lc  \int_{{\rm min} \lp x_0(Q^2),10^{-5}\rp }^1{\rm d}x \frac{{\rm d}^2\sigma_{hQ}^{\nu N}}{{\rm d}x{\rm d}Q^2}(x,Q^2,y)\rc \, ,
	\qquad \qquad \qquad \nonumber
\end{align}
\endgroup
separated by the threshold $Q^2_{\rm thr}$, 
in terms of structure functions computed with low- and high-$Q$ grids,
in analogy with Eq.~(\ref{eq:grid_matching}), and with $Q^2_{\rm min}=1.65$ GeV to ensure
that the high-$Q$ grid is not extrapolated to the non-perturbative region.
Depending on the $E_{\nu}$ value, one we can simplify Eq.~(\ref{eq:sigma_inclusive_separated}) to
\be
\label{eq:grid_matching_xsec}
  \sigma^{\nu N}(E_\nu)=
    \begin{cases}
     \simeq \sigma^{\nu N}_{lQ}(E_\nu)   & \text{if}\,  E_{\nu}\lsim 1~{\rm TeV} \, , \\[0.3cm]
     \simeq \sigma^{\nu N}_{hQ}(E_\nu)   & \text{if}\,  E_{\nu}\gsim 100~{\rm PeV} \, , \\[0.3cm]
     \sigma^{\nu N}_{lQ}(E_\nu)+\sigma^{\nu N}_{hQ}(E_\nu) \quad   & \text{otherwise} \, , 
    \end{cases}
    \ee
    since at low and high energies only the low-$Q$ and high-$Q$ structure function
    grids are required respectively, with the other term leading
    to a negligible contribution.
    We note that, as also indicated by  Fig.~\ref{fig:x_Q_Enu_coverage}, at high neutrino energies
    the high-$Q$ grid is also used for $Q \lsim Q_{\rm thr}$ to determine the contribution
    for $x\le 10^{-5}$ which is not covered by the low-$Q$ grid.

    Given the different statistical interpretation of the uncertainties
    associated to the low- and high-$Q$ structure function grids,
    the total uncertainty in Eq.~(\ref{eq:sigma_inclusive_separated}) is evaluated
    by adding in quadrature the uncertainties associated to the
    low- and high-$Q$ contributions, namely
    \be
    \label{eq:sigma_inclusive_separated_uncertainties}
\delta\sigma^{\nu N}(E_\nu)= \lp \lp \delta \sigma^{\nu N}_{lQ}(E_\nu)\rp^2 + \lp \delta \sigma^{\nu N}_{hQ}(E_\nu)\rp^2\rp^{1/2} \, ,
\ee
with $\delta \sigma^{\nu N}_{lQ}$ and $\delta \sigma^{\nu N}_{hQ}$ evaluated using
the uncertainty prescription for the low- and high-$Q$ structure function grids
respectively.
As opposed to the structure function case, the prescriptions
Eq.~(\ref{eq:sigma_inclusive_separated}), for central values,
and~(\ref{eq:sigma_inclusive_separated_uncertainties}), for the total uncertainties, lead
to smooth predictions for neutrino energies $E_{\nu}$ sensitive both
to low- and high-$Q$ structure functions.
Furthermore, the outcome of the cross-section calculation is stable upon moderate variations
of the threshold value $Q^2_{\rm thr}$.

\begin{figure}[t]
 \centering
 \includegraphics[width=1.00\linewidth]{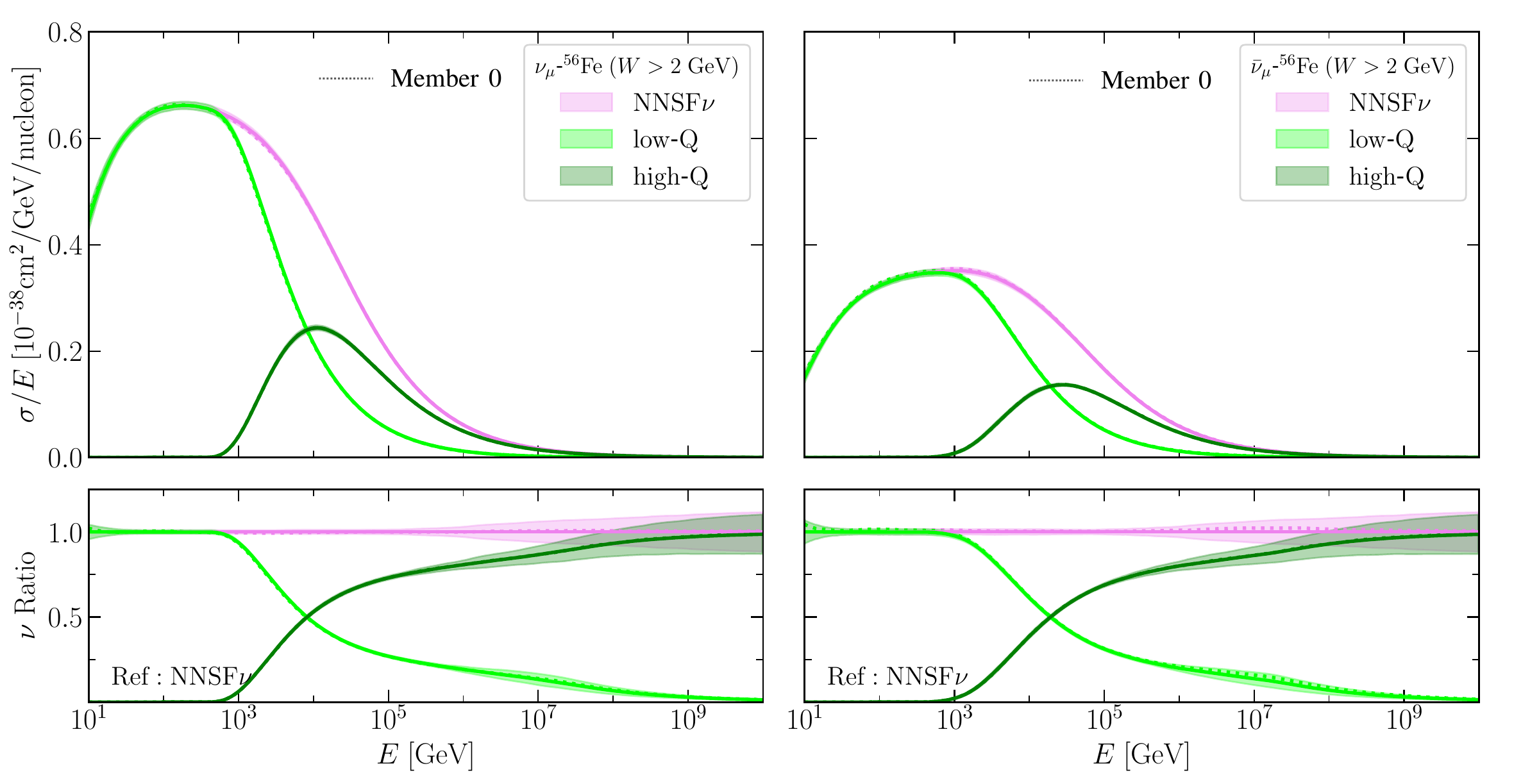}
 \caption{The NNSF$\nu$ predictions for
    inclusive cross sections for  neutrino (left)
    and antineutrino (right panel) scattering off an iron target.
    We indicate the individual contributions from the low-$Q$ and high-$Q$ structure function grids,
   denoted by $\sigma^{\nu N}_{lQ}(E_\nu)$ and $\sigma^{\nu N}_{hQ}(E_\nu)$ respectively in
   Eq.~(\ref{eq:sigma_inclusive_separated}), as well as their sum that defines
   the NNSF$\nu$ prediction.
   The NNSF$\nu$ uncertainty is evaluated by adding in quadrature the
   (uncorrelated) contribution from the two grids
   as indicated by Eq.~(\ref{eq:sigma_inclusive_separated_uncertainties}).
   The bottom panels display the results normalised to the central
   value of NNSF$\nu$.
 }    
 \label{fig:app-regions-inclxsec}
\end{figure}

The matching procedure described in this appendix to evaluate
inclusive neutrino cross sections 
in terms of the NNSF$\nu$ structure functions by means of the
prescriptions of Eq.~(\ref{eq:sigma_inclusive_separated}), for central values,
and Eq.~(\ref{eq:sigma_inclusive_separated_uncertainties}), for the total uncertainties,
is illustrated in Fig.~\ref{fig:app-regions-inclxsec}.
The NNSF$\nu$ predictions of the
    inclusive cross sections for  neutrino 
    and antineutrino  scattering off an iron target
    are compared to the corresponding individual contributions from the low-$Q$ and high-$Q$ structure function grids,
   denoted by $\sigma^{\nu N}_{lQ}(E_\nu)$ and $\sigma^{\nu N}_{hQ}(E_\nu)$ respectively in
   Eq.~(\ref{eq:sigma_inclusive_separated}).
For low ($E_{\nu}\lsim 1$ TeV) and high ($E_{\nu}\gsim 100$ PeV) values
of the neutrino energy, the calculation is fully dominated by the
contributions from the low- and high-$Q$ structure function grids
respectively, as indicated by Eq.~(\ref{eq:grid_matching_xsec}), while
for intermediate $E_{\nu}$ values between 1 TeV and 100 PeV
one must account for both contributions.
It is worthwhile noting that the
contribution from the low-$Q$ grid remains relevant up to rather high neutrino energies.
The smooth dependence in $E_\nu$ both for central values and uncertainties
further validates the NNSF$\nu$ matching which connects the various
energy regions.

\newcommand{\yadism}{{\sc\small YADISM}\xspace}
\renewcommand*{\thefootnote}{\arabic{footnote}}

\section{\yadism: DIS structure functions made easy}
\label{app:yadism}

Deep-inelastic structure functions can be evaluated with several
public codes such as {\sc\small APFEL}~\cite{Bertone:2013vaa} and
{\sc\small QCDNUM}~\cite{Botje:2010ay}.
Available DIS codes differ in the accuracy with which structure functions can be computed,
whether they are based on the $x$-space or the $N$-space formalism,
the treatment of heavy quark mass effects and of target mass corrections, the availability
of polarised and time-like coefficient functions, and
the possible inclusion of QED corrections among other considerations.

This appendix introduces \yadism, a new framework
for the evaluation of DIS structure functions from the same family
as the {\sc\small EKO}~\cite{Candido:2022tld} DGLAP evolution code.
The open source \yadism code can be obtained from its GitHub repository
\begin{center}
  \url{https://github.com/NNPDF/yadism/}
  \end{center}  
together with an detailed documentation, tutorials, and user-friendly examples
\begin{center}
\url{https://yadism.readthedocs.io/}
\end{center}  
One of the main advantages of \yadism is that it is integrated
with the fast interpolation grid toolbox {\sc\small PineAPPL}~\cite{Carrazza:2020gss},
and hence DIS structure functions can be treated on the same footing as hadronic observables
from the point of view of PDF fitting and related applications.
{\sc\small PineAPPL} provides a unique grid format, with application programming interfaces (APIs)
for different programming languages
and a command-line interface to manage the grid files.
Furthermore, \yadism implements the available N$^3$LO DIS coefficient functions,
which combined with (approximate) N$^3$LO evolution
and heavy quark matching conditions available in {\sc\small EKO} provide
theoretical calculations required to carry out a N$^3$LO PDF determination.
\yadism will be described in an upcoming publication~\cite{yadism},
here we summarise its main features, in particular those
relevant to the present study, and highlight benchmarking
studies carried out.

\paragraph{Grid formalism.}
As indicated by Eq.~(\ref{eq:sfs_pqcd}), in the perturbative
regime DIS structure functions are given by the factorised
convolution of process-dependent partonic scattering cross sections and
of process-independent parton distribution functions,
\be
\label{eq:sfs_pqcd_app}
F_i(x,Q^2) = \sum_{j}\int_x^1 \frac{dz}{z}\, C_{i,j}(z,\alpha_s(Q^2))f_j\lp \frac{x}{z},Q^2\rp \equiv
C_{j; i} \otimes f_j\, ,
 \ee
 where $j$ is an index that runs over all possible partonic initial states
 and $C_{i,j}$ is the process-dependent, but target-independent, coefficient function,
 given by an expansion in the QCD coupling $\alpha_s(Q^2)$.
 In the third term of Eq.~(\ref{eq:sfs_pqcd_app}) and in the following, sum over repeated indices is
 implicit.
 
As standard for fast interpolation techniques developed
in the context of PDF fits~\cite{Carli:2010rw,Carrazza:2020gss,Wobisch:2011ij,Bertone:2014zva},
the PDFs can be expanded
over an interpolation basis \be
f_j(\xi) = \sum_\alpha p_\alpha(\xi) f(\xi_\alpha) \equiv p_\alpha(\xi) f_\alpha \, ,
\qquad \xi = \frac x z \, ,
\ee
with $p_\alpha(x)$ some suitable polynomial basis.
This way the convolution in Eq.~(\ref{eq:sfs_pqcd_app}) can be replaced by a simple contraction
\be
\label{app:grid_formalism}
F_i = C_{j; i} \otimes f_j = C_{j \alpha; i} \cdot f_\alpha \, ,\qquad
C_{j \alpha; i} = C_{j; i} \otimes p_\alpha \, ,
\ee
in terms of PDFs evaluated at fixed grid points $\xi_\alpha$ and precomputed
coefficients $C_{j \alpha; i}$.
In \yadism the polynomial interpolation basis is  provided by
the {\sc\small EKO} modules.
The same grid structure can be generalised to accommodate extensions of
the basic structure function calculation in 
Eq.~(\ref{eq:sfs_pqcd_app}) such as heavy quark mass effects,
renormalisation and factorisation
scale variations~\cite{NNPDF:2019vjt,NNPDF:2019ubu},
and target mass corrections, among other effects.
Isospin modifications, required to evaluate the neutron, deuteron, or heavy nuclear
structure functions, can be accounted for either at the coefficient function
level or at the input PDF level.

The grid formalism summarised schematically in Eq.~(\ref{app:grid_formalism})
requires as input the corresponding DIS coefficient functions.
Table~\ref{tab:coefffuncs} provides an overview of the different types and accuracy of the DIS coefficient
functions currently implemented in \yadism.
For each perturbative order (NLO, NLO, and N3LO)
we indicate  the neutral-current
and charged-current light-to-light (``light''), light-to-heavy (``heavy''), heavy-to-light
and heavy-to-heavy (``intrinsic'') and ``asymptotic'' ($Q^2 \gg m_h^2$ limit)  coefficients functions
which have been implemented and benchmarked.
The NNLO heavy quark coefficient functions for CC scattering are currently available in a $K$-factor format,
and their implementation into the \yadism grid formalism is work in progress.
We note that the full calculation of the N$^3$LO NC heavy
coefficient functions is not available, but that an approximated expression can
be constructed from known partial results~\cite{niccolo,Kawamura:2012cr}.
Heavy quark structure functions can be evaluated in the FONLL
general-mass variable flavour number scheme (GM-VFN)~\cite{Forte:2010ta},
as well as in the fixed-flavour number (FFN) and zero-mass variable-flavour
number (ZM-VFN) schemes.
We point out that the list in Table~\ref{tab:coefffuncs} is going to be updated
as new features are added, and therefore the interested user is encouraged
to consult the online documentation for an up-to-date states
of available coefficient functions.

\newcommand{\grcell}{\cellcolor{green!25}}
\newcommand{\grokcell}{\cellcolor{green!25}\checkmark}
\newcommand{\blcell}{\cellcolor{blue!25}}
\newcommand{\ylcell}{\cellcolor{yellow!25}}
\newcommand{\rdcell}{\cellcolor{red!25}}
\newcommand{\rdxcell}{\cellcolor{red!25}\ding{55}}
\newcommand{\ftm}[1]{\footnotemark[#1]}

\renewcommand{\thefootnote}{\fnsymbol{footnote}}
\newcounter{numfootnote}
\setcounter{numfootnote}{\value{footnote}}
\setcounter{footnote}{0}

\newcounter{mysym}
\newcommand{\fnsym}[1]{\setcounter{mysym}{#1}$^{\fnsymbol{mysym}}$}

\begin{table}
  \label{tab:coefffuncs}
  \centering
   \renewcommand{\arraystretch}{1.60}
   \begin{tabular}{c | c c c c}
     \toprule
    $\qquad$ NLO$\qquad$ & $\qquad$light$\qquad$ & $\qquad$heavy$\qquad$ & $\qquad$intrinsic$\qquad$ &$\qquad$ asymptotic $\qquad$ \\
    \hline
    NC & \grokcell & \grokcell & \grokcell & \grokcell\\
    CC & \grokcell & \grokcell & \grokcell & \grokcell\\
    \midrule
    NNLO & & &\\
    \hline
    NC & \grokcell & \grokcell & \rdxcell & \grokcell\\
    CC & \grokcell & \ylcell tabulated\ftm{1} & \rdxcell & \grokcell\\
    \midrule
    N3LO & & &\\
    \hline
    NC & \grokcell &  \rdxcell\ftm{2} & \rdxcell & \rdxcell\ftm{3} \\
    CC & \grokcell &  \rdxcell\ftm{2} & \rdxcell & \rdxcell \\
    \bottomrule
   \end{tabular}
  {
    \footnotesize
    \renewcommand{\arraystretch}{1.2}
    \begin{tabular}{r l}
      \fnsym{1} & Already available as $K$-factors~\cite{Gao:2017kkx}, now being integrated in the grid format.\\
      \fnsym{2} & Full calculation not available but an approximated expression can
      be constructed from partial results~\cite{niccolo,Kawamura:2012cr}.\\
      \fnsym{3} & Calculation available, to be implemented.
    \end{tabular}
  }
  \vspace{0.2cm}
  \caption{Overview of the different types and accuracy of the DIS coefficient
    functions currently implemented in \yadism. For each perturbative order (NLO, NLO, and N3LO)
    we indicate  the light-to-light (``light''), light-to-heavy (``heavy''), heavy-to-light
    and heavy-to-heavy (``intrinsic'') and ``asymptotic'' ($Q^2 \gg m_h^2$ limit) coefficients functions
    which have been implemented and benchmarked.
    The NNLO heavy quark coefficient functions for CC scattering are available in $K$-factor format
    and are being implemented into the \yadism grid formalism.  \label{tab:coefffuncs}
  }
\end{table}

\renewcommand*{\thefootnote}{\arabic{footnote}}
\setcounter{footnote}{\value{numfootnote}}


\paragraph{Scale variations.}
As done by other public DIS tools, \yadism also provides
the option of varying the renormalisation and factorisation
scales in the calculation.
The code follows
the definitions of scale variations from~\cite{vanNeerven:2000uj,vanNeerven:2001pe},
which are consistent with the broader picture of scale
variations relevant for PDF fits from~\cite{NNPDF:2019ubu} where they also affect the
DGLAP evolution.
There are two kinds of scale variations: renormalization scale $Q_R$ dependence, related to
the ultraviolet renormalization scheme, and factorization scale $Q_F$ dependence, 
related to the subtraction of collinear logarithms in the adopted factorization scheme.
The factorization scale $Q_F$ sets  the boundary between
the coefficient functions and the DGLAP-evolved PDFs.
Scale variations at a given perturbative order can be constructed from combining
ingredients already present at the previous perturbative order, and hence
for this reason they
represent a suitable predictor of potentially unknown missing higher orders.
Within \yadism, the scale variation contributions to the DIS structure
functions are stored in separate grids such that the values of the scale ratios
$\mu_F^2=Q_F^2/Q^2$ and $\mu_R^2=Q_R^2/Q^2$ can be evaluated a posteriori.

As described  in Sect.~\ref{sec:fitting}, in the NNSF$\nu$ analysis the \yadism
structure functions enter the fit to constrain the
neural network parametrisation in Region II, with an error
function defined in terms of a theory covariance matrix.
This covariance matrix accounts both for the PDF and MHO theory uncertainties, the latter
evaluated from scale variations using the 9-point prescription.
The calculation of scale variations provided by \yadism and the subsequent
determination of the MHOU theory covariance matrix has been benchmarked
with the results of~\cite{NNPDF:2019ubu}.

\paragraph{Benchmarking with {\sc\small APFEL}.}
The calculations of DIS structure functions
and reduced cross-sections provided by \yadism
have been thoroughly benchmarked  with those provided by {\sc\small
  APFEL} and {\sc\small QCDNUM}.
Specifically, we have verified that \yadism  reproduces
the {\sc\small APFEL} predictions for those of the DIS coefficient functions listed
in Table~\ref{tab:coefffuncs} which are also available in the latter.
Excellent agreement is found in all cases considered, with some residual
differences understood as will be discussed in more detail in~\cite{yadism}.

\begin{figure}[!t]
 \centering
 \includegraphics[width=\textwidth]{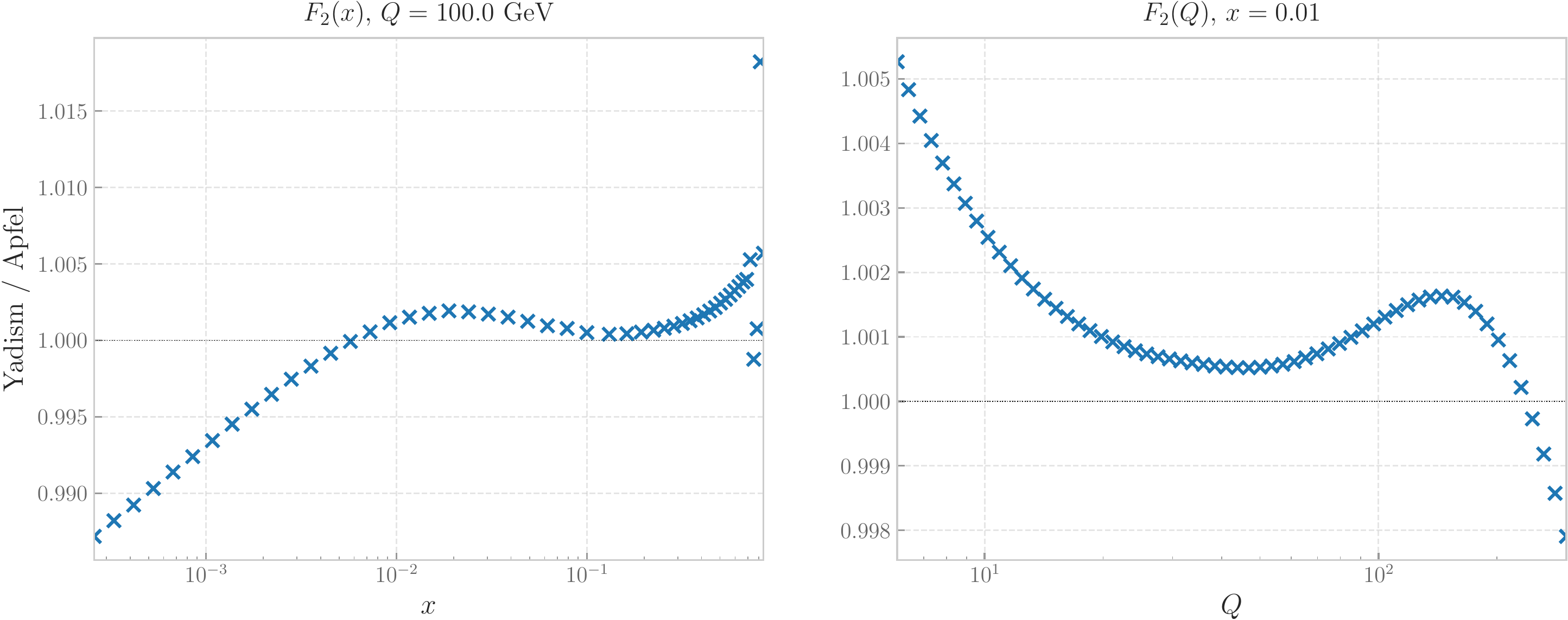}
 \includegraphics[width=\textwidth]{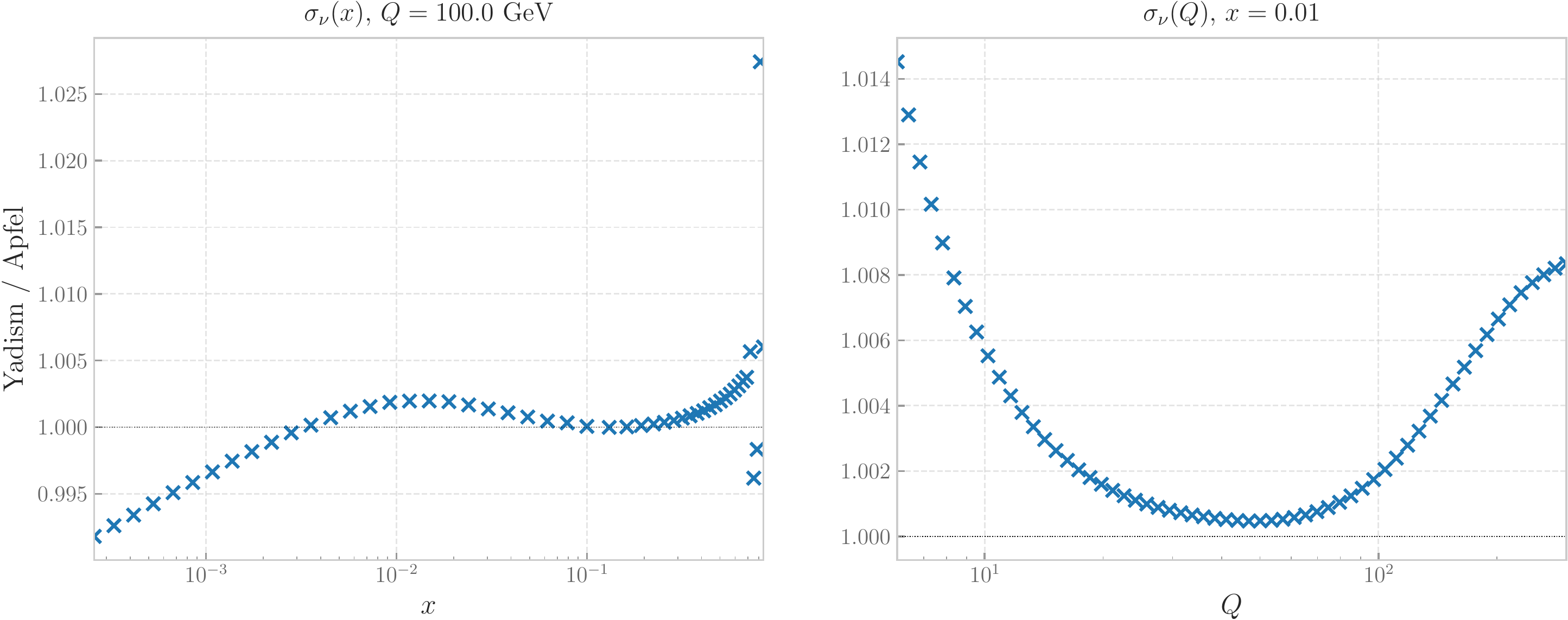}
 \caption{The ratio between the {\sc\small YADISM}  and {\sc\small APFEL}
   calculations of the neutrino-initiated structure function $F_2^{\nu A}(x,Q^2)$
   (top) and of the corresponding double-differential cross-sections (bottom panels)
   at NNLO
   for the same choice of input PDFs and theory settings.
   The benchmark comparison is presented both for fixed $Q=100$ GeV (left)
   and for fixed $x=0.01$ (right panels).
   A similar level of agreement is obtained for other DIS observables
   as well as for other regions of $(x,Q^2)$ relevant for phenomenology.
   }
 \label{fig:benchmark-apfel-yadism}
\end{figure}

To illustrate this agreement,
Fig.~\ref{fig:benchmark-apfel-yadism} displays
the ratio between the {\sc\small YADISM}  and {\sc\small APFEL}
   calculations of the neutrino-initiated structure function $F_2^{\nu A}(x,Q^2)$
    and of the corresponding double-differential cross-sections 
   at NNLO on a proton target
   for the same choice of input PDFs and theory settings.
   Specifically, we use in both cases the central replica
   of  NNPDF4.0 NNLO with $\alpha_s(m_Z)=0.118$
and FONLL-C for heavy quark mass effects.
    The benchmark comparison is presented both for fixed $Q=100$ GeV 
   and for fixed $x=0.01$.
   The benchmark shows agreement at the \textperthousand \ level between the two calculations
   for the whole kinematic region in $(x,Q^2)$ relevant for this study.
A similar level of agreement is obtained for other DIS observables.

\section{Nuclear effects in neutrino scattering}
\label{app:nuclear}

As compared to baseline predictions on a proton target, there
are two effects that modify the neutrino scattering rates
on nuclear targets.
The first is related to the different content of up and down quarks, since
for an isoscalar nuclear target the average nucleon has on 25\% less (more)
up (down) valence quarks than in a proton target, an effect which is
more marked for  non-isoscalar nuclei where  $(A/2)>Z$.
The second is associated to the modifications of the bound nucleon
structure that take place in heavy nuclei when compared with  their free-nucleon counterparts,
and in the DIS region is quantified by the nuclear PDFs.
In this appendix we illustrate the impact of these two effects
on neutrino inelastic structure functions.

For a general nuclear target, the $F_2$ structure functions per nucleon  at LO in terms of nPDFs are
\bea
 F_2^{\nu A}(x,Q^2) &=& 2x\lp f^{(A)}_{\bar{u}} + f^{(A)}_{d} + f^{(A)}_{s} + f^{(A)}_{\bar{c}} \rp(x,Q^2) \, , \nonumber  \\
 F_2^{\bar{\nu} A}(x,Q^2) &=& 2x\lp f^{(A)}_u + f^{(A)}_{\bar{d}} + f^{(A)}_{\bar{s}} + f^{(A)}_c \rp(x,Q^2) \, ,
 \label{app:npdfs_1}
\eea
with similar expressions for $xF_3^{\nu A}$, and we assume that $Q^2$ is large enough so that
QCD factorisation holds.
For an isoscalar nucleus ($Z=A/2$) and neglecting nPDF effects, we can write
\bea
 F_2^{\nu A}(x,Q^2) &=& 2x\lp \lp f^{(p)}_{\bar{u}} + f^{(p)}_{d}  + f^{(p)}_{\bar{d}} + f^{(p)}_{u}\rp/2 + f^{(p)}_{s} + f^{(p)}_{\bar{c}} \rp(x,Q^2) \, , \nonumber  \\
 F_2^{\bar{\nu} A}(x,Q^2) &=& 2x\lp \lp f^{(p)}_u + f^{(p)}_{\bar{d}}+ f^{(p)}_d + f^{(p)}_{\bar{u}}\rp/2 + f^{(p)}_{\bar{s}} + f^{(p)}_c \rp(x,Q^2) \simeq F_2^{\nu A}(x,Q^2) \, ,
 \label{app:npdfs_2}
\eea
in terms of the free proton PDFs, and there the equality in the second equation holds neglecting the small
charm and strange asymmetries.
Comparing Eqns.~(\ref{app:npdfs_1}) with~(\ref{eq:neutrinoSFs}),
in the valence region 
 $F_2^{\nu A}  > F_2^{\nu p} $ and conversely for antineutrino scattering
as a consequence of $f^{(p)}_{u}>f^{(p)}_{d}$.
These differences cancel out when taking the sum of neutrino and antineutrino
structure functions
\be
\label{eq:nuclear_ratios_sum}
 F_2^{(\nu+\bar{\nu}) A}(x,Q^2) = 2x\lp f^{(p)}_{\bar{u}} + f^{(p)}_{d}  + f^{(p)}_{\bar{d}} + f^{(p)}_{u} + f^{(p)}_{s} + f^{(p)}_{\bar{c}}+f^{(p)}_{\bar{s}} + f^{(p)}_c  \rp = F_2^{(\nu+\bar{\nu}) p}(x,Q^2) \, ,
\ee
which coincides with the result for scattering on a free proton target.

Once isospin effects are corrected for, neutrino structure functions on nuclear targets
can still differ from those associated to free nucleons due to genuine nuclear modifications.
These are typically quantified by nuclear modification ratios of the form
\be
\label{eq:nuclear_ratios}
R_{F_i}^{(A)}(x,Q^2) = \frac{A F_i^{\nu A}(x,Q^2)}{\lp Z F_i^{\nu p}(x,Q^2) + (A-Z)F_i^{\nu n}(x,Q^2)  \rp} \, ,
\ee
which are equal to unity if the bound-nucleon structure functions equal the free-nucleon ones.
In the DIS region we can express these nuclear modification ratios in terms of the nPDFs
\be
\label{eq:nuclear_ratios_2}
R_{f_q}^{(A)}(x,Q^2) = \frac{A f_q^{(N/A)}(x,Q^2)}{\lp Zf_q^{(p)}(x,Q^2) + (A-Z)f_q^{(n)}(x,Q^2)  \rp } \, ,
\ee
where $f^{(N/A)}$ indicates the PDFs of the average nucleon $N$ bound within a nuclei
with $Z$ protons and $(A-Z)$ neutrons.
Alternatively one can define
\be
\label{eq:nuclear_ratios_v2}
\widetilde{R}_{f_q}^{(A)}(x,Q^2) = \frac{f_{q}^{(N/A)}(x,Q^2)}{ f_q^{(p)}(x,Q^2)} \, ,\qquad
\widetilde{R}_{F_i}^{(A)}(x,Q^2) = \frac{F_{i}^{\nu A}(x,Q^2)}{ F_i^{\nu p }(x,Q^2)} \, ,
\ee
which differ from unity due to both bound-nucleon modifications and due to the different
valence quark content between the numerator and the denominator.

\begin{figure}[t]
 \centering
 \includegraphics[width=0.99\linewidth]{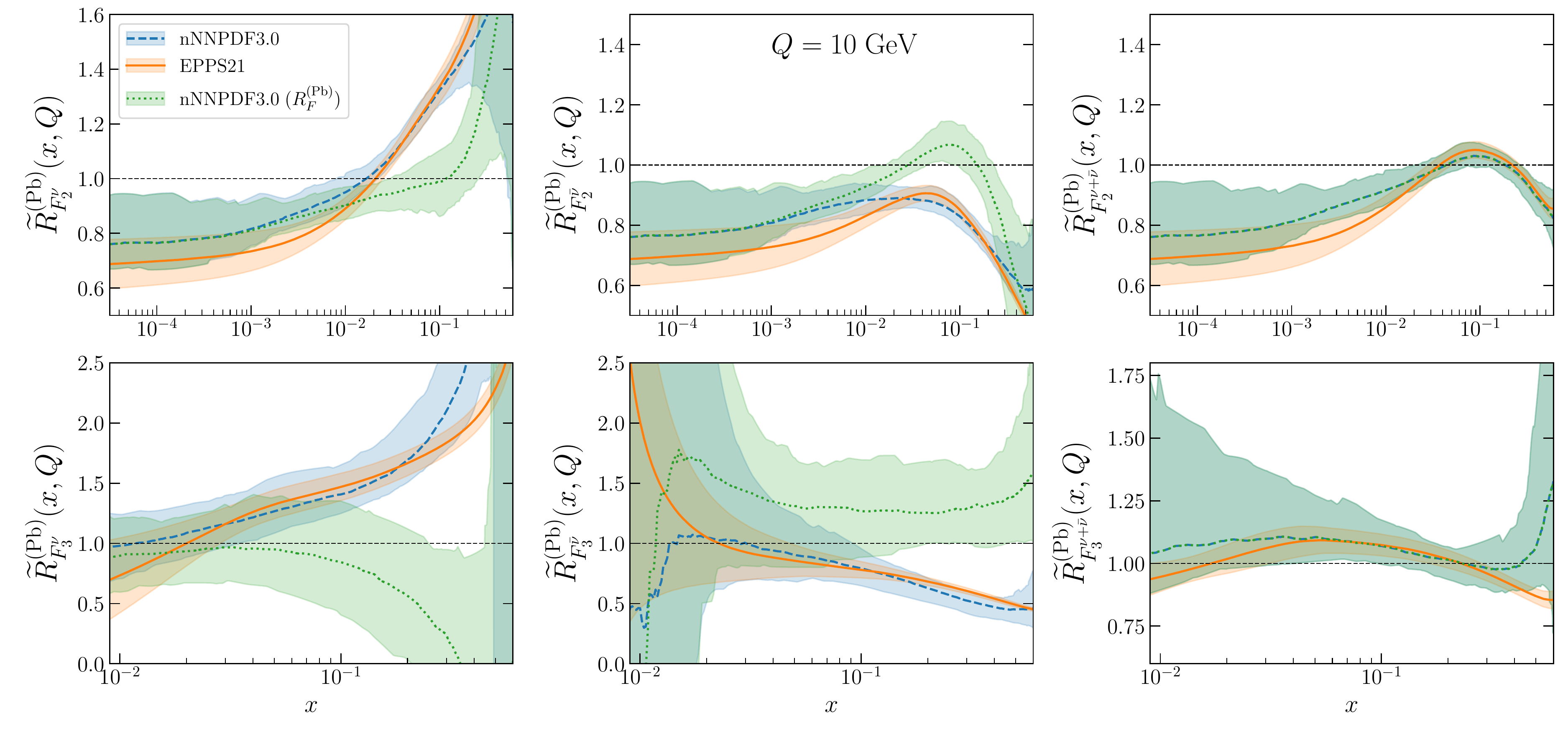}
 \caption{The nuclear structure function ratio
 Eq.~(\ref{eq:nuclear_ratios_v2}) for $F_2$ (upper) and $xF_3$ (lower panels) evaluated
 at LO on a lead target for nNNPDF3.0 and EPPS21 at $Q=10$ GeV.
  We display separately the ratios for the neutrino and
  antineutrino scattering, as well as for their sum. 
In the case of nNNPDF3.0, we also display the nuclear modification ratio
defined as in Eq.~(\ref{eq:nuclear_ratios})
where isospin effects are subtracted.
The bands indicate the 90\% CL
   intervals.
 }    
 \label{fig:nuclearcorr-neutrinoSFs_q10gev}
\end{figure}

Fig.~\ref{fig:nuclearcorr-neutrinoSFs_q10gev} displays the nuclear structure function ratio
 Eq.~(\ref{eq:nuclear_ratios_v2}) for $F_2$  and $xF_3$ at $Q=10$ GeV,
  where we show separately the ratios for the neutrino and
  antineutrino scattering as well as for their sum.
  Results are shown for two recent global nPDF determinations,
  nNNPDF3.0 (baseline in this work) and EPPS21~\cite{Eskola:2021nhw},
  which are in good agreement within uncertainties.
In the case of nNNPDF3.0, we also display the nuclear modification ratio
defined as in Eq.~(\ref{eq:nuclear_ratios})
where isospin effects are subtracted.
The bands indicate the 90\% CL
   intervals evaluated using the corresponding prescription for each nPDF set.

As compared to a proton target, in the valence region ($x\gsim 0.01$) one finds that the
$F_2$ and $xF_3$ structure functions in lead are enhanced (suppressed) for
neutrino (antineutrino scattering), mostly as a consequence of the increased content
of down quarks in lead as compared to the proton.
These differences cancel out in the sum of neutrino and antineutrino structure functions
due to Eq.~(\ref{eq:nuclear_ratios_sum}), and there the ratio Eq.~(\ref{eq:nuclear_ratios_v2})
reduces to Eq.~(\ref{eq:nuclear_ratios}), namely the genuine nuclear modifications
of bound nucleons and compared to free nucleons.
Bound-nucleon modification ratios, Eq.~(\ref{eq:nuclear_ratios}), are small
in the valence peak region, except at rather large-$x$ where structure functions
are suppressed, and become more significant for $F_2$ in the sea quark region,
$x \lsim 0.01$, leading to a suppression of up to 25\% as compared to the free-nucleon baseline.
Therefore,  from Fig.~\ref{fig:nuclearcorr-neutrinoSFs_q10gev} one concludes that in the valence region
the dominant nuclear effects are the isospin-related ones, while in the sea region instead
bound-nucleon modifications are important and need to be accounted for.
These nuclear effects propagate to the inclusive cross section via Eq.~(\ref{eq:sigma_inclusive}),
e.g. for neutrino scattering on lead one has an enhancement as compared to proton targets
at low and intermediate energies and a suppression at high and ultra-high energies.

\begin{figure}[ht]
 \centering
 \includegraphics[width=0.99\linewidth]{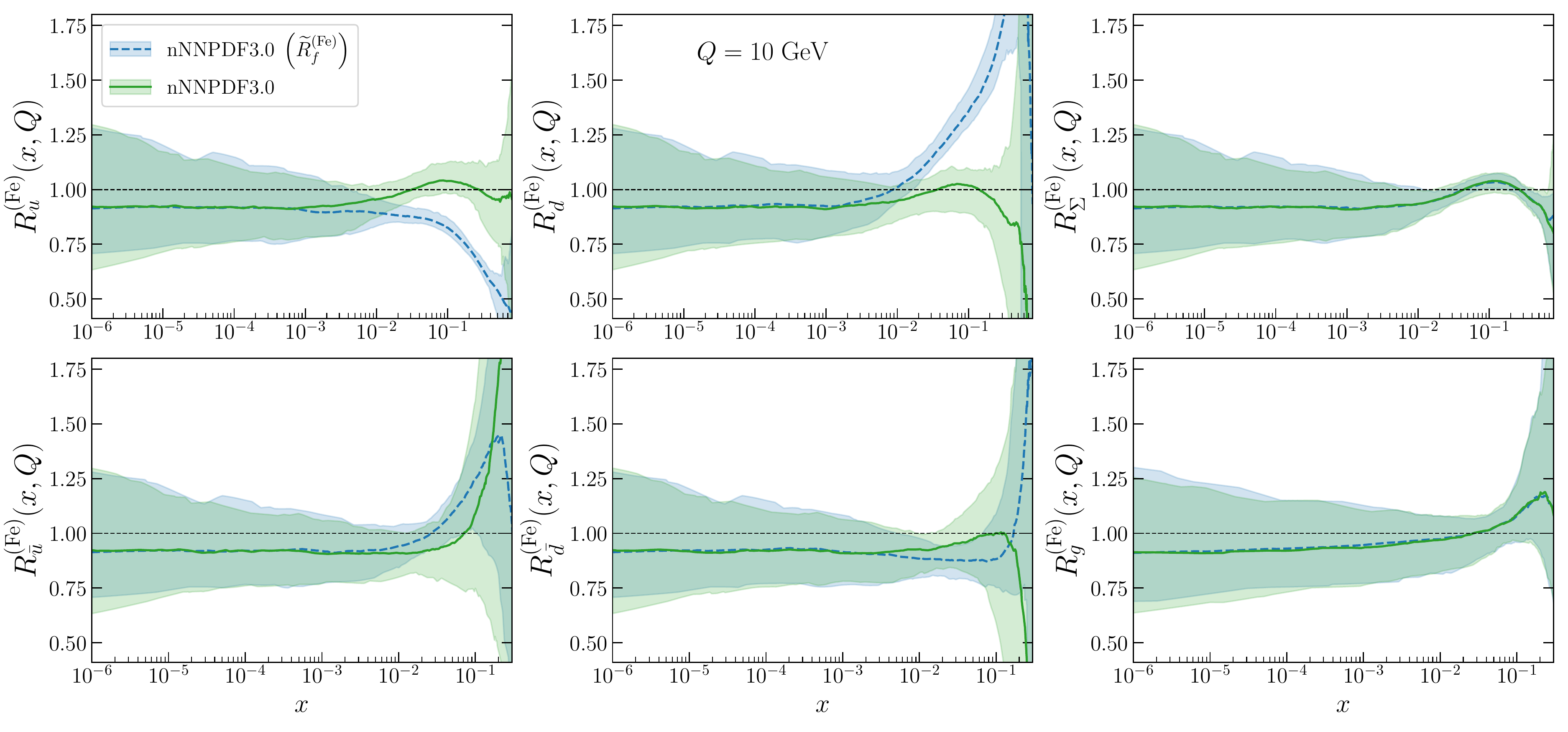}
 \includegraphics[width=0.99\linewidth]{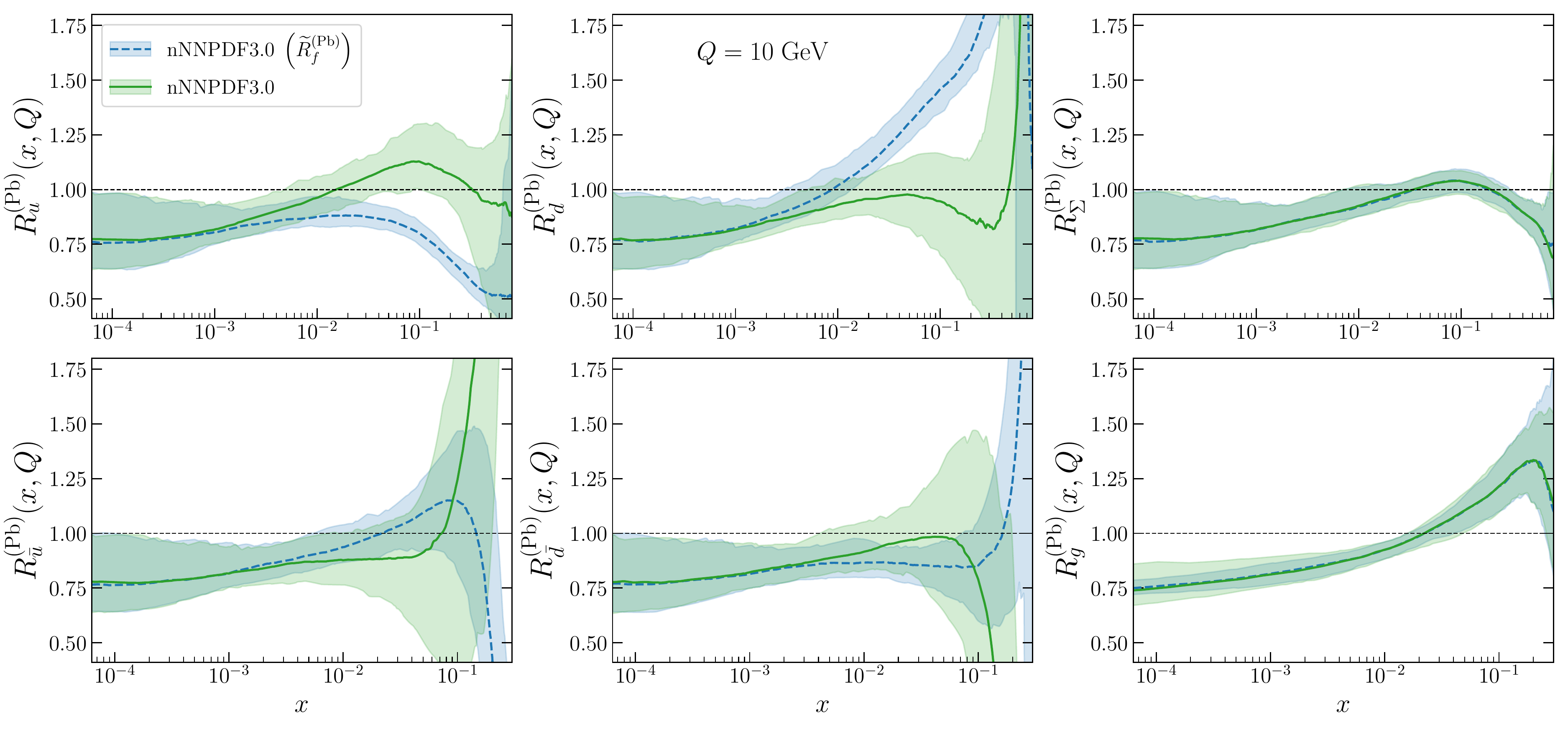}
 \caption{The nuclear modification ratios in nNNPDF3.0 at $Q=10$ GeV for iron (upper)
 and lead (lower panels) targets.
 The bands indicate the 90\% CL intervals.
 From left to right and from top to bottom we show the nuclear ratios
corresponding to the up, down, singlet, antiup, antidown quark and the gluon nPDFs.
 }    
 \label{fig:nuclearcorr-neutrinoPDFs_q10gev}
\end{figure}

For completeness, we display in Fig.~\ref{fig:nuclearcorr-neutrinoPDFs_q10gev}
the corresponding nuclear modification ratios at the nPDF level
in the case of the nNNPDF3.0 determination for both iron (Fe) and lead (Pb) targets.
We compare the two definitions, Eqns.~(\ref{eq:nuclear_ratios_2})
 and~(\ref{eq:nuclear_ratios_v2}), differing in that isospin effects are accounted
 for in the former but not in the latter.
 As in Fig.~\ref{fig:nuclearcorr-neutrinoSFs_q10gev}, the bands indicate the 90\% CL intervals.
 In the valence region, one observes the expected suppression (enhancement) for up (down)
 quark nPDFs when using the ratio defined in Eq.~(\ref{eq:nuclear_ratios_v2}).
 For isoscalar nPDF combinations such as the total quark singlet $\Sigma$
 and the gluon the two nuclear ratio definitions coincide. 
From Fig.~\ref{fig:nuclearcorr-neutrinoPDFs_q10gev} we observe the quark shadowing
in the sea region also reported by the structure function ratios in the case of lead target,
while for iron targets in this region the nuclear ratios are consistent with unity
within uncertainties.

\clearpage
\providecommand{\href}[2]{#2}\begingroup\raggedright\endgroup

\end{document}